\documentclass[twocolumn]{aa}
\usepackage{txfonts}
\usepackage{graphicx}
\usepackage{natbib}

\newcommand{\excs}{\extracolsep{\fill}} 
\newcommand{\mjup}{~M$_{\rm{Jup}}$ }
\begin{document}
   \title{Probing long-period companions to planetary hosts}
   \subtitle{VLT and CFHT near infrared coronographic imaging surveys}

\author{
        G. Chauvin\inst{1}
	\and A.-M. Lagrange\inst{2}
	\and S. Udry \inst{3}  
	\and T. Fusco\inst{4}
	\and F. Galland\inst{2}
	\and D. Naef\inst{1}
	\and J.-L. Beuzit\inst{2}
	\and M. Mayor \inst{3}}

 \offprints{Ga\"el Chauvin \email{gchauvin@eso.org}}
\institute{
$^{1}$European Southern Observatory, Casilla 19001, Santiago 19, Chile\\
$^{2}$Laboratoire d'Astrophysique, Observatoire de Grenoble, 414, Rue de la piscine, Saint-Martin d'H\`eres, France\\
$^{3}$Observatoire de Gen\`eve, 51 Ch. des Maillettes, 1290 Sauverny, Switzerland\\
$^{4}$ONERA, BP52, 29 Avenue de la Division Leclerc Ch\^atillon Cedex, France \\
}

   \date{Received; accepted }

 \abstract
   {}
   {We present the results of a deep imaging survey of stars surrounded by planets detected with the radial velocity technique. The purpose is to search for and to characterize long-period stellar and substellar companions. The sample contains a total of 26 stars, among which 6 exhibit additional radial velocity drifts. }
   {We used NACO, at the ESO Very Large Telescope, and PUEO-KIR, at the Candian French Hawaiian Telescope, to conduct a near-infrared coronographic survey with adaptive optics of the faint circumstellar environment of the planetary hosts. The domain investigated ranges between 0.1$~\!''$ to $15~\!''$ (i.e. about 3 to 500~AU, according to the mean distance of the sample). The survey is sensitive to companions within the stellar and the substellar domains, depending on the distance to the central stars and on the star properties.}
   {The images of 14 stars do not reveal any companions once the field objects are removed. 8 stars have close potential companions that need to be re-observed within 1--2 years to check for physical companionship. 4 stars are surrounded by faint objects which are confirmed or very probable companions.  The companion to HD\,13445 (Gl\,86) is already known. The HD\,196885 star is a new close visual binary system with a high probability of being bound. The 2 newly discovered companions, HD\,1237\,B and HD\,27442\,B, share common proper motions with the central stars. Orbital motion is detected for HD\,1237\,B. HD\,1237\,B is likely a low-mass M star, located at 70~AU (projected distance) from the primary. HD\,27442\,B is most probably a white dwarf companion located at about 240 AU (projected distance).}
   {}

 \keywords{Stars: low-mass, brown dwarfs, planetary systems -- Instrumentation: adaptive optics}

   \maketitle
%
\section{Introduction}

Since the discovery of 51\,Peg\,b \cite{mayor95}, more than 160 extrasolar giant planets (EGPs) have been found through radial velocity (RV) measurements. Their physical and orbital characteristics (mass, separation, eccentricity) have challenged the scenario of planetary formation. The observed properties of the period-mass distribution show tendencies that need to be explained \cite{udry03}. There is a lack of very light ($\leq$\,0.75\,M$_{\rm Jup}$) planets at long ($\geq$\,100\,days) periods, which cannot be explained by the observational bias inherent to the RV technique. There is a lack of massive ($\geq$\,2\,M$_{\rm Jup}$) planets with short ($\leq$\,100\,days) periods and a lack of brown dwarfs (BDs) at short ($\leq$\,4\,AU) separations, known as the so-called ``BD desert''. Whether this BD desert extends to larger separations is presently unknown. The present multiplicity studies in the substellar domain are limited by small statistical samples (Gizis et al. 2001, McCarthy \& Zuckerman 2004).

To understand the way(s) EGPs and BDs form and evolve, observational studies of their physical and orbital characteristics must be extended to longer periods and to lower masses. In addition, if EGPs or BDs are orbiting at large separations, they will gravitationally affect the formation and dynamical evolution of any possible inner planets (see Rivera \& Lissauer 2000, Eggenberger et al. 2004). In particular, they will constrain the stability domain of the inner planetary system and may be responsible for the large eccentricities observed for several planets detected with the RV technique.
\begin{table*}
\caption{Sample of planetary hosts observed with PUEO-KIR at CFHT and with NACO at VLT. The main properties of the systems are summarized. The telescopes, the number of companion candidates detected and their status are given with the number of observing epochs for each system.}             
\label{tab:sample}      
\centering          
\begin{tabular*}{\textwidth}{@{\excs}llllllllllll}
\hline\hline  \noalign{\smallskip}      
Name      & Other Name &V       &    K    &   d & SpT    &  $b$   & Telescope  & Nb cand. & Epochs & Status & References \\
          &            &(mag)   & (mag)   & (pc)         &        &($^o$)  &            &            &        &            \\
\noalign{\smallskip} \hline\noalign{\smallskip}                     
HD\,1237  &GJ\,3021     & 6.7    & 4.86    & 17.6         &  G6V   &  -37.1 & VLT        & 1          & 3& C & 1a           \\
HD\,13445 &Gl\,86       & 6.17   & 4.13    &  10.9        & K1V    & -62.0  & VLT        & 1          & 3& C & 2a -- 1b, 3b, 4b \\ 
HD\,17051 &iot\,Hor     & 5.4    & 4.14    & 17.2         & G0V    &  -58.3 & VLT        & 1          & 2& B & 3a           \\
HD\,27442              && 4.4    & 1.75    & 18.2         & K2IV   & -42.1  & VLT        & 1          & 3& C & 4a           \\
HD\,28185              && 7.81   & 6.19    & 39.6         & G5     & -36.0  & VLT        & 0          & 1& N    & 5a           \\   
HD\,52265              && 6.3    & 4.95    & 28.1         & G0III-IV& -0.5  & VLT        & 7          & 2& B & 6a           \\
HD\,89744              && 5.74   & 4.45    & 39.0         & F7V     & +56.4 & CFHT       & 1          & 1& U & 7a           \\
HD\,92788              && 7.1    & 5.72    & 32.3         & G5      & +47.4 & CFHT       & 0          & 1& N    & 8a           \\ 
HD\,95128 &47\,Uma      & 5.10   &  3.75   & 14.1         & G1V     & +63.4 & CFHT       & 0          & 1& N    & 9a, 10a           \\       
HD\,106252             && 7.36   &  5.93   & 37.4         & G0V     & +70.7 & CFHT       & 0          & 1& N    & 11a           \\ 
HD\,114762             && 7.30   &  5.81   & 40.1         & F9V     & +79.3 & CFHT       & 1          & 1& U & 12a           \\
HD\,121504             && 7.6    & 6.12    & 44.4         & G2V     & +5.7  & VLT        & 56         & 1& U & 13a           \\ 
HD\,130322             && 8.05   & 6.23    & 29.7         & K0III   & +51.0 & CFHT/VLT   & 1          & 2& B & 14a           \\
HD\,141937             && 7.25   &  5.76   & 33.5         & G2/G3V  & +26.8 & CFHT       & 1          & 1& U & 15a -- 2b         \\
HD\,154857             && 7.25   &  5.51   & 68.5         &    G5V  & -10.0 & VLT        & 28         & 1& U & 16a           \\

HD\,160691&$\mu$\,Arae  & 5.2   & 3.68    & 15.3         & G3IV-V  & -11.5 & VLT        & 5          & 3& B & 16a, 17a, 18a           \\
HD\,162020             && 9.18   &  6.54   &   31.2       &   K2V   & -6.7  & VLT        & $>100$        & 1& U & 19a           \\
HD\,168443             && 6.92   & 5.21    & 37.9         & G5      & +2.5  & CFHT       & $>100$ (5)   & 1& U (B) & 19a -- 2b         \\
HD\,179949             && 6.25   &  4.94   & 27.0         & F8V     & -15.8 & CFHT       & 5          & 1& U & 21a           \\
HD\,183263             && 7.86   &  6.42   &    52.7      & G2IV    & -4.3  & VLT        & 57         & 1& U & 22a           \\ 
HD\,186427 &16\,Cyg\,B & 6.20   &  4.65   & 21.4         & G3V     & +13.2 & CFHT       & 0          & 1& N & 23a -- 2b    \\
HD\,187123             && 7.86   & 6.34    &   47.9       & G5      & +4.7  & CFHT/VLT   & 5     & 2& B & 24a -- 2b\\
HD\,192263             && 8.1    & 5.54    & 19.9         & K2V     & -18.7 & VLT        & 8      & 3& B & 25a\\
HD\,196885             && 6.4    & 5.07    &  33          &F8IV     & -18.0 & VLT        & 3     & 1& U & no reference, 26a\\
HD\,202206             && 8.1    & 6.49    & 46.3         & G6V     & -40.5 & VLT        & 3     & 2& B & 19a, 20a\\
HD\,217107             && 6.2    & 4.54    & 19.7         & G*IV    & -53.3 & VLT        & 0     & 1& N       & 27a            \\ 
\noalign{\smallskip} \hline                    
\end{tabular*}
\begin{list}{}{}
\item[\scriptsize{- RADIAL VELOCITIES REFERENCES:}] \scriptsize{(1a) Naef et al. 2001, (2a) Queloz et al. 2000, (3a) Kurster et al. 2000, (4a) Butler et al. 2001, (5a) Santos et al. 2001, (6a) Butler et al. 2000, (7a) Korzennik et al. 2000, (8a) Fischer et al. 2000, (9a) Butler \& Marcy (1996), (10a) Fischer et al. (2002a), (11a) Fischer et al. (2002b), (12a) Latham et al. (1989), (13a) Mayor et al. 2004, (14a) Udry et al. 2000, (15a) Udry et al. 2002, (16a) McCarthy et al. 2004, (17a) Butler et al. 2001, (18a) Santos et al. 2004, (19a) Udry et al. 2002, (20a) Correia et al. 2004, (21a) Tinney et al. 2000, (22a) Marcy et al. 2005, (23a) Cochran et al. 1997, (24a) Butler et al. 1998, (25a) Santos et al. 1999, (26a) http://exoplanets.org/esp/hd196885/hd196885.shtml, (27a) Fischer et al. 1999}\\
\item[\scriptsize{- IMAGING REFERENCES:}] \scriptsize{(1b) Els et al. 2001, (2b) Luhman \& Jayawardhana 2002, (3b) Mugrauer \& Neuh\"auser 2005, (4b) Lagrange et al. 2006} \\
\item[\scriptsize{- NOTES:}] \scriptsize{(C) confirmed comoving object, (B) stationary background objects, (U) undefined sources and (N) no faint objects detected}
\end{list}
\end{table*}
\begin{figure*}[t]
   \centering
\includegraphics[width=15.5cm]{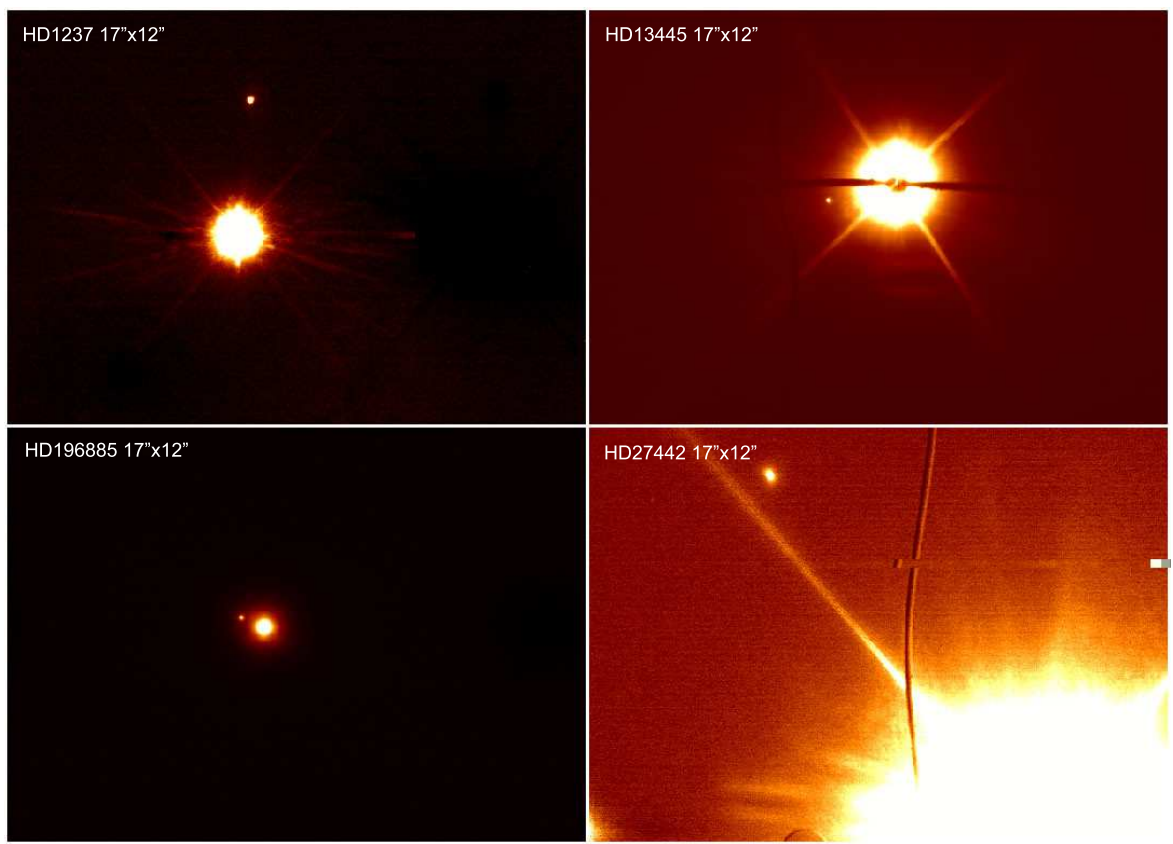}
\caption{VLT/NACO images of the companions to HD\,1237, HD\,13445 (Gl\,86), HD\,27442 and the very probable one to HD\,196885. Orbital motion is detected for the companion to HD\,13445 (Gl\,86), known to be bound (Els et al. 2001, Mugrauer \& Neuh\"auser 2005, Lagrange et al. 2006). Follow-up observations indicate that the companion candidates HD\,1237 and  HD\,27442 are comoving objects. We also detect the orbital motion of HD\,1237\,B relative to A, which confirms its companionship. Finally,  HD\,196885 is a close visual binary system with a high probability for being bound (see Section 4.2).}
\label{fig:imagelie}
\end{figure*}

Up to now, the best method to detect close ($\leq$\,4\,AU) substellar companions has been the RV technique. However, this method is currently insensitive to larger separations. With the development of high contrast and high angular resolution imaging, resulting from large telescopes equipped with adaptive optics (AO) and coupled to dedicated focal devices, such as coronagraphs or differential imagers, BDs and very massive EGPs can now be rapidly probed at large separations (typically $\geq$\,50--100\,AU). Since the discovery of Gl\,229\,B \cite{naka95}, coronagraphic imaging techniques have been successfully used, from space and from the ground, to detect BD companions around nearby field stars: HR\,7672\,B \cite{liu02}, HD\,130948\,B \cite{potter02},  Gl\,494\,B (Beuzit et al. 2004),  G\,239-25 (Forveille et al. 2004) or, more recently, HD\,49197\,B (Metchev et al. 2004). 

In order to push the detection performances down to the planetary mass regime, systematic searches have recently focused on young ($1-50$~Myr), nearby associations. EGPs and BDs are actually hotter, brighter and therefore easier to image when they are young. These surveys led to the detection of very low mass substellar companions of 10 to 30~M$_{\rm Jup}$: HR\,7329\,B (Lowrance et al. 2000), TWA5\,B (Lowrance et al. 1999), GSC\,08047-00232\,B (Chauvin et al. 2003) and, more recently AB\,Pic\,B (Chauvin et al. 2005). However, the most important results obtained so far are the discoveries of the possible planetary mass companions to the young BD 2MASSWJ\,1207334-393254 (Chauvin et al. 2004, 2005) and the young star GQ\,Lup (Neuh\"auser et al. 2005).

In the context of older ($0.5-10$~Gyr) stars with planets detected through RV measurements, the detection capabilities are currently limited to massive BD companions. The discovery of the companion Gl\,86\,B (see Els et al. 2001) is a good illustration. Mugrauer \& Neuh\"auser (2005) have recently demonstrated that this object is likely to be a cool white dwarf, although its photometry is compatible with a T dwarf. In recent years, systematic searches of BDs orbiting planetary hosts have been conducted in the northern hemisphere (Oppenheimer et al. 2001, Luhman \& Jayawardhana 2002, Patience et al. 2002). Of these, Luhman \& Jayawardhana (2002) were the most sensitive one. However, none of the 25 stars that they observed revealed bound companions between a few tens and a few hundreds of AU of the primary.

In 2003, we commenced deep coronographic imaging survey of 26 planetary hosts, using PUEO-KIR, at CFHT, and NACO, at VLT. We report in Section 2 the sample, observing strategy, instrument set-ups and the data reduction and analysis. We present in Section 3 the results of these observations and the detection performances achieved. Section 4 discusses in more detail those cases exhibiting RV drifts and for which no companions are known at long-periods. Finally, the discovery of the 2 companions to the stars HD\,1237 and HD\,27442, as well as the possible companion to the star HD\,196885 is detailed. 

%
\section{Observations}

\subsection{Sample and observing strategy}

Table~\ref{tab:sample} lists the main characteristics (distance, spectral type, VK magnitudes and galactic latitude) of the 26 stars observed. The observing strategy was similar at CFHT and at VLT. The first step was to obtain deep coronographic images to search for faint and close companion candidates in the circumstellar environment. Then, a non-saturated classical image was immediately recorded to obtain an accurate position of the star on the detector, to measure the relative photometry between the companion candidates and the star and to look for a very close binary. The detection survey was performed in K-band where the AO correction is more reliable and the contrast between stars and substellar companions is optimal. 

In cases of positive detection, the companion candidates were re-observed at a second epoch to check if the faint objects shared a common proper motion with the central star. Depending on the object proper motions, this required an interval of 1-2 years between successive observations. When true companions were detected, new images were also recorded in J and H bands for direct comparison with the photometry predicted by evolutionary models of stellar and substellar objects. Typical total exposure times were 5-10 minutes. However, only the highest quality exposures were retained, leading to an effective exposure time of about 5 minutes per target. 
\begin{table*}[t]
\caption{Detection performances of PUEO-KIR at CFHT and of NACO at VLT. The detection limits are converted in term of mass, based on the predictions of the BA98 (Baraffe et al. 1998) evolutionary model in the stellar regime and the COND03 (Baraffe et al. 2003) model in the substellar regime }             
\label{tab:perf}      
\centering          
\begin{tabular*}{\textwidth}{@{\excs}llllllll}
\hline\hline\noalign{\smallskip}        
&             &                 &                               &                          \multicolumn{4}{c}{\underline{\hspace{2.8cm}\scriptsize{EVOLUTIONARY MODEL PREDICTIONS}\hspace{2.8cm}}}  \\
Instrument    &   Separation    &   $\Delta K$ (6$\sigma$)      &   M$_{K}$ (6$\sigma$)     &  Mass (0.5 Gyr)  &  Mass (1.0 Gyr)  &  Mass (5.0 Gyr) &  Mass (10.0 Gyr) \\
              & (arcseconds)    & (mag)                         & (mag)                     & (M$_{\rm{Jup}}$) & (M$_{\rm{Jup}}$) & (M$_{\rm{Jup}}$)& (M$_{\rm{Jup}}$)  \\
\noalign{\smallskip} \hline\noalign{\smallskip} 
CFHT/PUEO-KIR     &   1.0           & 8                             & 10--12                    &  70              & 75               & 80              & 80             \\
              &   2.0           & 11                            & 13--15                    &  30              & 40               & 60              & 70             \\
VLT/NACO      &   1.0           & 12                            & 14--16                    &  25              & 30               & 60              & 70             \\
              &   2.0           & 13.5                          & 16--18                    &  15              & 20               & 40              & 50            \\
\hline                    
\end{tabular*}
\end{table*}

\subsection{Instruments and astrometric calibration}

\subsubsection{PUEO-KIR at CFHT}

The AO observations were carried out on May 20, 21 and 22 2003 at CFHT, using the 1024 $\times$ 1024 near-infrared camera KIR\footnote{\emph http://www.cfht.hawaii.edu/Instruments/Detectors/IR/KIR/}, coupled to the AO curvature system PUEO (see Roddier et al, 1991; Arsenault et al, 1994) (located at the Cassegrain focus). We used the KIR upgrade coronographic mode (named GRIF) that provides an occulting mask with a diameter of $0.8~\!''$.  The coronographic observations were obtained with the K$~\!'$ ($\lambda=2.12~\mu m$, $\Delta\lambda=0.34~\mu m$) filter for all the CFHT sources. The direct imaging observations were obtained with the Br$\gamma$ ($\lambda=2.166~\mu m$, $\Delta\lambda=0.022~\mu m$) narrow band filter. 

No astrometric calibrators were observed to  properly determine the platescale and the detector orientation of the KIR camera.  Therefore, we have used conservative values for the detector orientation and the platescale of respectively $0.0\pm2^o$ and $34.8\pm0.2$\,mas/pix. The total field of view (FoV) was then $35.6~\!'' \times 35.6~\!''$. Follow-up observations were done for only two stars (HD\,130322 and HD\,187123) with such significant proper motions that the companion candidates could be unambiguously identified as background objects.

\subsubsection{NACO at VLT}

The NACO observations were obtained during multipurpose runs of guaranteed time observations (GTO) and open time observations\footnote{These GTO runs were scheduled between May 21 and June 7 2003 (GTO-071.C-0507), on September 7 2003 (GTO-71.C-0462), and between November 11 and 17 2003 (GTO-072.C-0624). We also obtained open time observations on September 22 2004 (073.C-0468) and on July 28 and 29 2005 (075.C-0825).}. The NACO\footnote{\emph http://www.eso.org/instruments/naos/} instrument is equipped with an AO system (Rousset et al. 2002) that provides diffraction limited images in the near infrared (nIR) and illuminates the CONICA camera (Lenzen et al. 2002), equipped with a $1024\times1024$ pixel Aladdin InSb array. Note that in May 2004, the CONICA detector was changed for a more efficient one. The coronographic masks used with NACO have diameters of 0.7$~\!''$. and 1.4$~\!''$. All NACO targets were observed in coronography, using the CONICA Ks ($\lambda=2.18~\mu m$, $\Delta\lambda=0.35~\mu m$) filter. In direct imaging, we used the Br$\gamma$ ($\lambda=2.166~\mu m$, $\Delta\lambda=0.023~\mu m$) narrow band filter, as well as a neutral density filter, providing a transmissivity factor of 0.014, for the brightest stars. In order to correctly sample the PSF (better than Nyquist), we used the S27 objective, which gives a FoV of $28~\!'' \times28~\!''$ centered on the stars. 

The $\Theta_1$ Ori C astrometric field (McCaughrean \& Stauffer 1994) was observed (on November 12 2003, September 22 2004 and July 29 2005) to enable detector plate scale and orientation calibration. The orientation of true north of the S27 camera was found respectively at $-0.06^o$, $0.0^o$ and $-0.05^o$ east of the vertical with an uncertainty of $0.20^o$ and the plate scale was $27.01\pm0.05$\,mas/pix. On June 3 2003, the grade 2 astrometric calibration binary WDS21579-5500 from the Washington Visual Double Star Catalog (Worley \& Douglass 1996) was observed and the S27 camera was found at  $0.01\pm0.20^o$ east of the vertical. The plate scale remains stable at $27.01\pm0.05$\,mas.
\begin{figure*}[t]
\centering
\includegraphics[width=5.9cm]{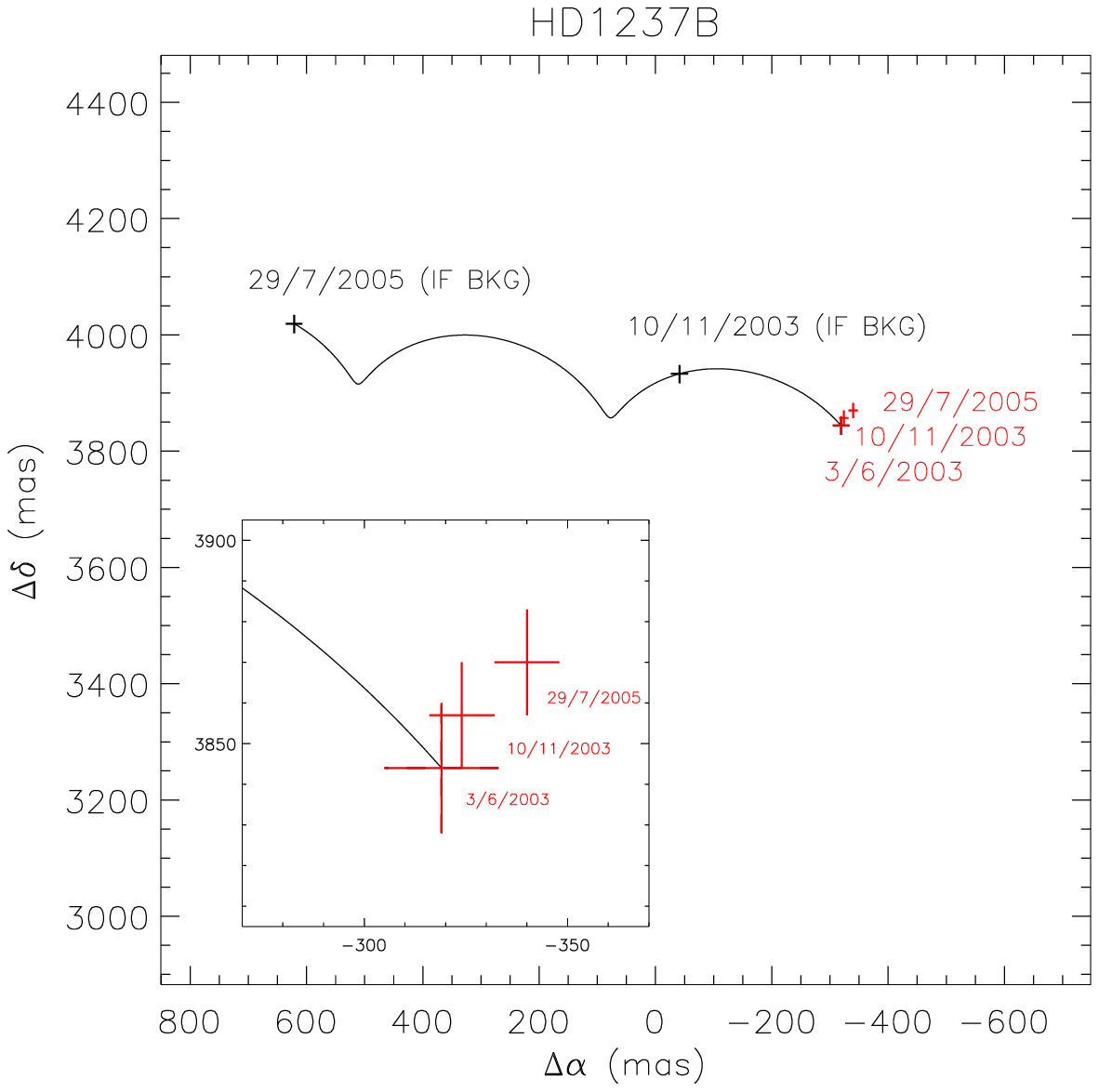}  \includegraphics[width=5.9cm]{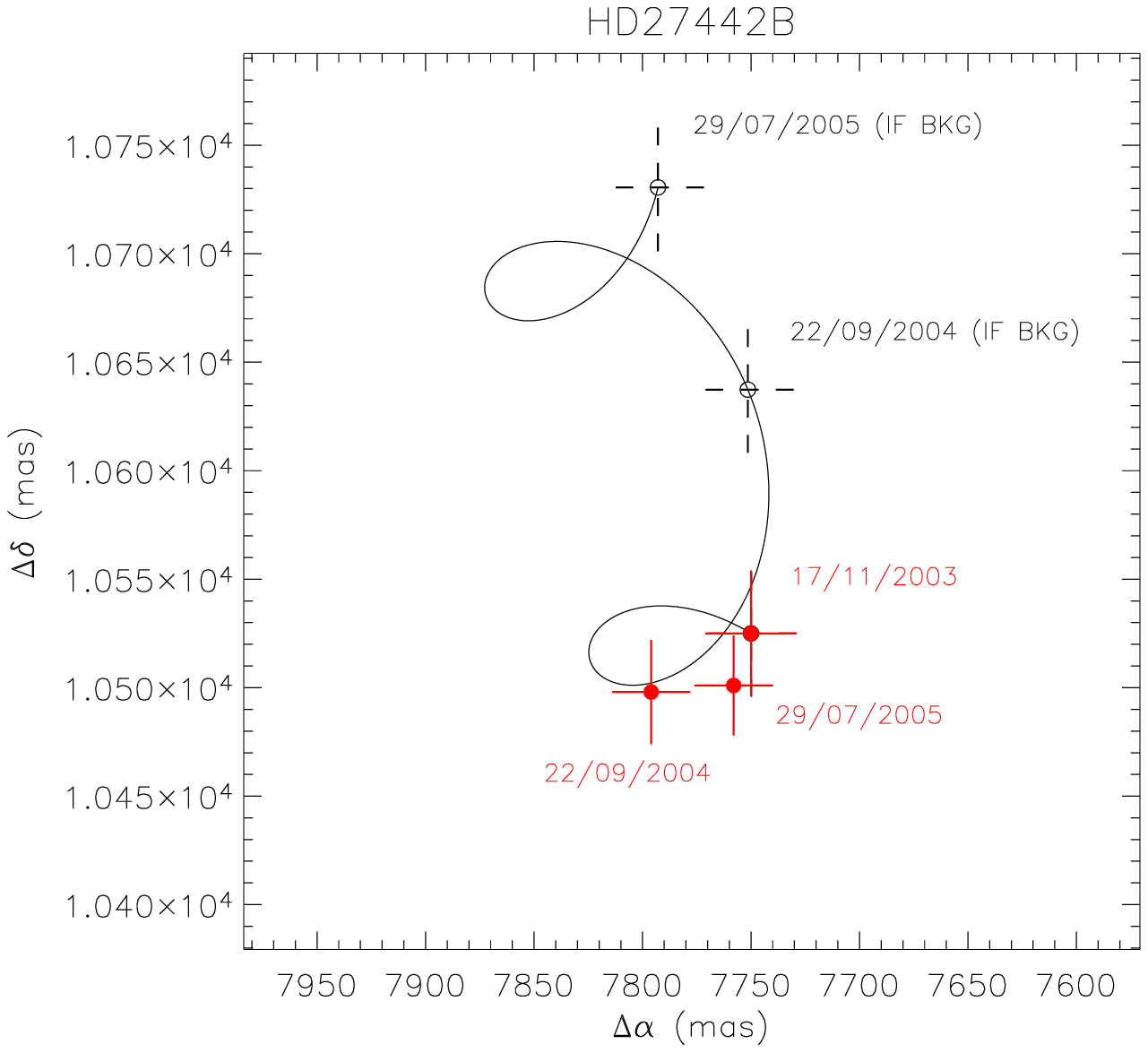}\includegraphics[width=5.9cm]{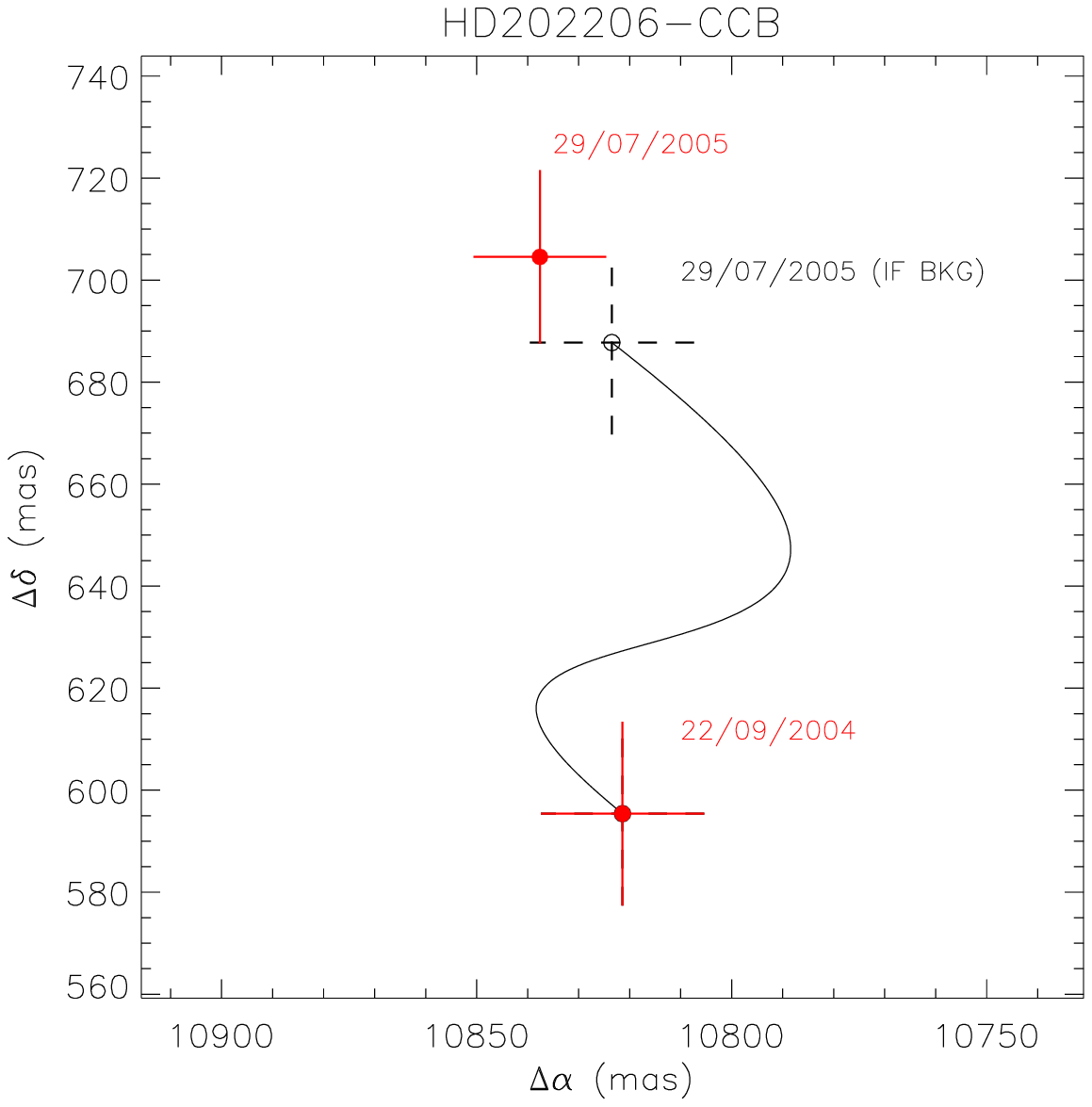}
\caption{VLT/NACO Measurements (\textit{full circles} with uncertainties) of the offset positions of the companion HD\,1237\,B to A (\textit{\textbf{left}}), of the companion HD\,27442\,B to A (\textit{\textbf{middle}}) and of the companion candidate CC-B relative to HD\,202206 (\textit{\textbf{right}}). For each diagram, the expected variation of offset positions, if B is a background object, is shown (\textit{solid line}), based on a distance and on a primary proper motion, as well as the initial offset position of B (or CC-B) from A. The \textit{empty circles} give the corresponding expected offset positions of a background object for the different epochs of observations of HD\,1237\,A, HD\,27442\,A and HD\,202206, with total uncertainties. In both cases, HD\,1237\,B and HD\,27442\,B are confirmed to be comoving objects. the candidate companion CC-B to HD\,202206 has been rejected as comoving and is most likely a stationary background object. In addition, the NACO astrometric measurements clearly show an orbital motion of HD\,1237\,B relative to A, which confirms the companionship of this low mass star.}
\label{figastro}  
 \end{figure*}

\subsection{Data reduction and analysis}

The data were flat-fielded, sky substracted and cross-correlated using the \textit{eclipse} software \citep{devi97}. Companion candidates were identified using the ``peak'' \textit{eclipse} algorithm. The relative positions between the companion candidates and  the central star were determined using the deconvolution algorithm of V\'eran et al. (1998). These were double-checked using cross-correlation functions. The position of the parent star was measured on the direct image that followed the coronagraphic one. The shifts ($\leq 1$ pixel) induced between images taken in different filters have been accounted for. For NACO, these shifts were measured using a sequence of pinhole images recorded with the respective instrumental settings. For PUEO-KIR, direct images of a single star in narrow and broad band filters were recorded to determine them. 

In order to estimate the relative flux between the star and any companion candidates, we used aperture photometry. The results for the two companions HD\,1237\,B and HD\,27442\,B are reported in Table~3. In the case of the close binary HD\,196885, we used the myopic deconvolution algorithm \textit{MISTRAL} (Conan et al. 2000) to obtain relative JHK photometry and relative astrometry (see Table~\ref{tab:photometry}). For HD\,1237 and HD\,27442, the transformation between the K$_{s}$ filter of NACO and the K filter used by CTIO-2MASS was found to be negligible ($\leq0.03$~mag) with respect to uncertainties of our relative photometry. The detection limits were obtained, using the direct and the coronagraphic imaging data. The method consists of reducing the level of the PSF wing and in measuring the detection limit. For a given angular sector (typically 20--30$^o$), centered on the star, we compute the azimuthal average of the PSF that we subtract from any angular direction in the sector. We then measure the 6 $\sigma$ detection limit, for a given radius from the star, in a sliding box of 5$\times$5 pixels.

\section{Results}

\subsection{Identification of companion candidates}

We did not detect any companion candidates around the 6 stars HD\,28185, HD\,92788,  HD\,95128 (47\,Uma), HD\,106252, HD\,186427 (16\,Cyg\,B) and HD\,217107. HD\,95128 (47\,UMa) and HD\,186427 (16\,Cyg\,B) had already been observed by Luhman \& Jayawardhana (2002), using enhanced detection capabilities with the Keck II Telescope in conjunction with KCam (but with a search radius of $3.3~\!''$). They did not detect any fainter candidates. 

Of the remaining 20 stars, at least one faint companion candidate was detected. The number of companion candidates and their status are summarized in Table~\ref{tab:sample}.
\begin{itemize}
\item[-] The stars HD\,168443 and HD\,187123 are in the line of sight of the galactic plane ($b<15^o$) and several faint sources were detected. For HD\,168443, we have detected more than 100 companion candidates among which 5 were identified by Luhman \& Jayawardhana (2002) as background objects. For HD\,187123, five companion candidates were detected including 2 identified by Luhman \& Jayawardhana (2002). They all appear to be background objects.
 
\item[-] A faint companion candidate was detected for HD\,89744\,A. Using Sofi at the ESO New Technology Telescope, Mugrauer et al. (2004) identified this source as a background object. Although PUEO-KIR detection performances were better than Sofi, no additional faint companion candidates were detected. Note that a stellar companion exists around HD\,89744\,A, but at a larger separation than the KIR FoV (see Mugrauer et al. 2004).
 
\item[-] 8 stars (HD\,52265, HD\,114762, HD\,121504, HD\,141937, HD\,154857 HD\,162020, HD\,179949 and HD\,183263) have so far only been observed at one epoch. Additional follow-up observations are needed to identify possible comoving companions. 

\item[-] 6 other stars (HD\,17051, HD\,160691, HD\,192263, HD\,202206, in addition to the CFHT candidates HD\,130322 and HD\,187123) were found with companion candidates that happened to have different proper motions from their central stars. An example of the analysis of the proper and parallax motions is provided in Fig.~\ref{figastro} (\textit{right}). We show respectively the expected offset positions and the observed offset positions of the companion candidate relative to HD\,202206. 

\item[-] Close to the star HD\,196885, direct images reveal a close bright object (see Fig.~\ref{fig:imagelie}). Eventhough we do not have second epoch measurements, the probability that this object is a background object is extremely small due to its proximity  to the star HD\,196885 and its relative brightness (see Table~\ref{tab:photometry}). Its status is detailed in Section~4.

\item[-] Finally, in 3 cases, the faint object shares a common proper motion with the central star: HD\,1237, HD\,13445 (Gl\,86) and HD\, 27442 (see Fig.~\ref{fig:imagelie}). Gl\,86\,B is a known companion \cite{els01} and is mentioned just for completeness. HD\,1237\,B and HD\,27442\,B are bound companions (see Fig.~\ref{figastro}). Astrometric measurements and photometric properties of the companions are summarized in Table~\ref{tab:astrometry} and \ref{tab:photometry}. The results obtained for these two objects are discussed in Section~4.
\end{itemize}
\begin{table*}[t]
\caption{Characteristics of the planetary hosts exhibiting additional linear RV trends.}
\centering
\begin{tabular*}{\textwidth}{@{\excs}lllllllllll}     
\hline\hline\noalign{\smallskip} 
             &                &              &                     &                                                   &                     &             &               &        &\multicolumn{2}{c}{\underline{\hspace{0.3cm}\scriptsize{ADDITIONAL COMPANION?}\hspace{0.3cm}}}      \\
Name         &        SpT     & Age          & M$_{\rm{A}}$        &   $\rm{M}_{b;}$ $_{\rm{c}}\,\rm{sin}\textit{i}$   & P                   &  $e$        & RV drift      & Ref.   & $a_{\rm{min}}$\,$(0.5~\rm{M}_{\odot})$   &    $a_{\rm{min}}$\,$(0.1~\rm{M}_{\odot})$    \\
             &                & (Gyr)        & (M$_{\odot}$)  &   ($\rm{M}_{\rm{Jup}}$)              & (days)              &             & (m.s$^{-1}$.yr$^{-1}$)  &       &   (AU)        & (AU)                  \\
\noalign{\smallskip} \hline\noalign{\smallskip} 
HD\,28185    &  G5            & 5.5     & 1.0            & 5.7                      & 383                 & 0.07                    & 11.0                    & 1, 9, 10 & $\leq6$     & $\leq25$        \\
HD\,154187   &  G5            & 5.0          & 1.2            & 1.8                      & 398                 & 0.51                    & -14.0                   & 2       & $\leq15$    & $\leq70$        \\
HD\,187123   &  G3V           & 5.0          & 1.0            & 0.48                     & 3.096               & 0.43                    & -7.0                    & 3, 4     & $\leq8$     & $\leq40$        \\
HD\,202206   &  G6V            & 5.0          & 1.15               & 17.4; 2.44               & 255.9; 1383.4      & 0.43; 0.267             & 42.9                    & 5, 6     & \multicolumn{2}{l}{explained by the 2$^{\rm{nd}}$ planet}                \\
HD\,217107   & G8IV           & 8.0          &  0.98              & 1.37; $\leq13$           & 7.1; $\geq3000$    &0.13;  $\sim0.5$         & 38.0                    & 7, 8     & \multicolumn{2}{l}{explained by the 2$^{\rm{nd}}$ planet}   \\
\noalign{\smallskip} \hline
\end{tabular*}
\label{tab:egp}
\begin{list}{}{}
\item[\scriptsize{REFERENCES:}] \scriptsize{(1) Santos et al. 2001, (2) McCarthy et al. 2004, (3) Butler et al. 1998, (4) Vogt et al. 2000, (5) Udry et al. 2001, (6) Correia et al. 2005, (7) Fischer et al. 1999, (8) Vogt et al. 2005, (9) Schaerer et al. 1993, (10) Strassmeier et al. 2000}    
\end{list}
\end{table*}

\subsection{Survey performances}

\begin{figure*}

   \centering
   \includegraphics[width=8cm]{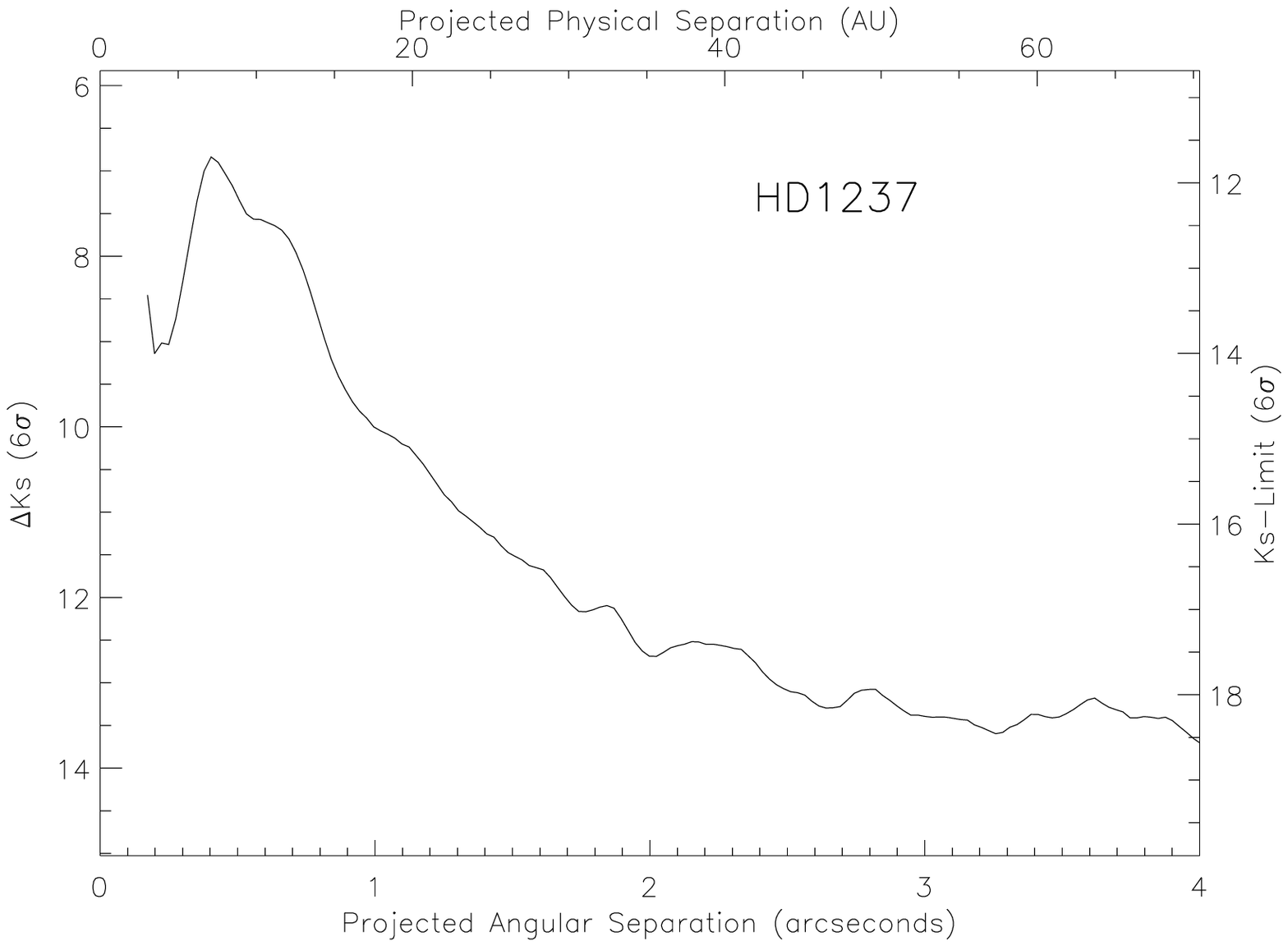}
\hspace{.6cm}
   \includegraphics[width=8cm]{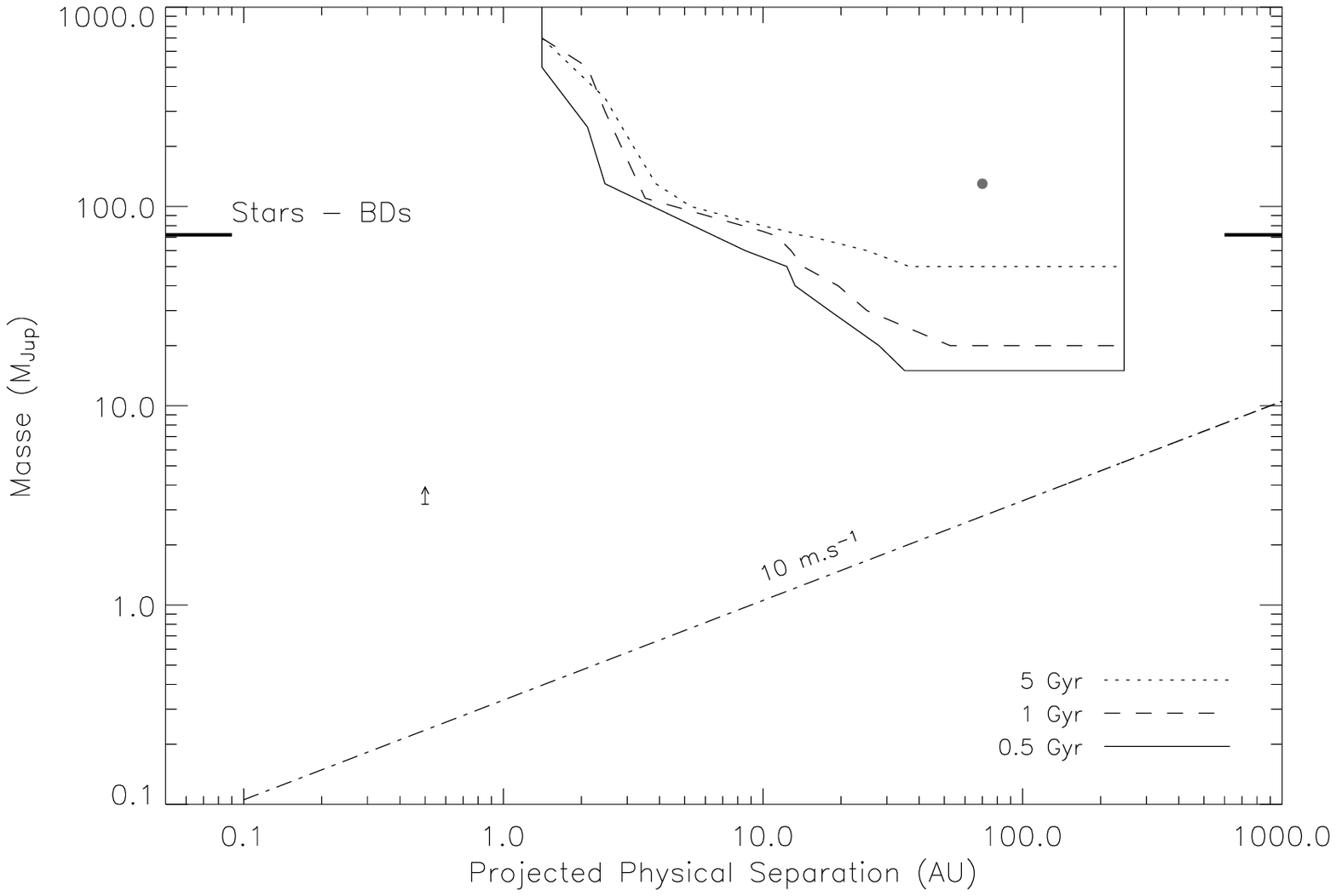}\hspace{.6cm}\\

   \includegraphics[width=8cm]{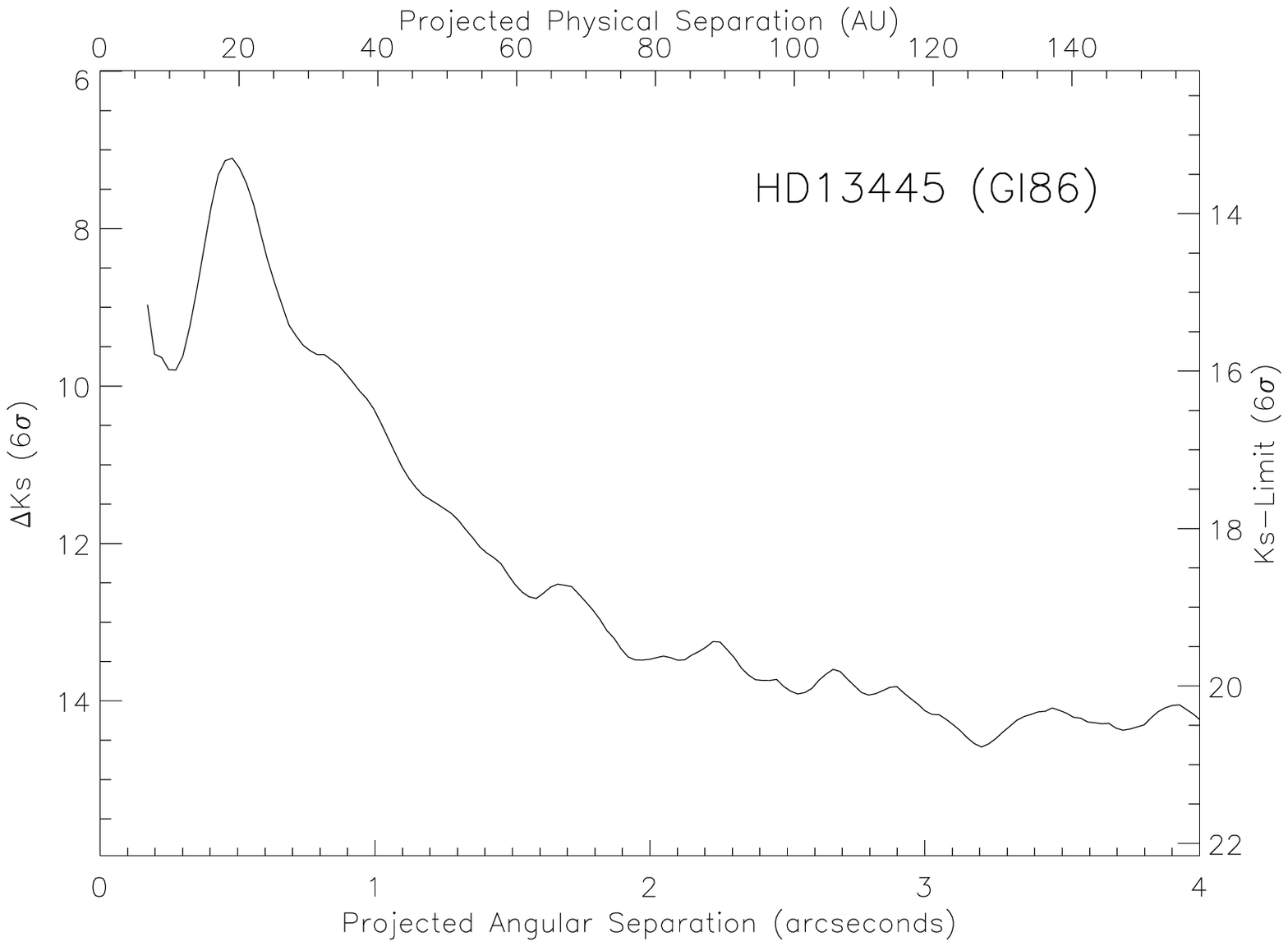}
\hspace{.6cm}
   \includegraphics[width=8cm]{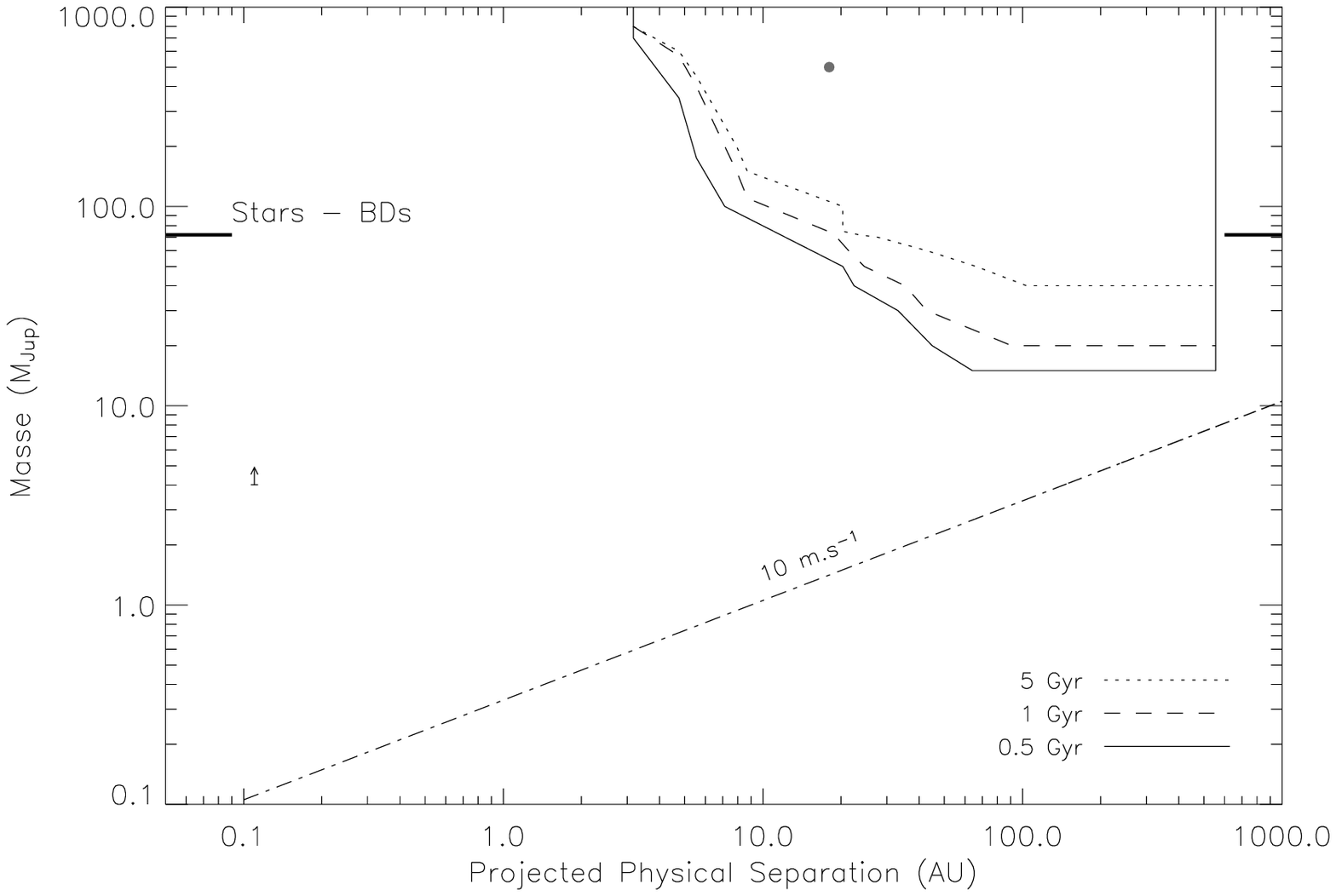}\hspace{.6cm}\\

   \includegraphics[width=8cm]{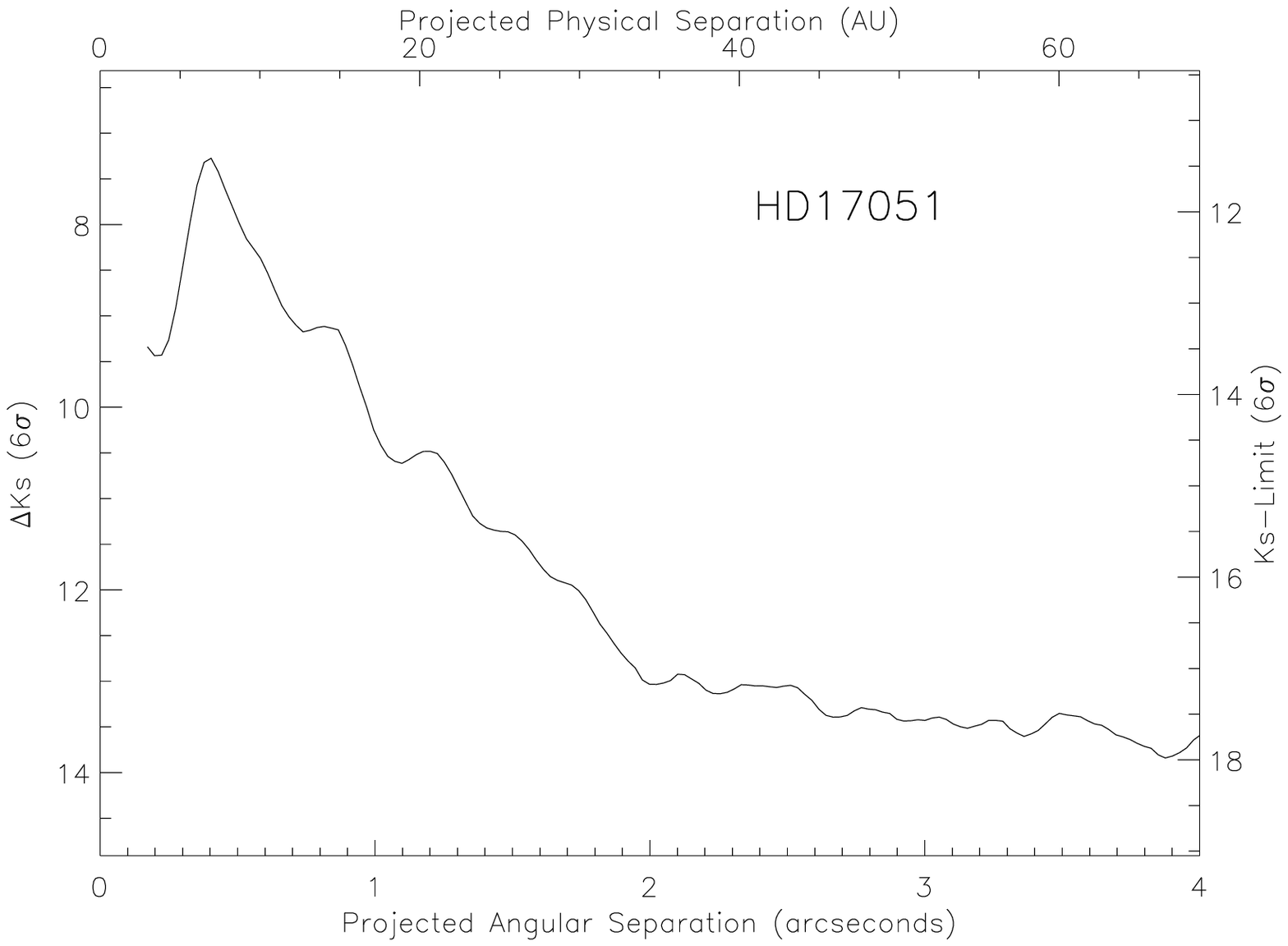}
\hspace{.6cm}
   \includegraphics[width=8cm]{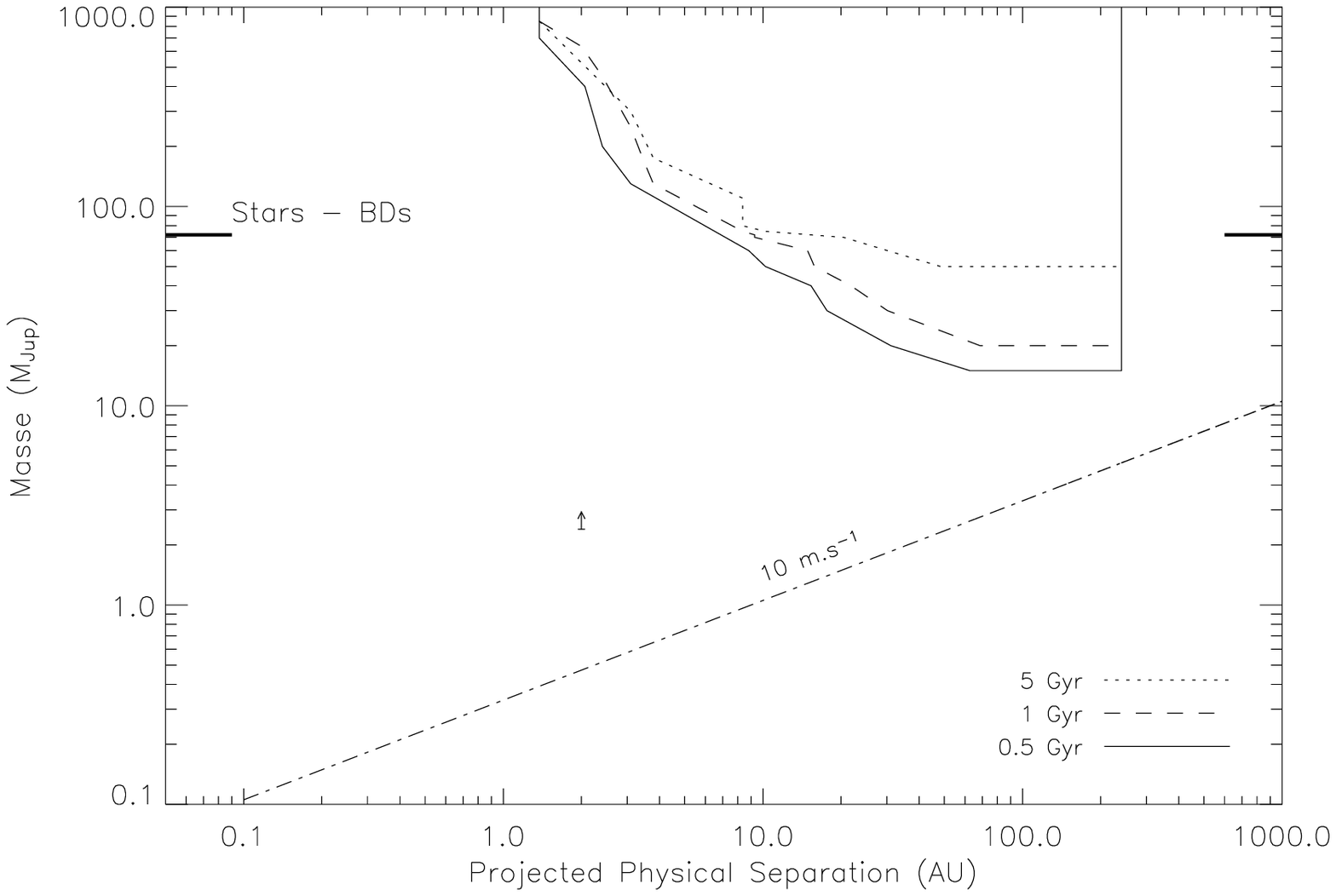}\hspace{.6cm}\\

   \includegraphics[width=8cm]{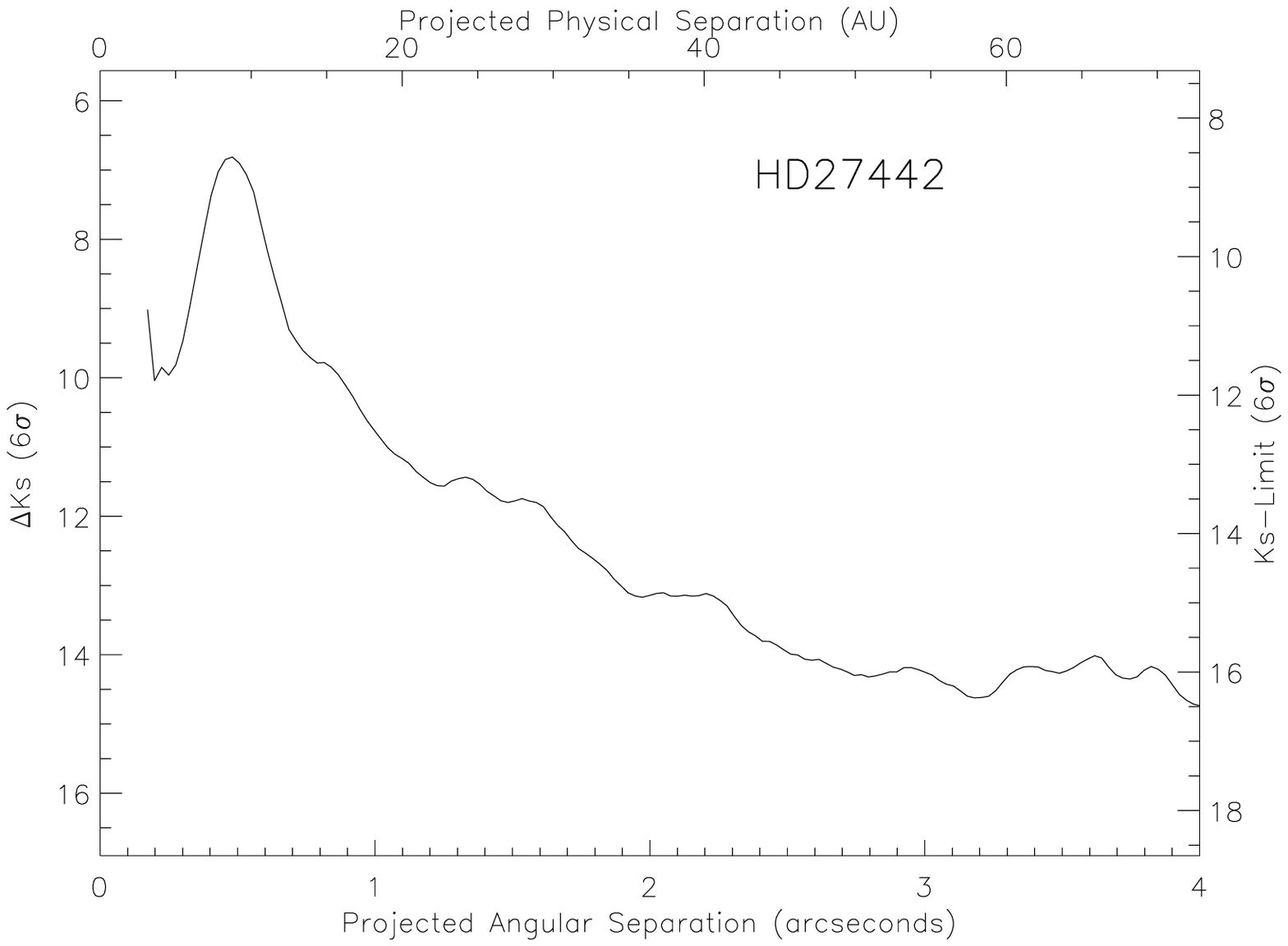}
\hspace{.6cm}
   \includegraphics[width=8cm]{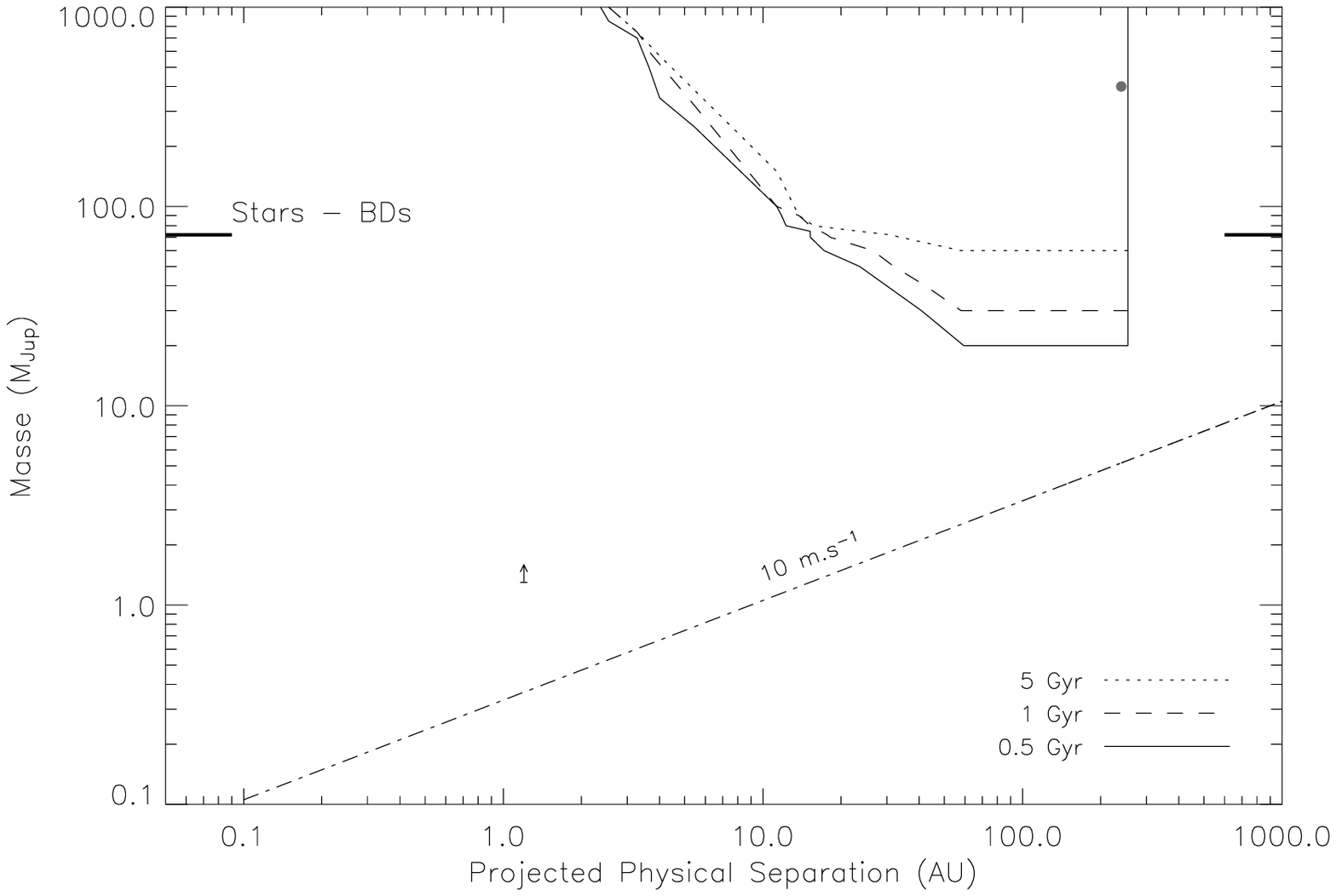}\hspace{.6cm}\\
\caption{Detection Limits of HD\,1237 (November 2003, VLT/NACO, total exposure time of 300s), HD\,13445 (Gl\,86) (July 2005, VLT/NACO total exposure time of 270s), HD\,17051 (November 2003, VLT/NACO total exposure time of 360s) and HD\,27442 (July 2005, VLT/NACO total exposure time of 105s). See detail of the detection limit estimation in Section 2.2}
\label{lalim}
   \end{figure*}

\begin{figure*}

\centering

   \includegraphics[width=8cm]{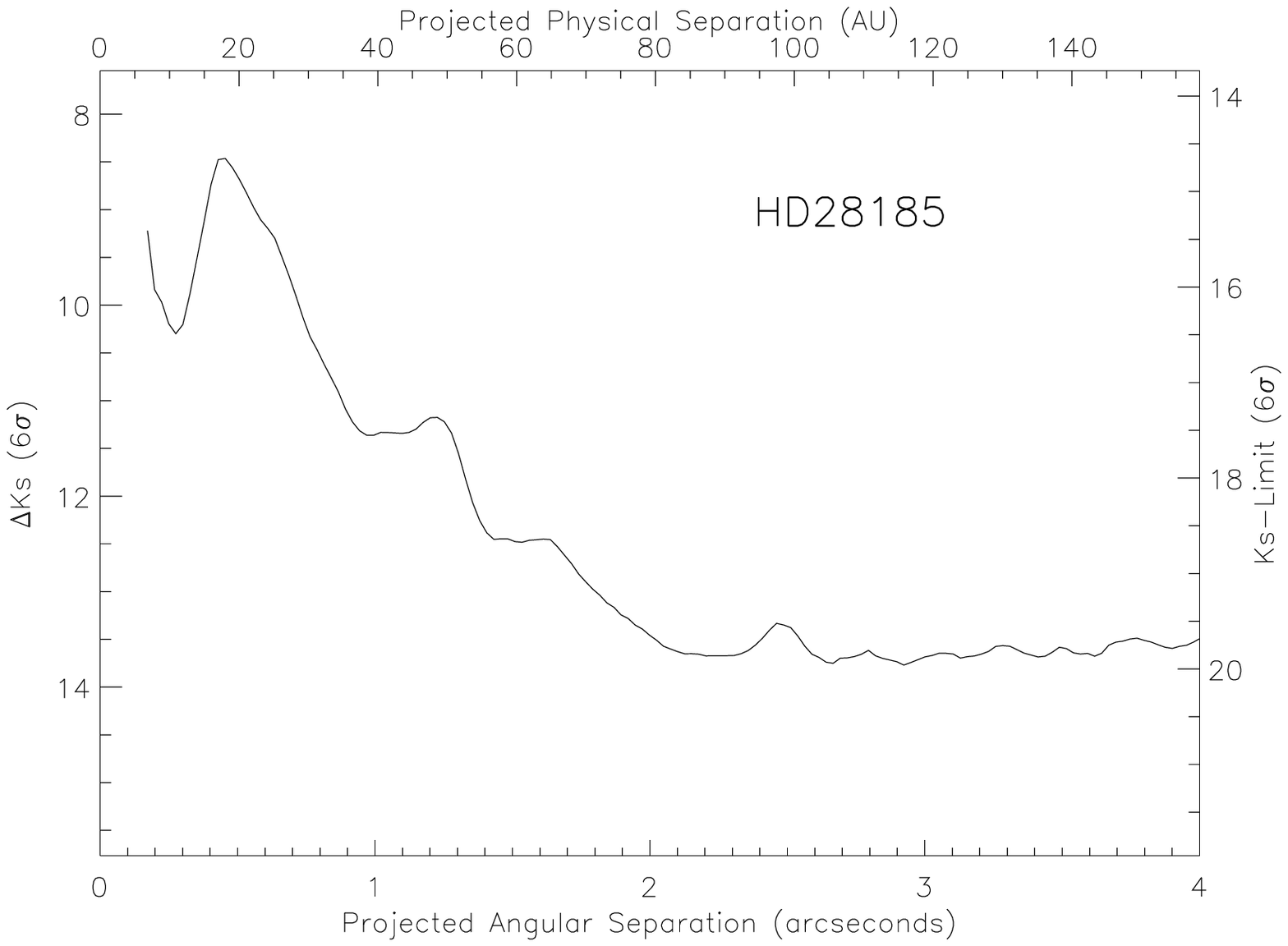}
\hspace{.6cm}
   \includegraphics[width=8cm]{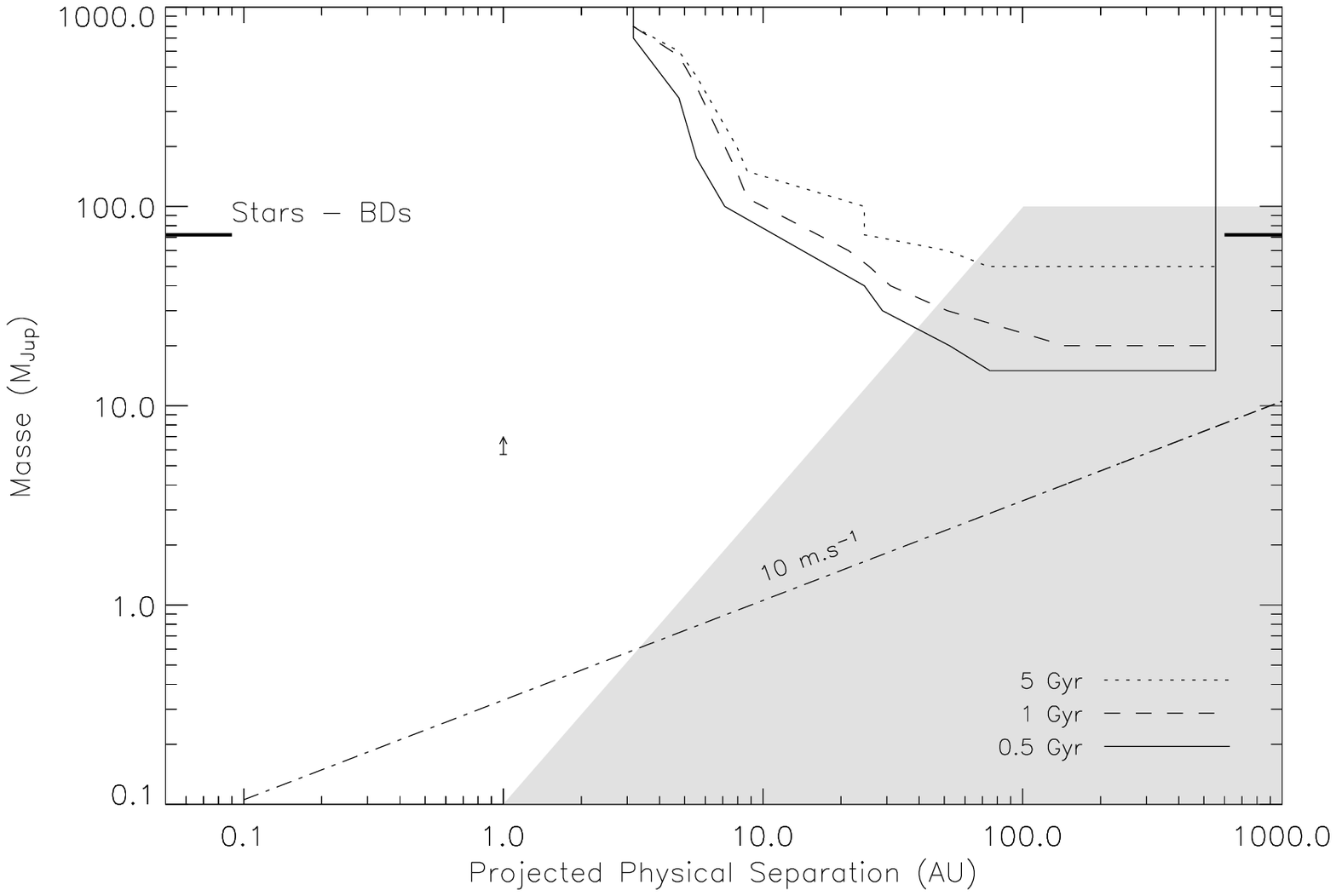}\hspace{.6cm}\\

   \includegraphics[width=8cm]{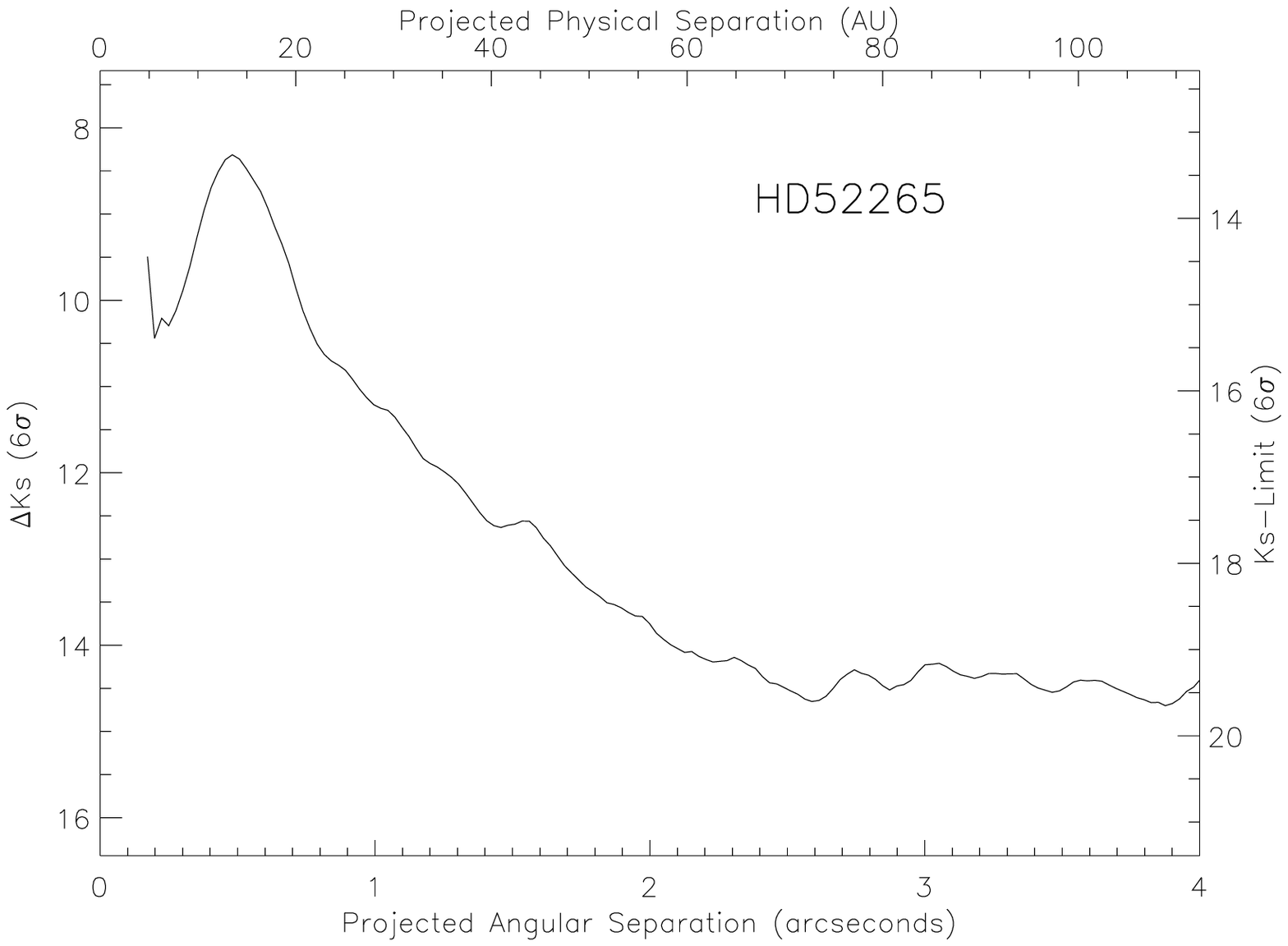}
\hspace{.6cm}
   \includegraphics[width=8cm]{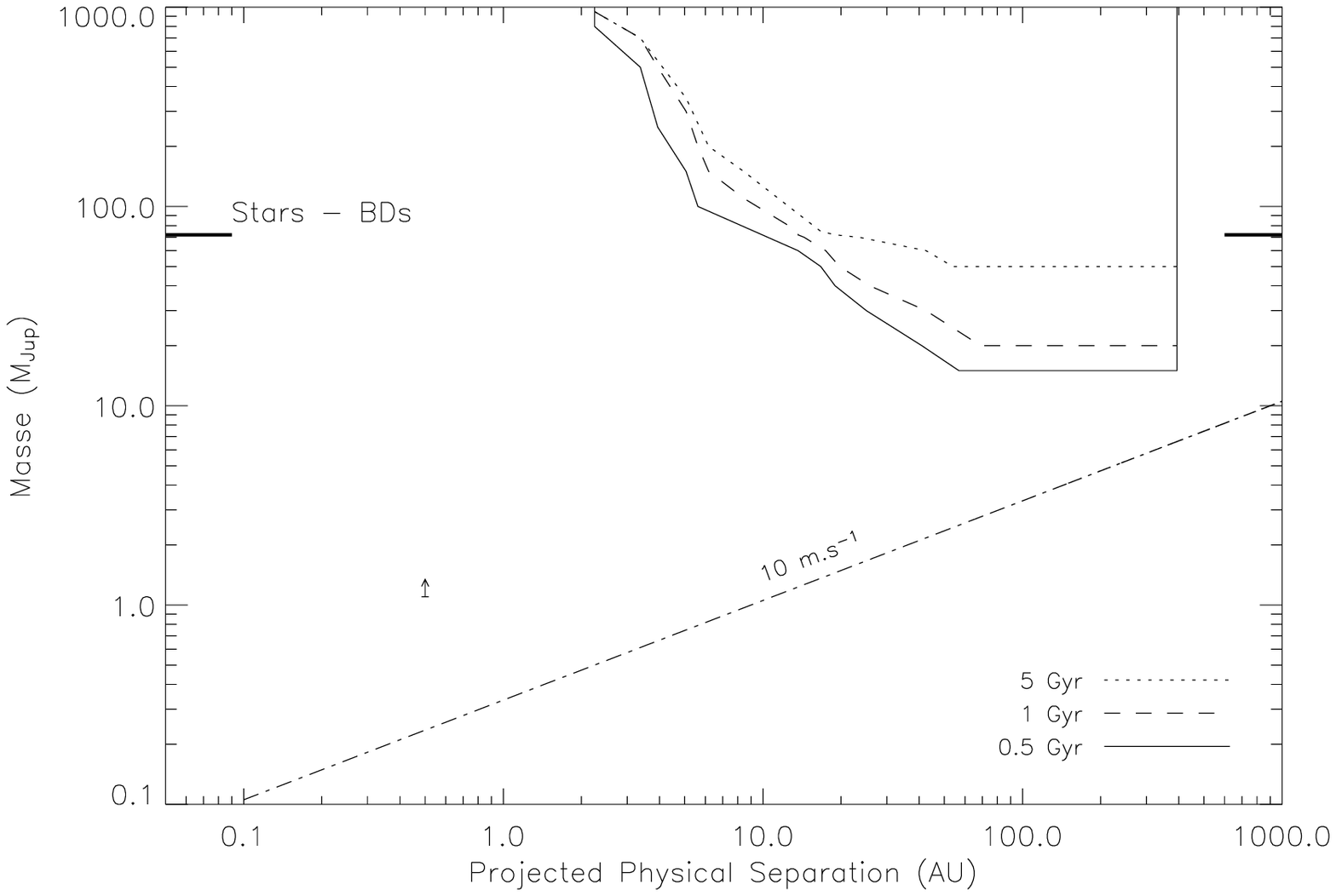}\hspace{.6cm}\\

   \includegraphics[width=8cm]{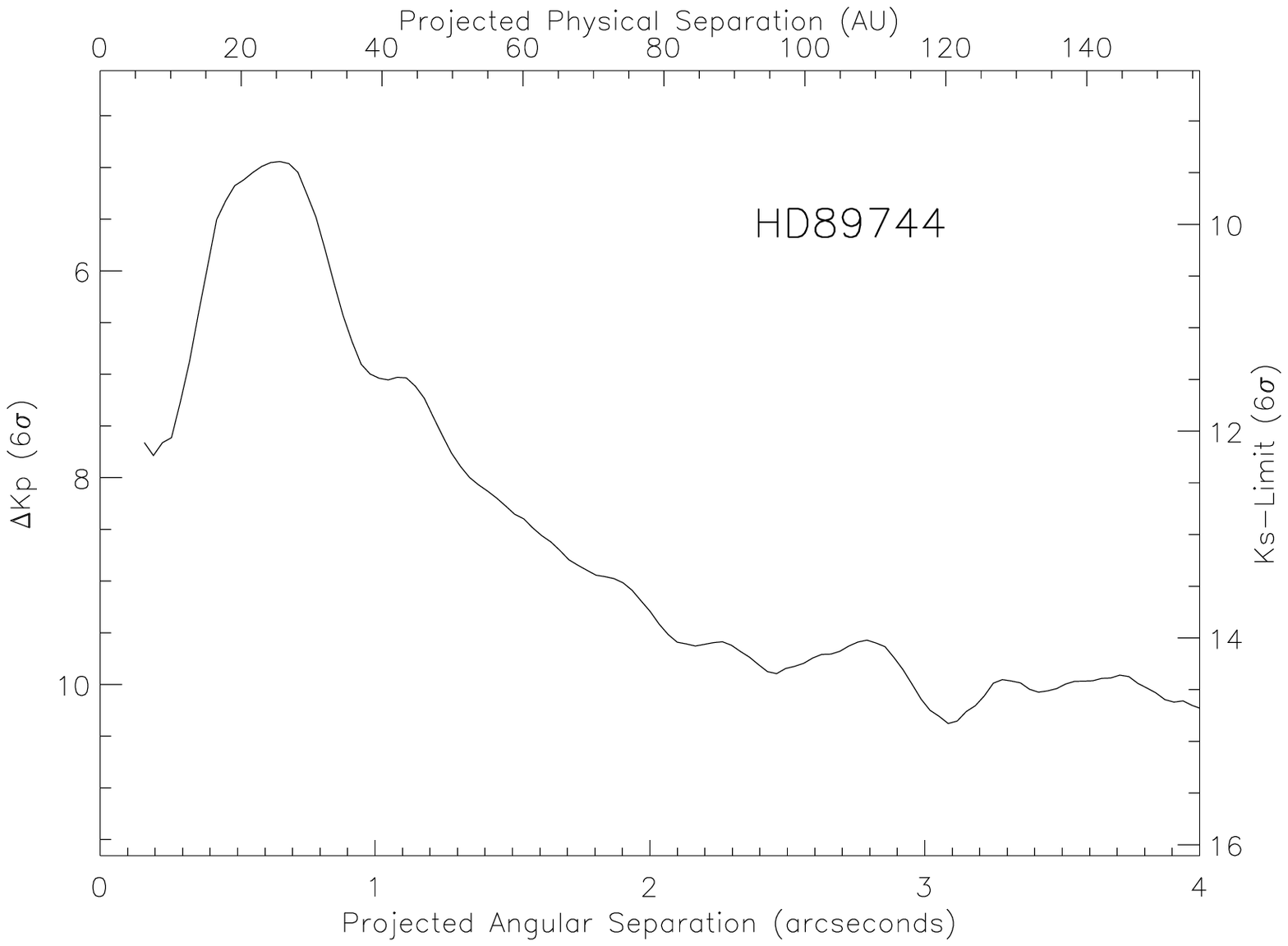}
\hspace{.6cm}
   \includegraphics[width=8cm]{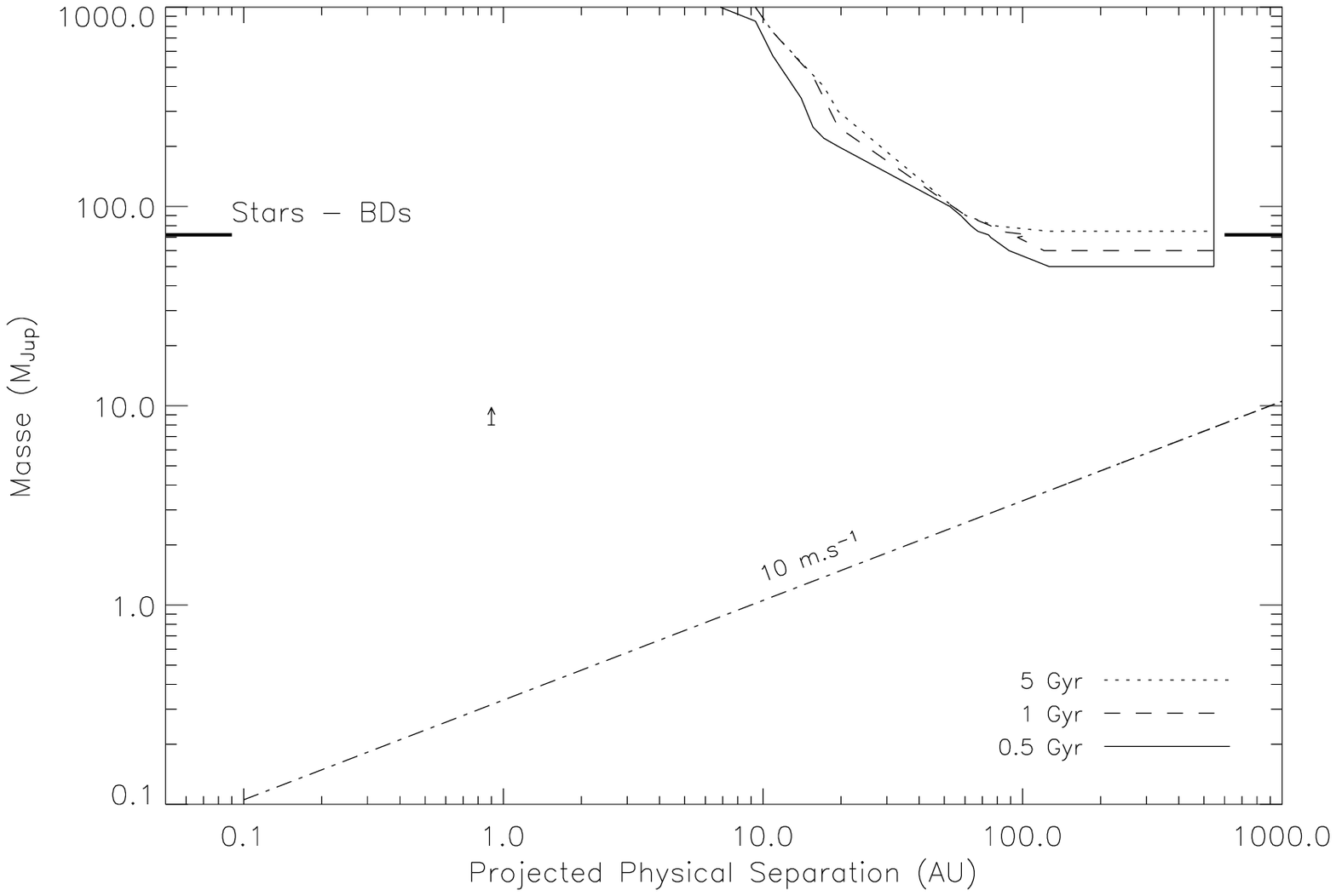}\hspace{.6cm}\\

   \includegraphics[width=8cm]{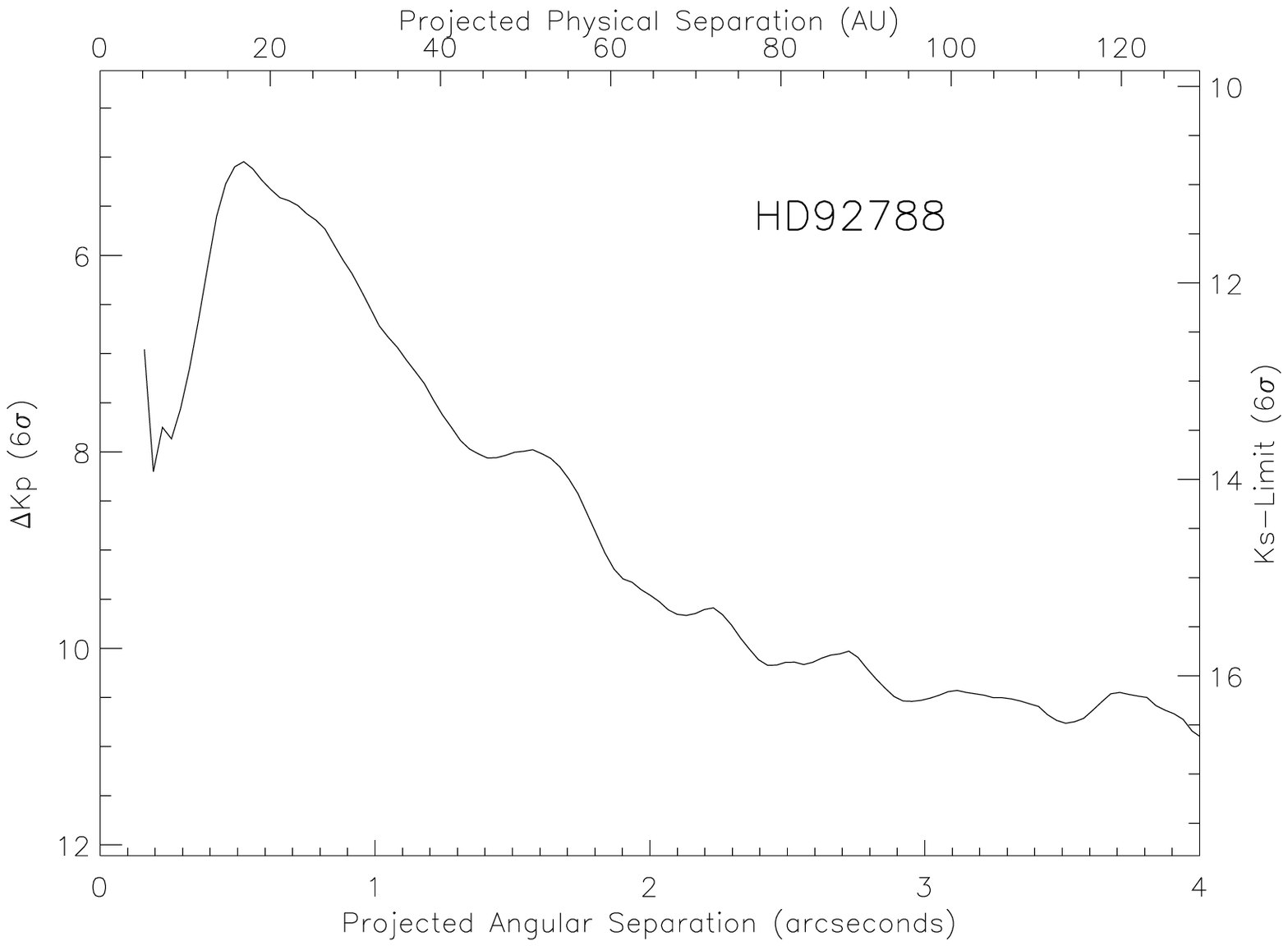}
\hspace{.6cm}
   \includegraphics[width=8cm]{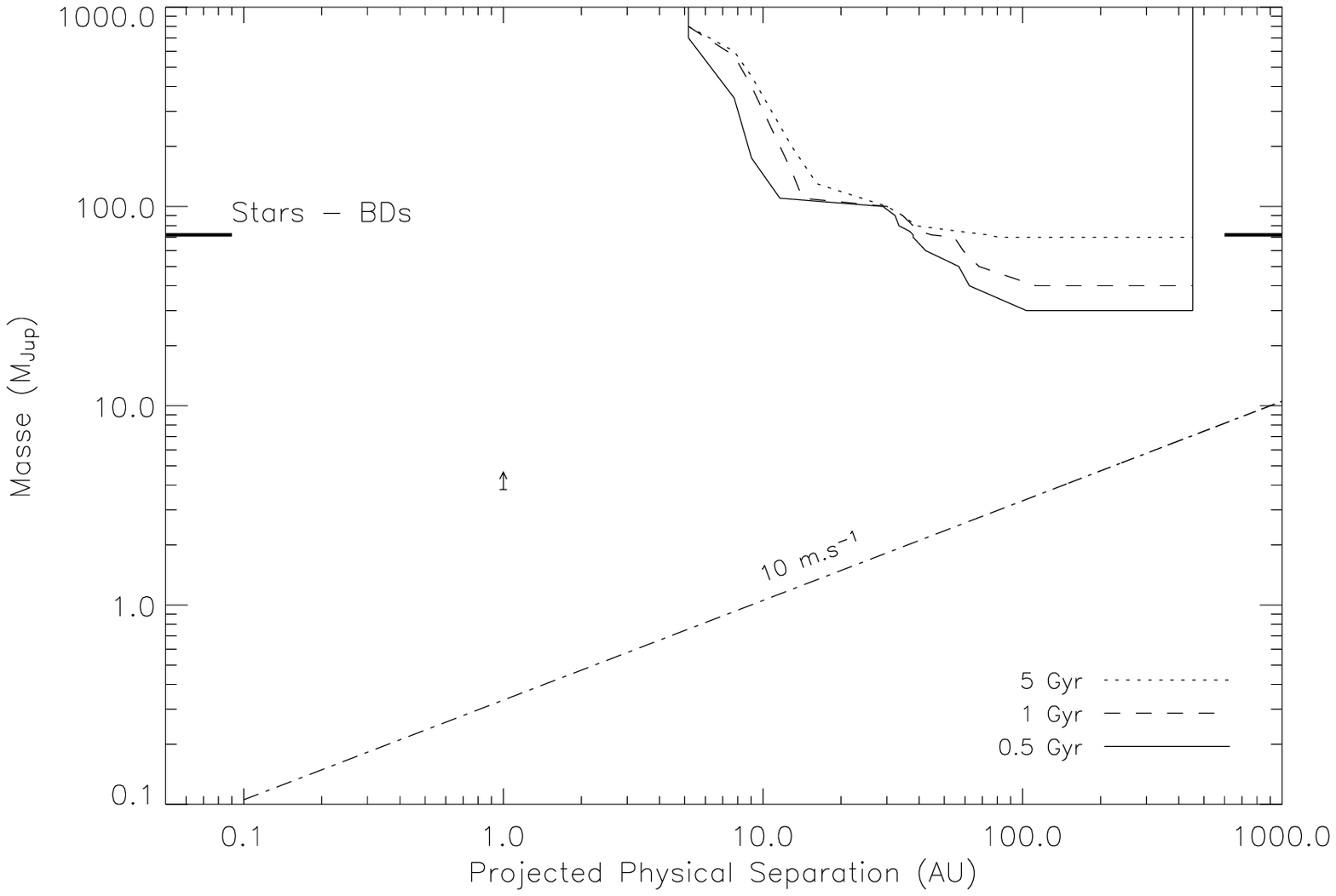}\hspace{.6cm}\\
 \caption{Detection Limits of HD\,28185 (July 2005, VLT/NACO, total exposure time of 160s), HD\,52265 (November 2003 VLT/NACO,, total exposure time of 300s), HD\,89744 (May 2003, CFHT/PUEO-KIR, total exposure time of 200s) and HD\,92788 (May 2003, CFHT/PUEO-KIR, total exposure time of 300s). See detail of the detection limit estimation in Section 2.2}
\label{figlim2}
   \end{figure*}

\begin{figure*}

\centering

   \includegraphics[width=8cm]{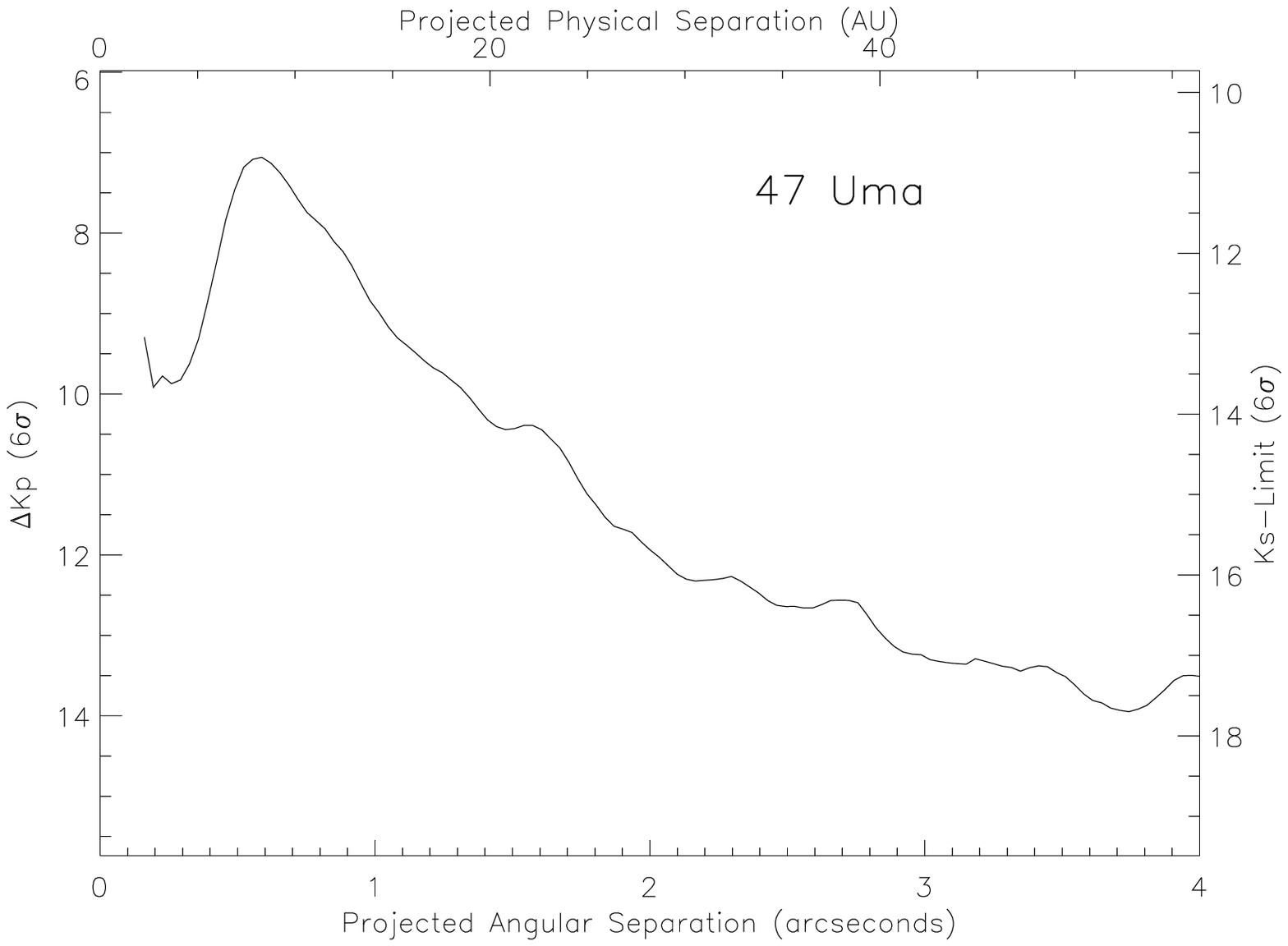}
\hspace{.6cm}
   \includegraphics[width=8cm]{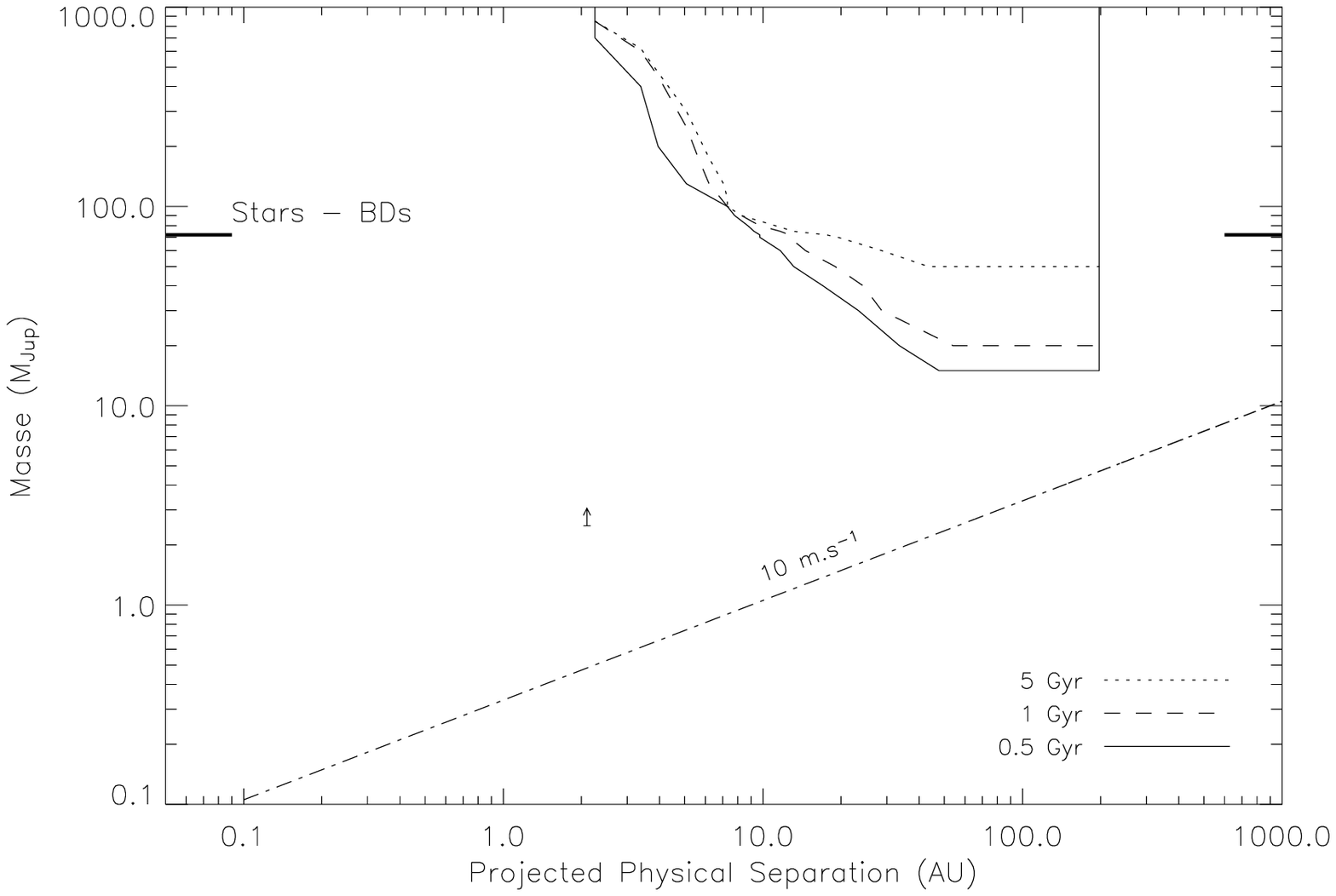}\hspace{.6cm}\\

   \includegraphics[width=8cm]{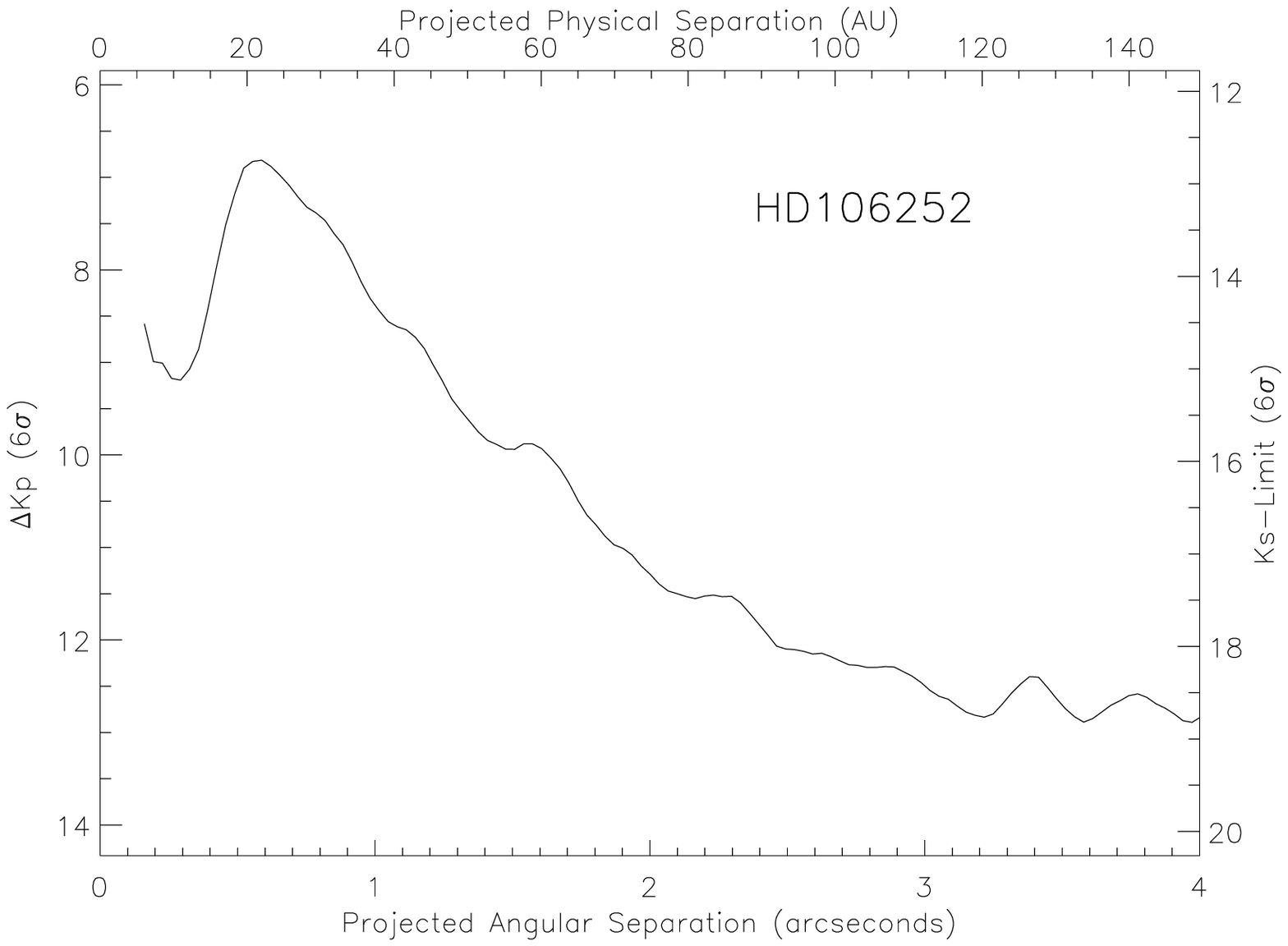}
\hspace{.6cm}
   \includegraphics[width=8cm]{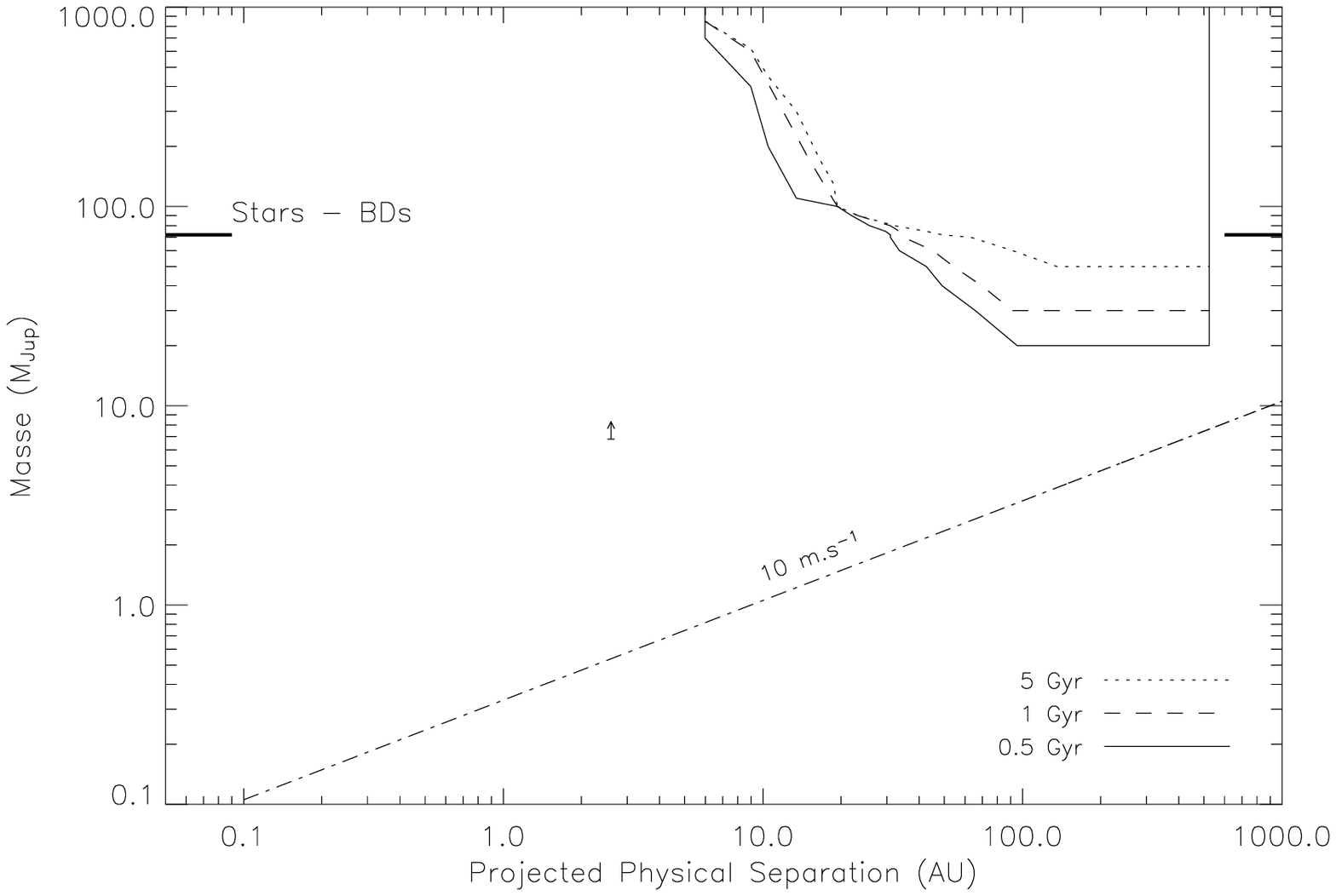}\hspace{.6cm}\\

   \includegraphics[width=8cm]{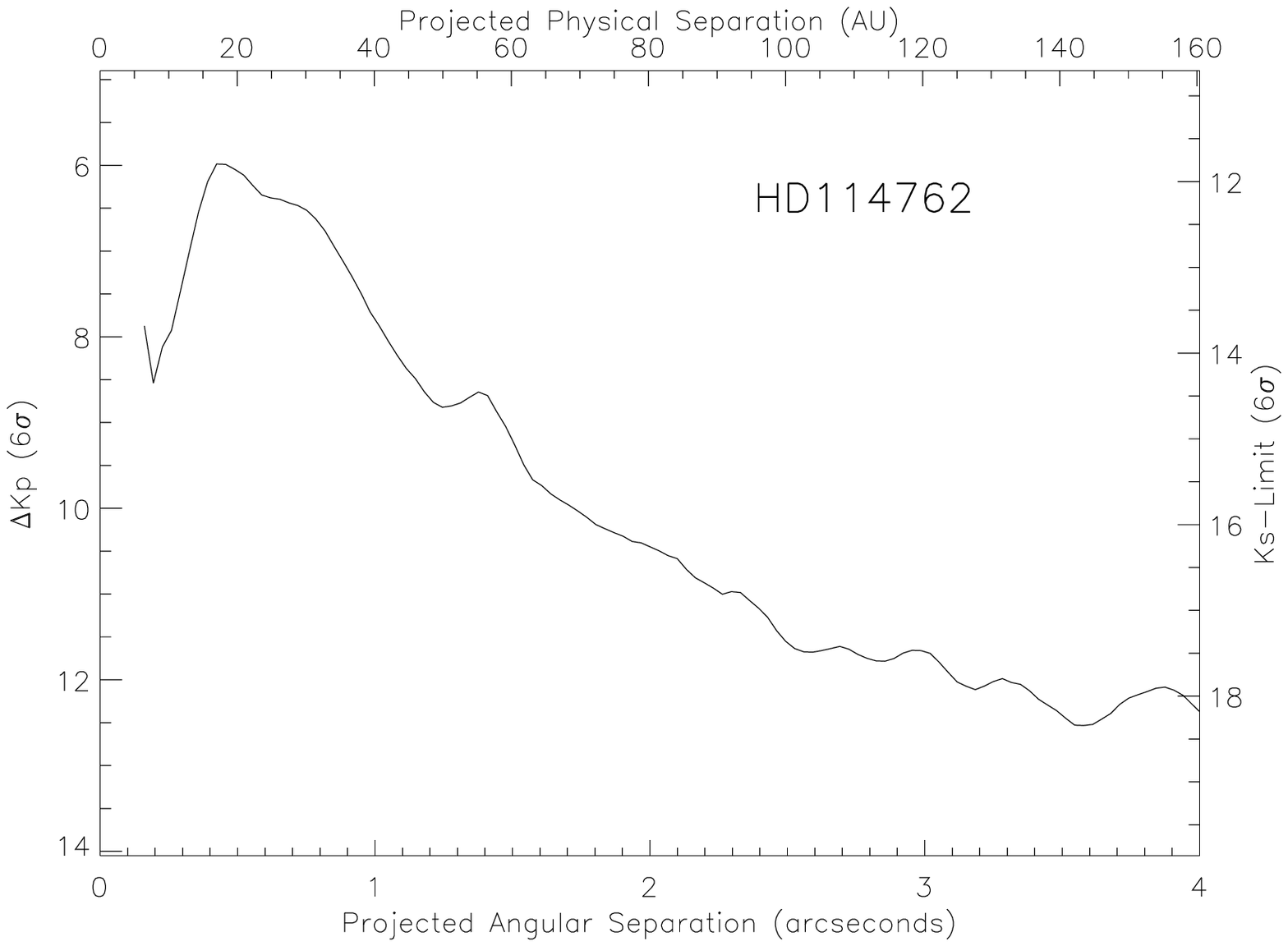}
\hspace{.6cm}
   \includegraphics[width=8cm]{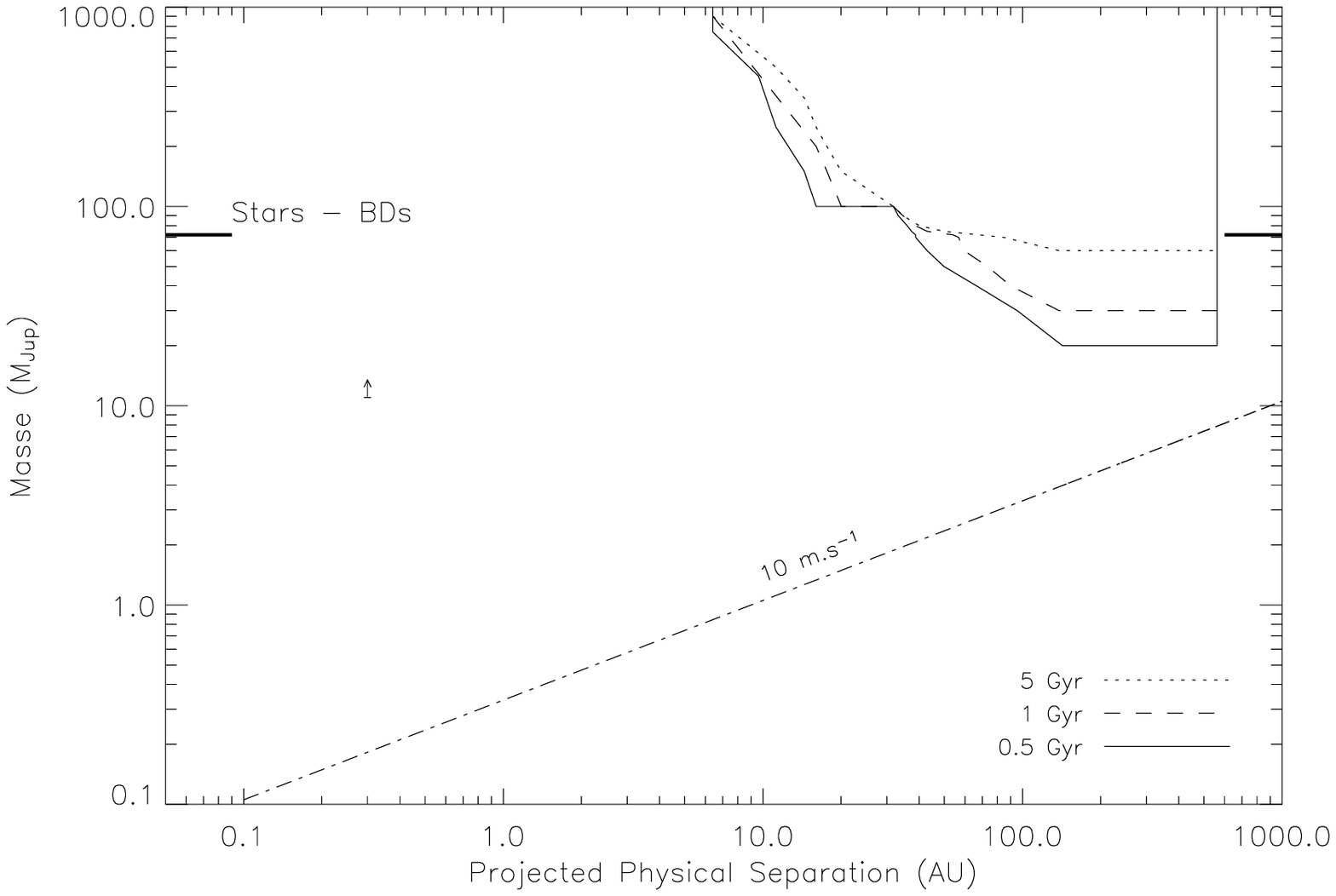}\hspace{.6cm}\\

   \includegraphics[width=8cm]{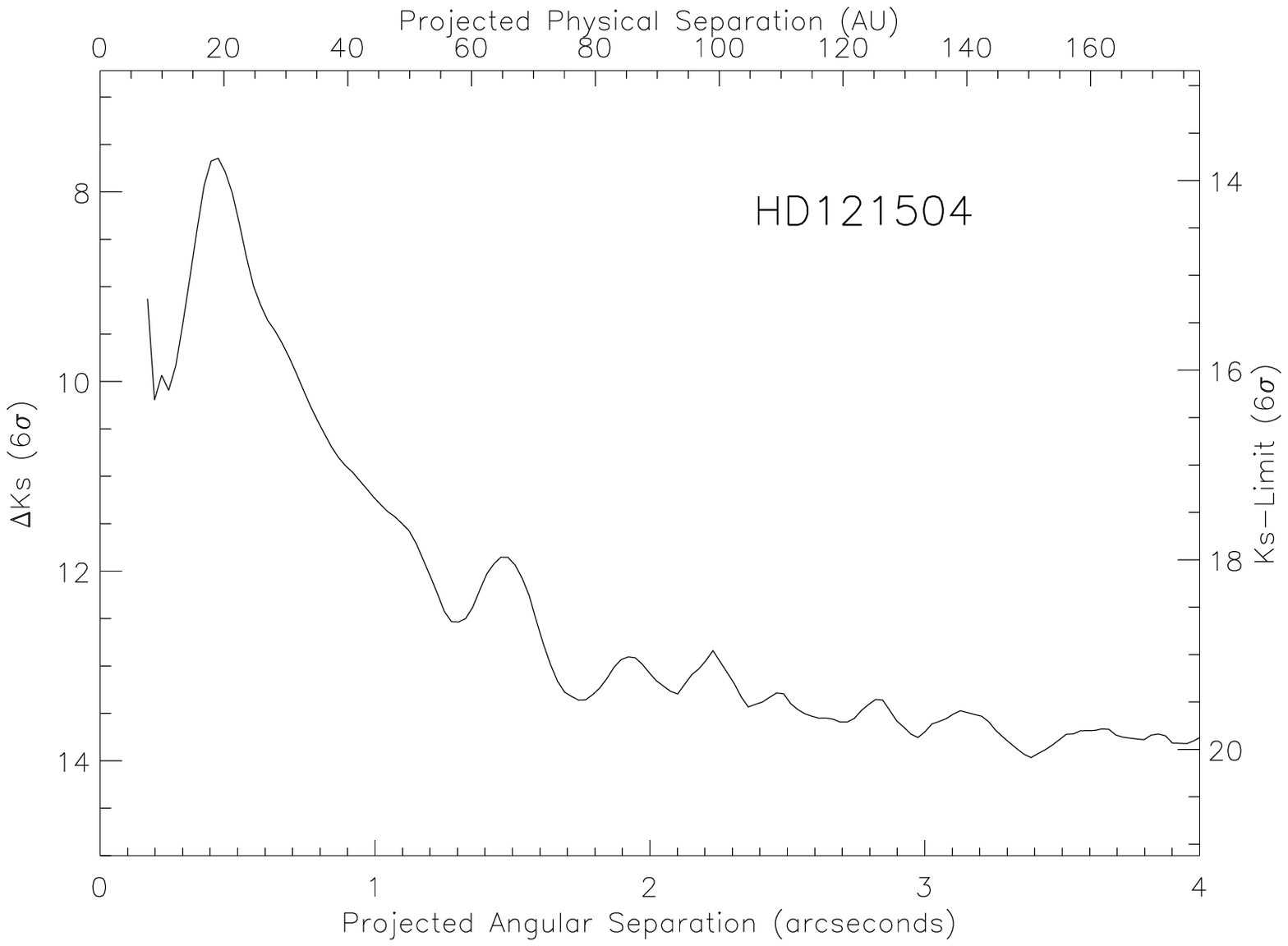}
\hspace{.6cm}
   \includegraphics[width=8cm]{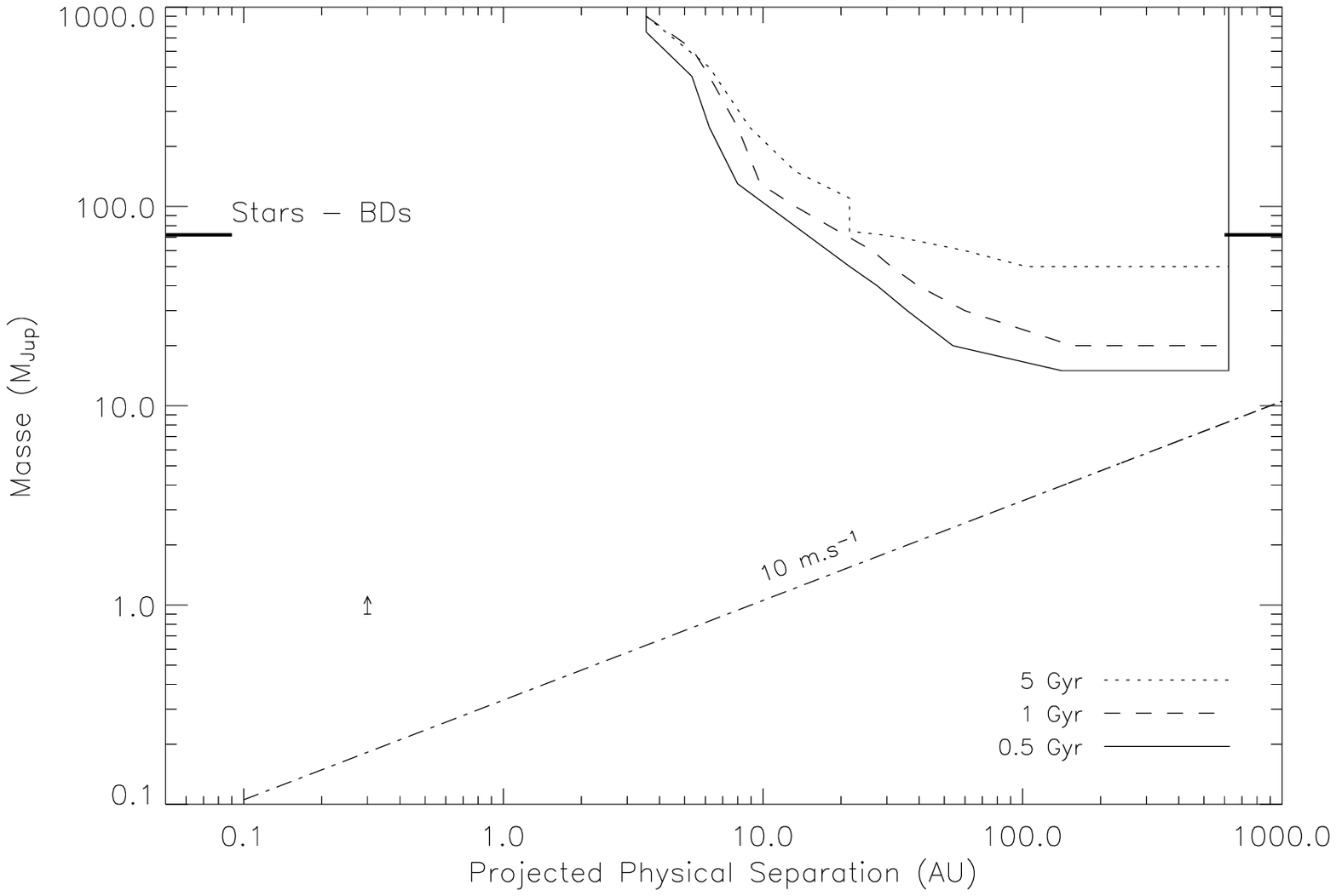}\hspace{.6cm}\\

\caption{Detection Limits of HD\,95128 (47\,Uma) (May 2003, CFHT/PUEO-KIR, total exposure time of 140s), HD\,106252 (May 2003, CFHT/PUEO-KIR, total exposure time of 300s) and HD\,114762 (May 2003, CFHT/PUEO-KIR, total exposure time of 300s) and HD\,121504 (July 2005, VLT/NACO, total exposure time of 300s). See detail of the detection limit estimation in Section 2.2}   
\label{figlim3}
   \end{figure*}

\begin{figure*}

   \centering
   \includegraphics[width=8cm]{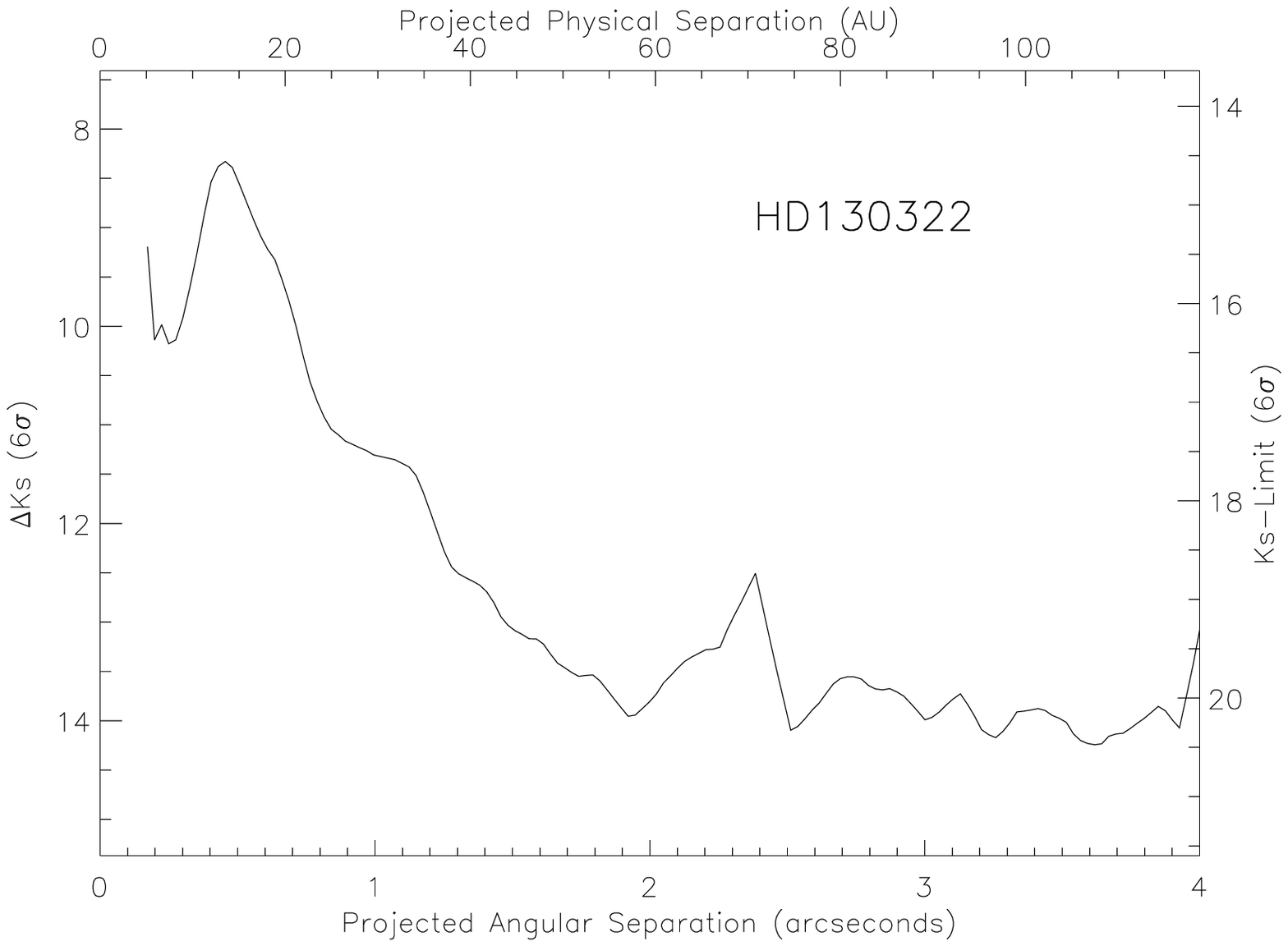}
\hspace{.6cm}
   \includegraphics[width=8cm]{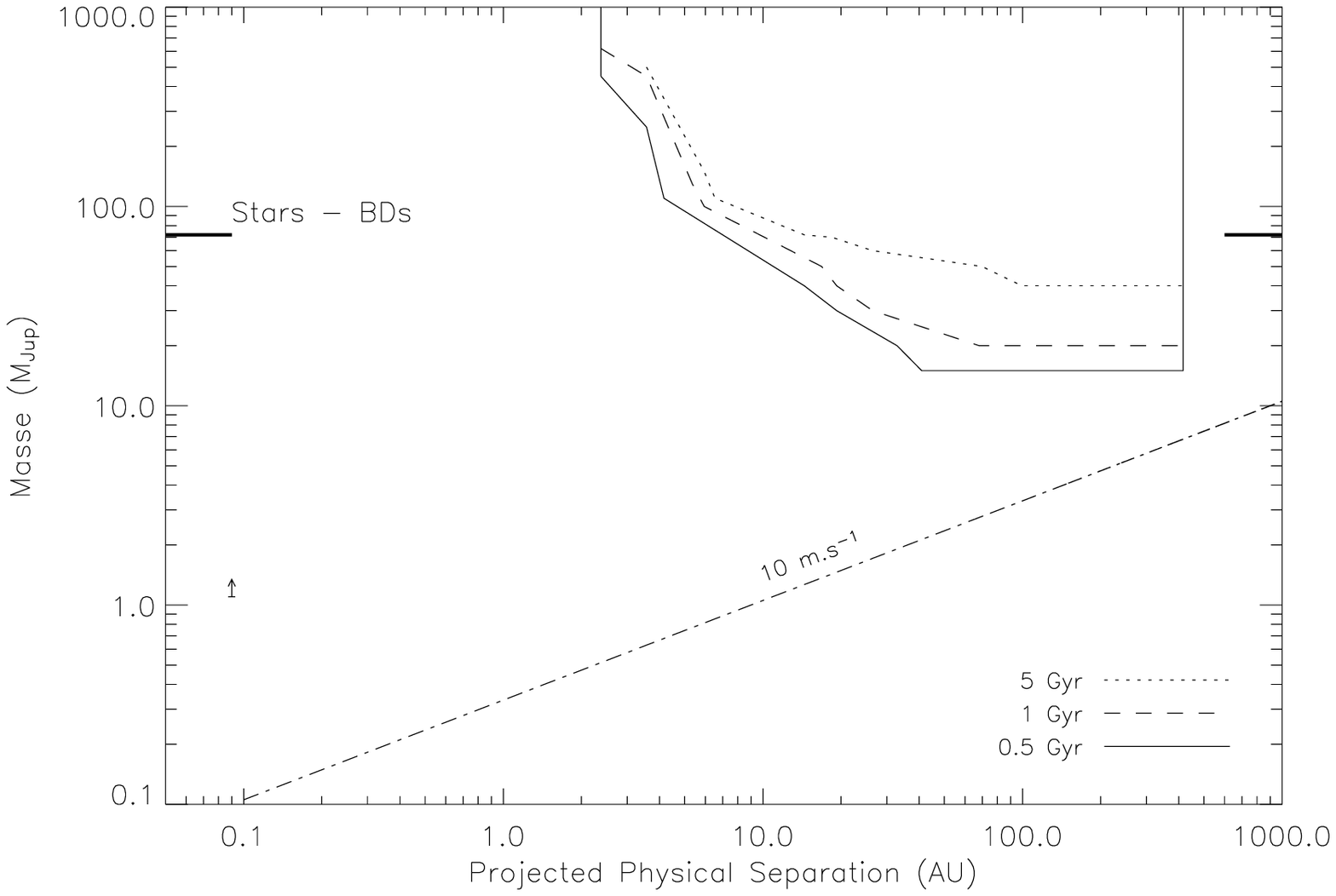}\hspace{.6cm}\\

   \includegraphics[width=8cm]{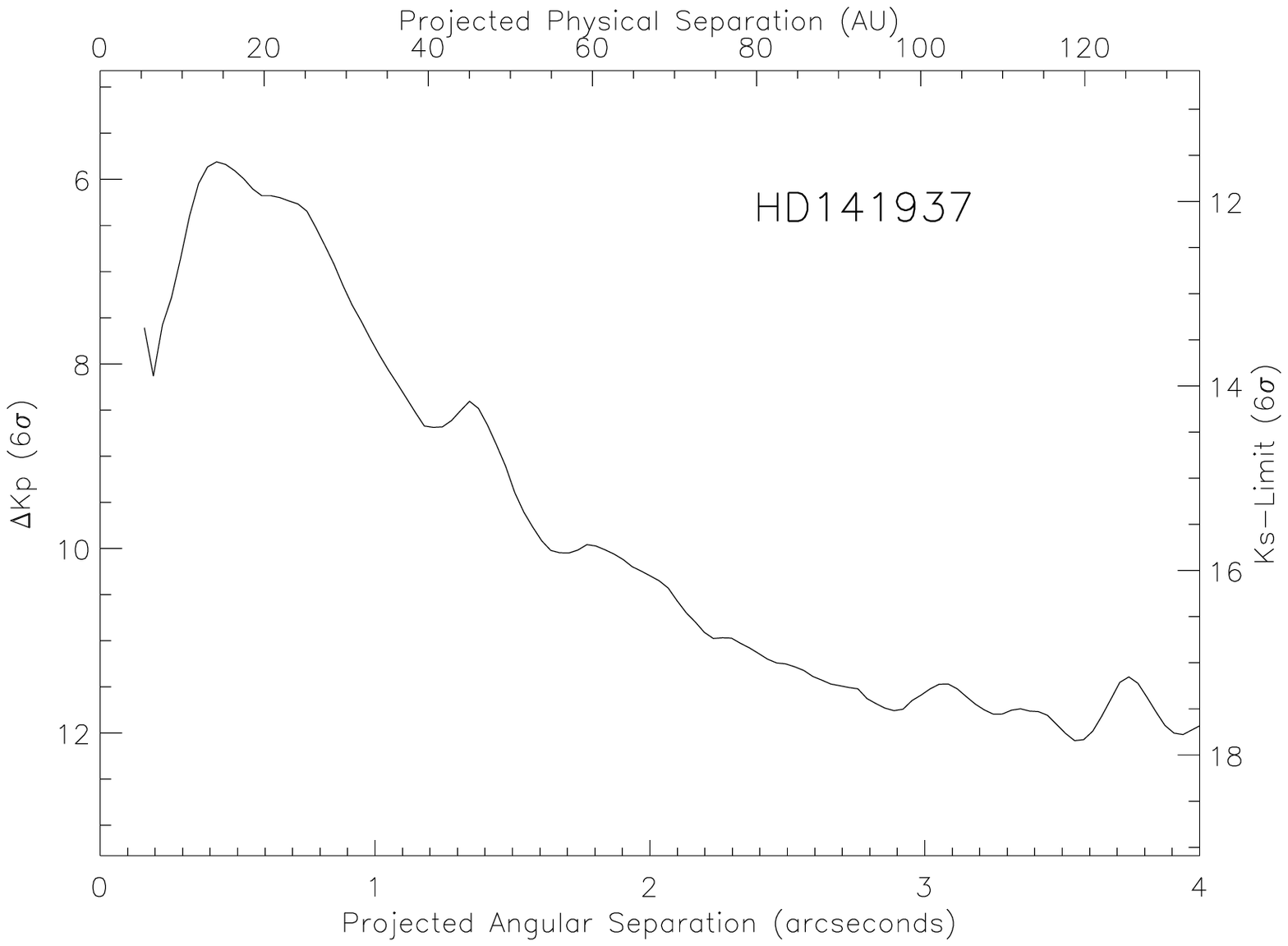}
\hspace{.6cm}
   \includegraphics[width=8cm]{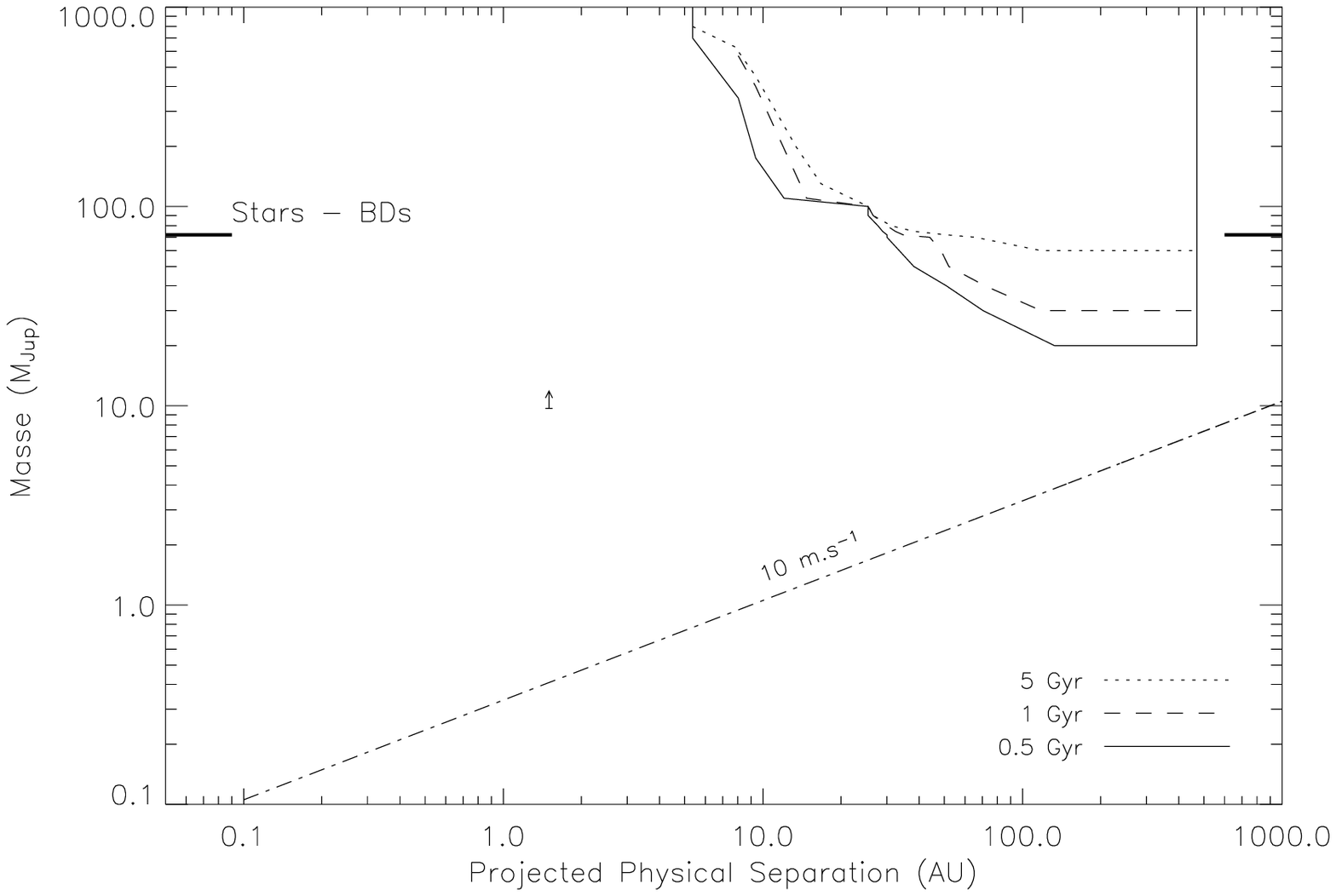}\hspace{.6cm}\\

   \includegraphics[width=8cm]{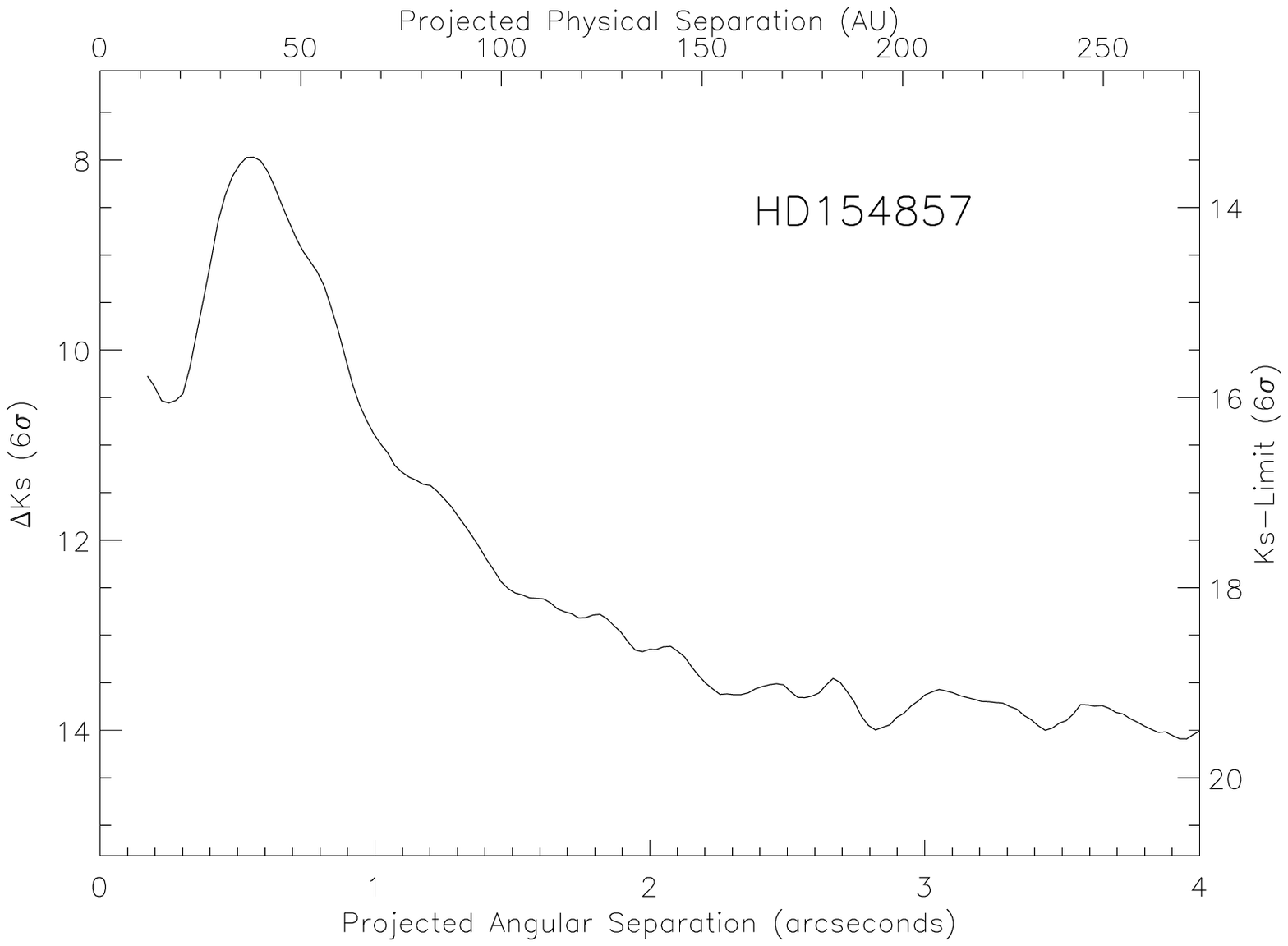}
\hspace{.6cm}
   \includegraphics[width=8cm]{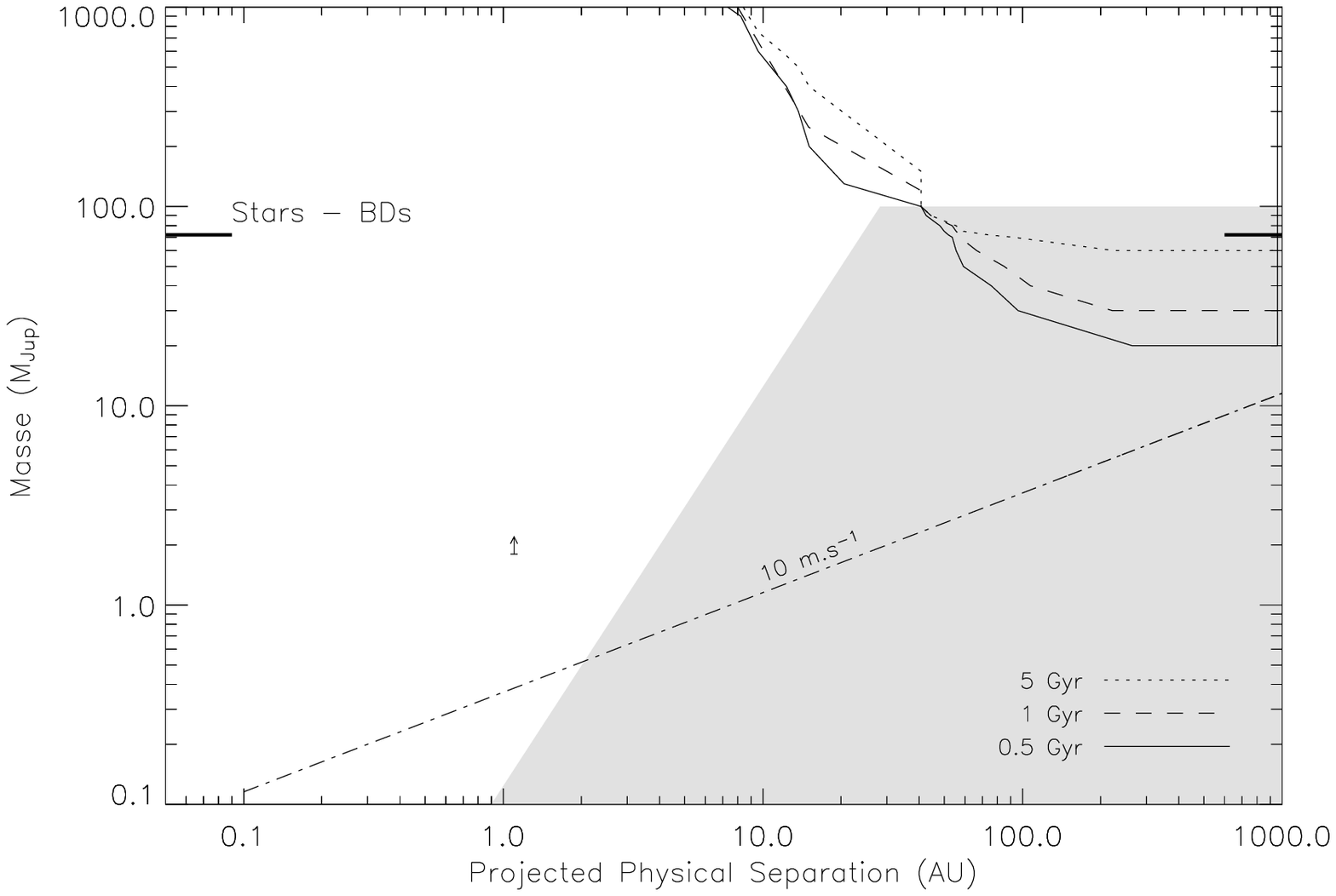}\hspace{.6cm}\\

   \includegraphics[width=8cm]{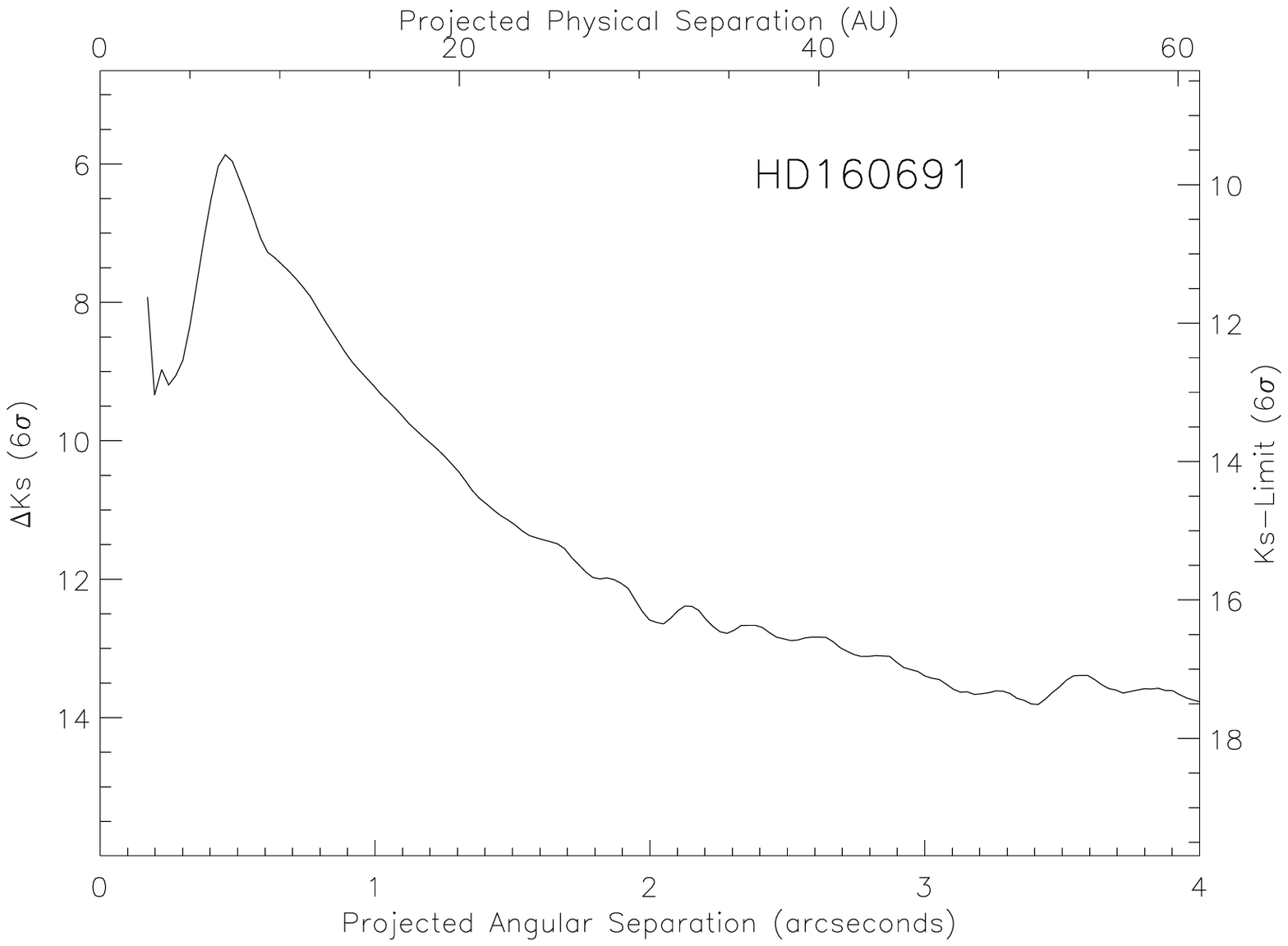}
\hspace{.6cm}
   \includegraphics[width=8cm]{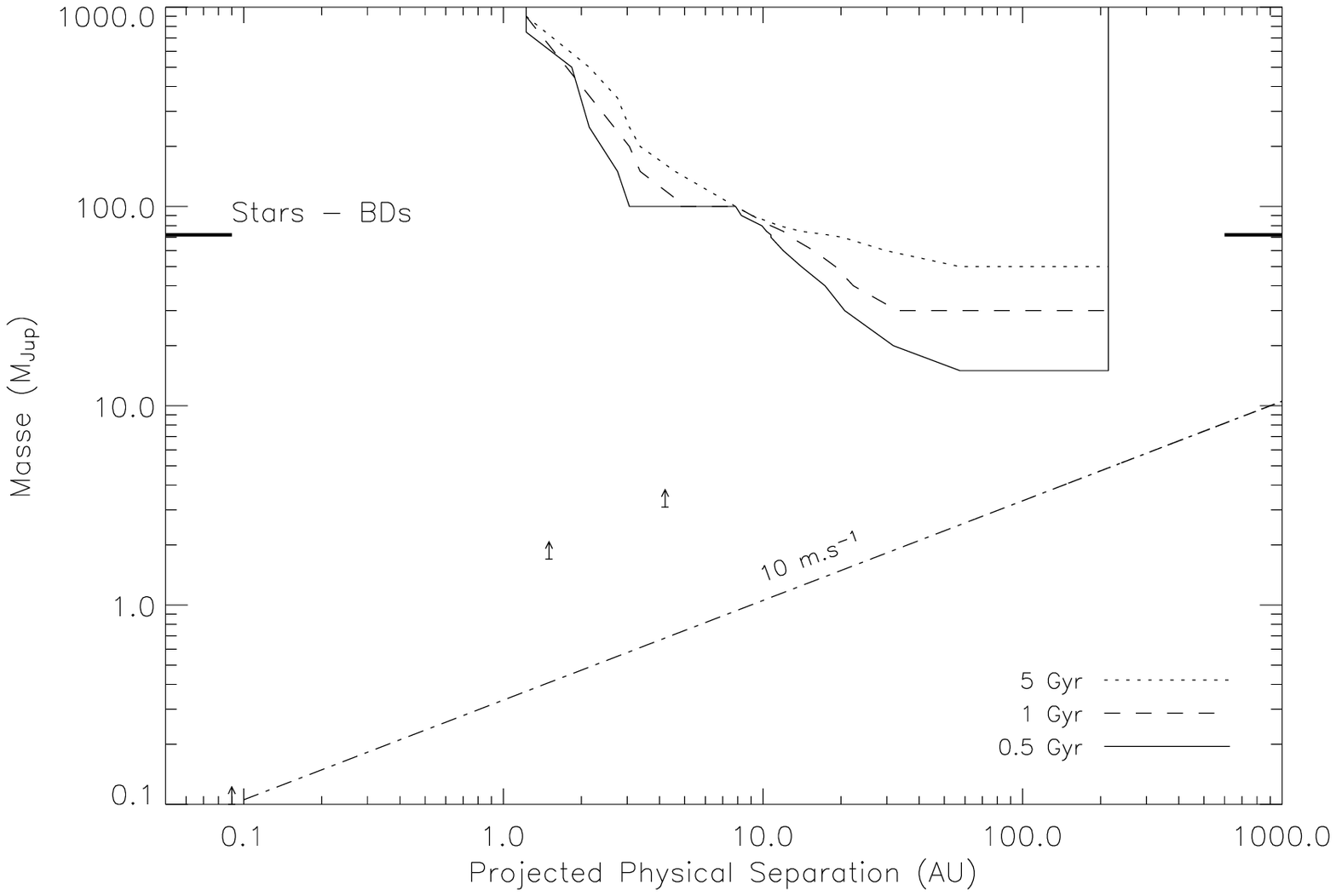}\hspace{.6cm}\\
\caption{Detection Limits of HD\,130322 (July 2005, VLT/NACO, total exposure time of 300s), HD\,141937 (May 2003, CFHT/PUEO-KIR, total exposure time of 200s), HD\,154857 (July 2005, VLT/NACO, total exposure time of 300s) and HD\,160691 (July 2005, VLT/NACO, total exposure time of 240s). See detail of the detection limit estimation in Section 2.2}
\label{figlim4}
\end{figure*}

\begin{figure*}

   \centering

   \includegraphics[width=8cm]{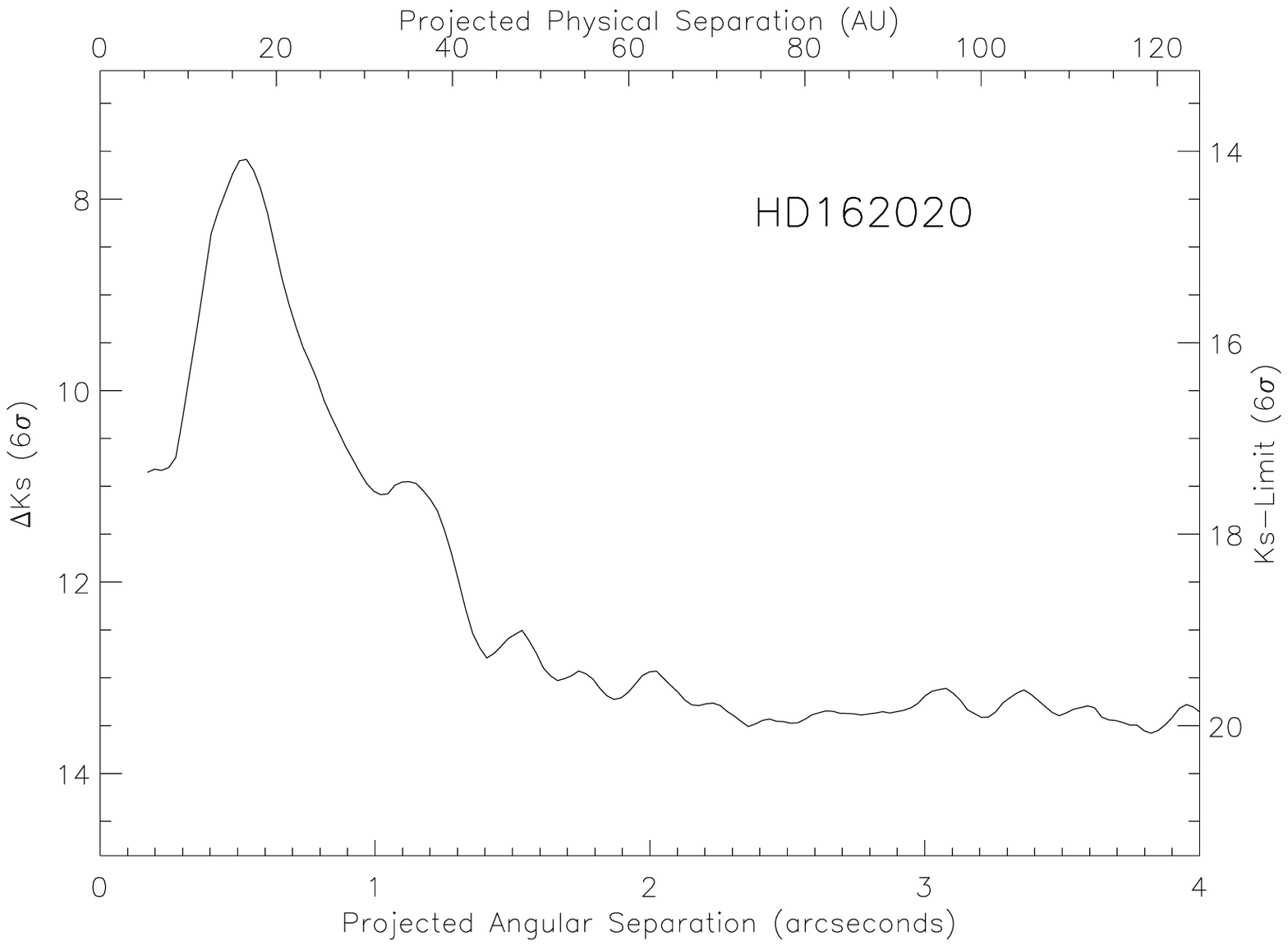}
\hspace{.6cm}
   \includegraphics[width=8cm]{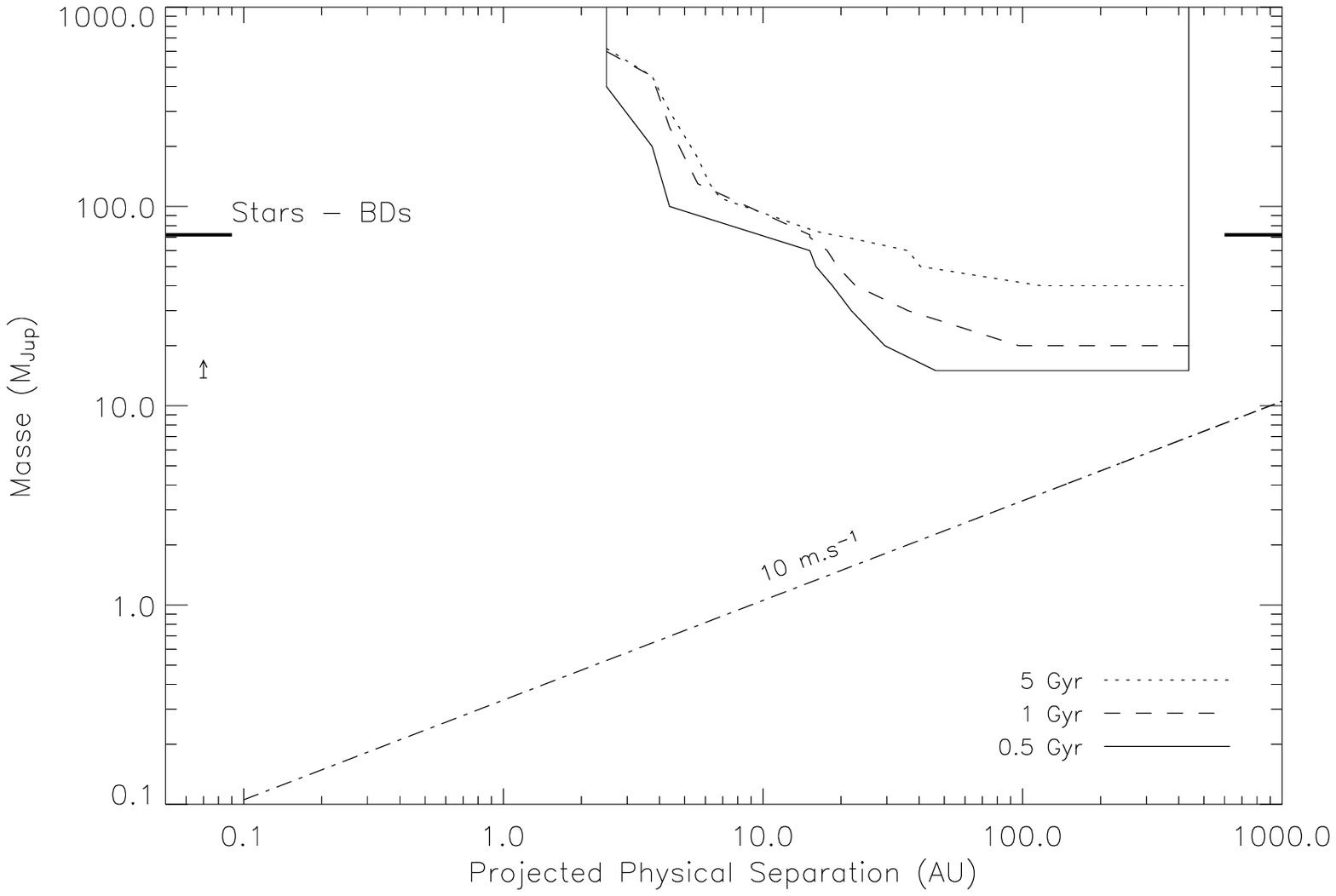}\hspace{.6cm}\\

   \includegraphics[width=8cm]{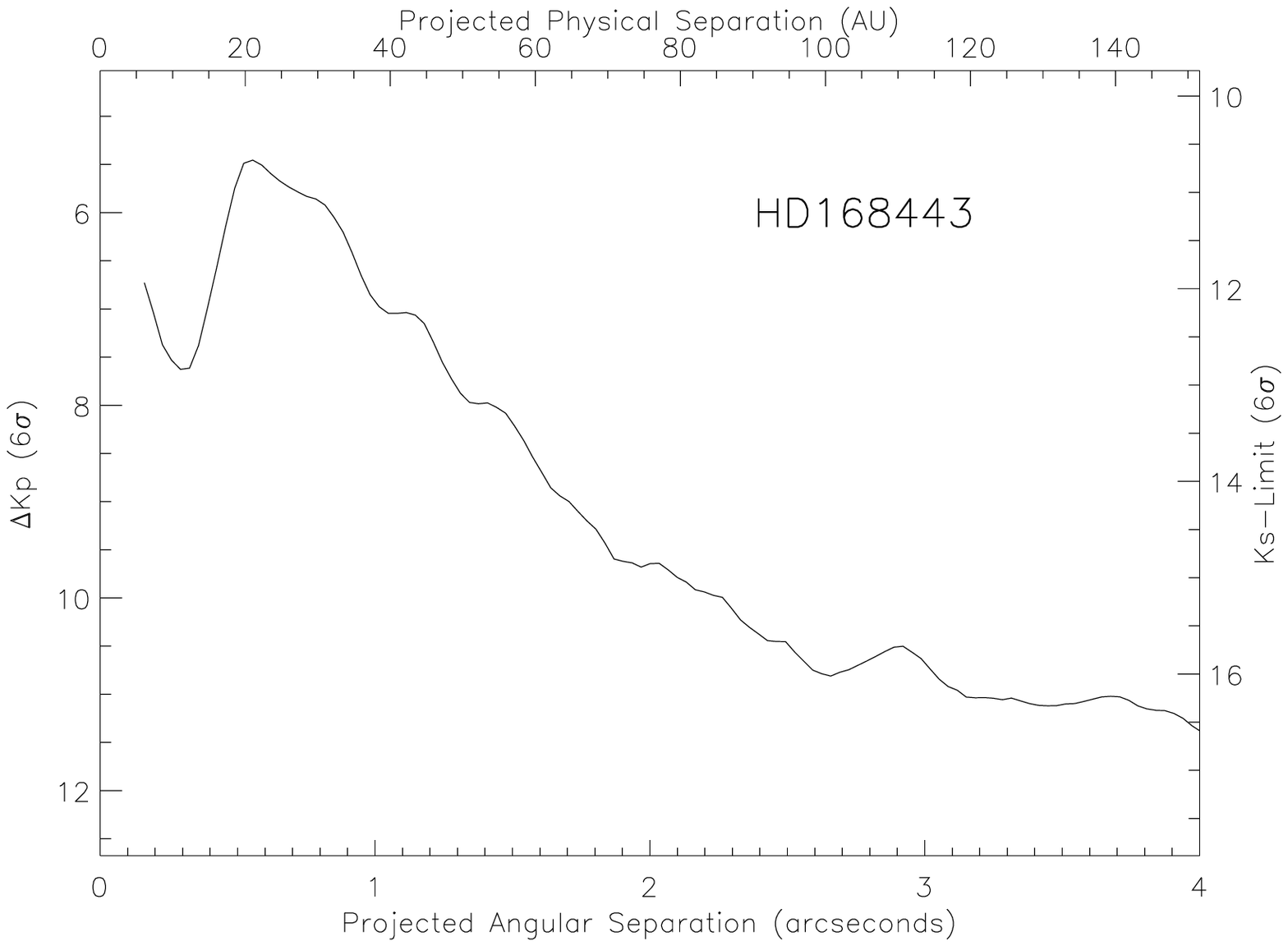}
\hspace{.6cm}
   \includegraphics[width=8cm]{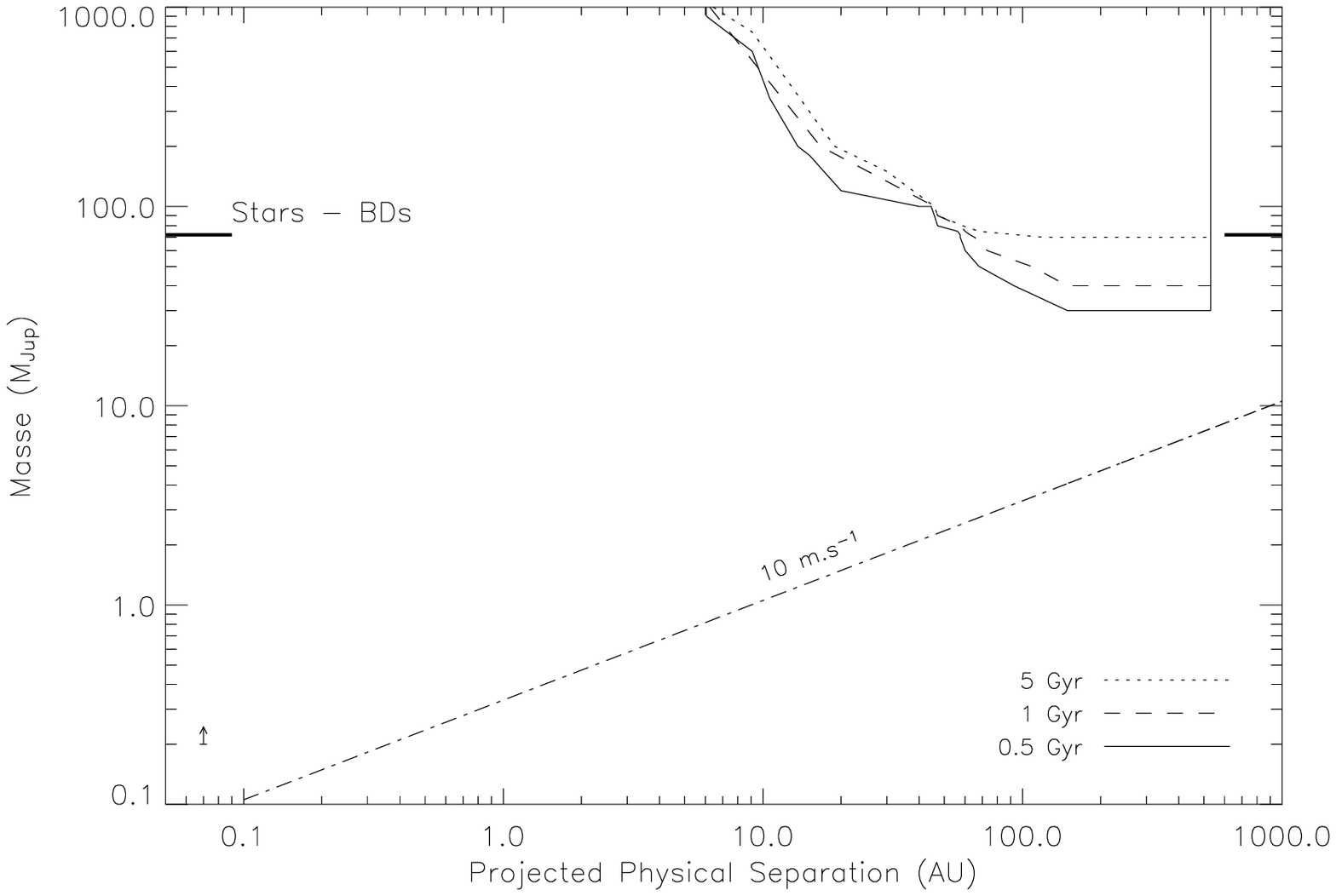}\hspace{.6cm}\\

   \includegraphics[width=8cm]{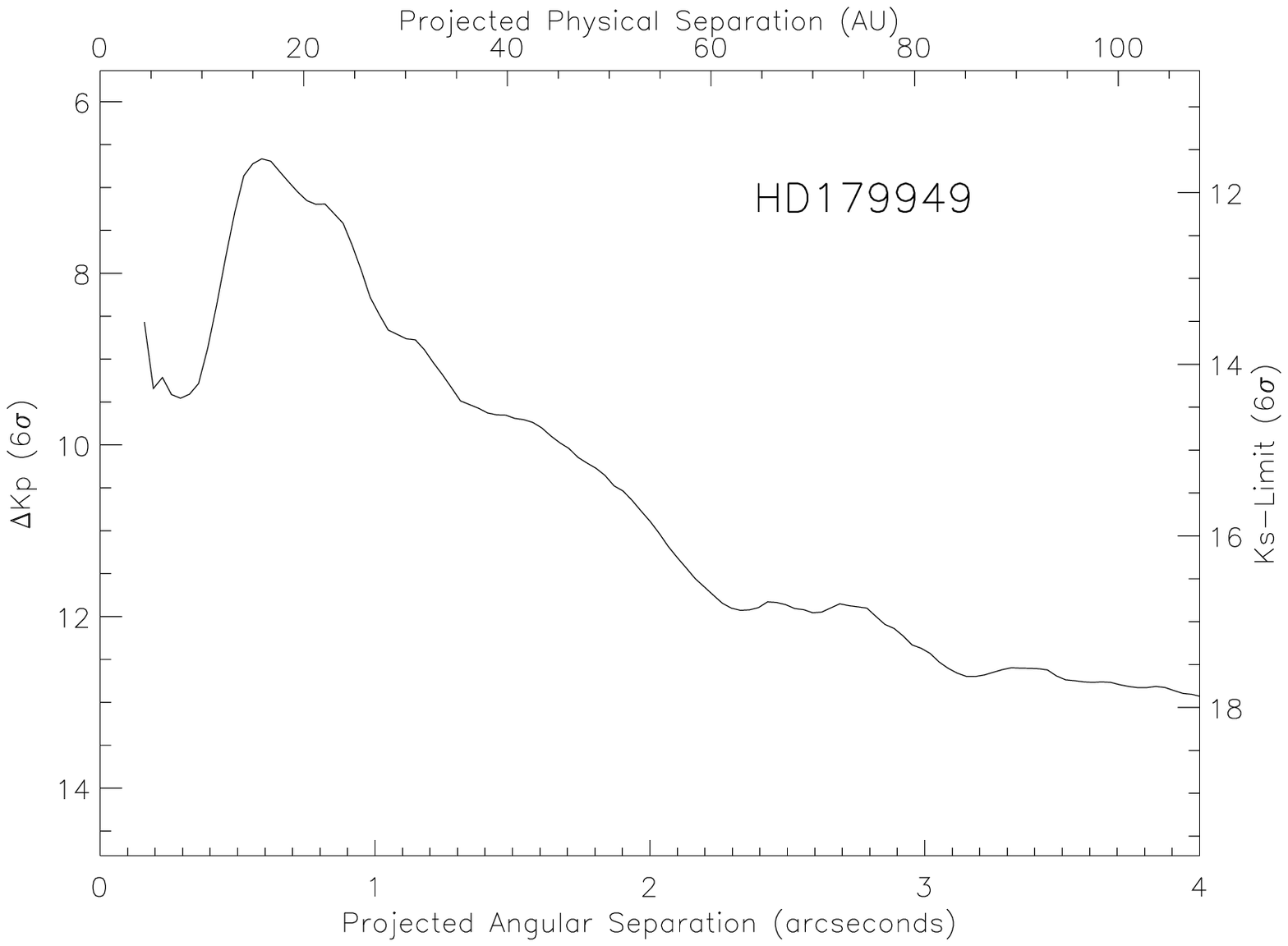}
\hspace{.6cm}
   \includegraphics[width=8cm]{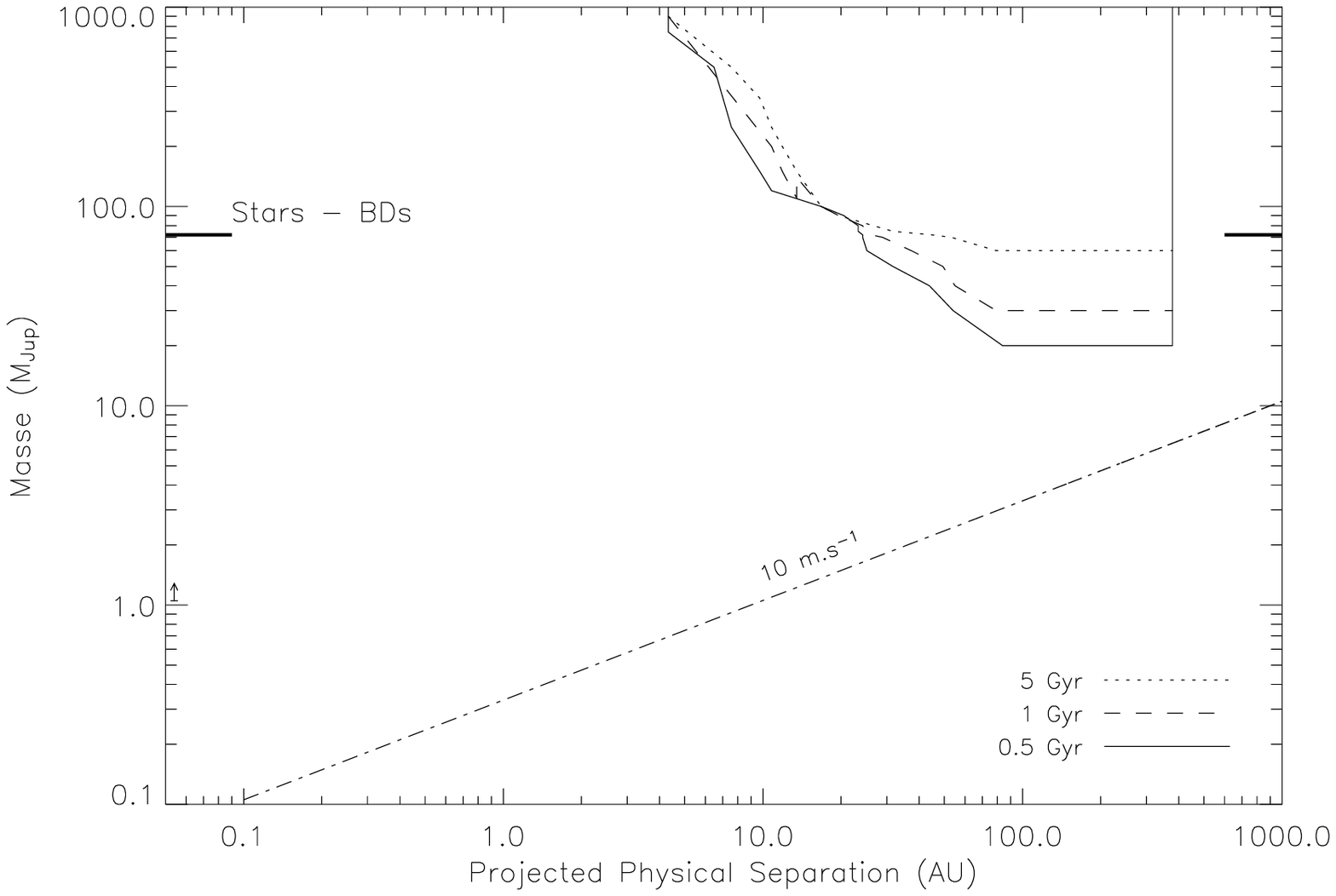}\hspace{.6cm}\\

   \includegraphics[width=8cm]{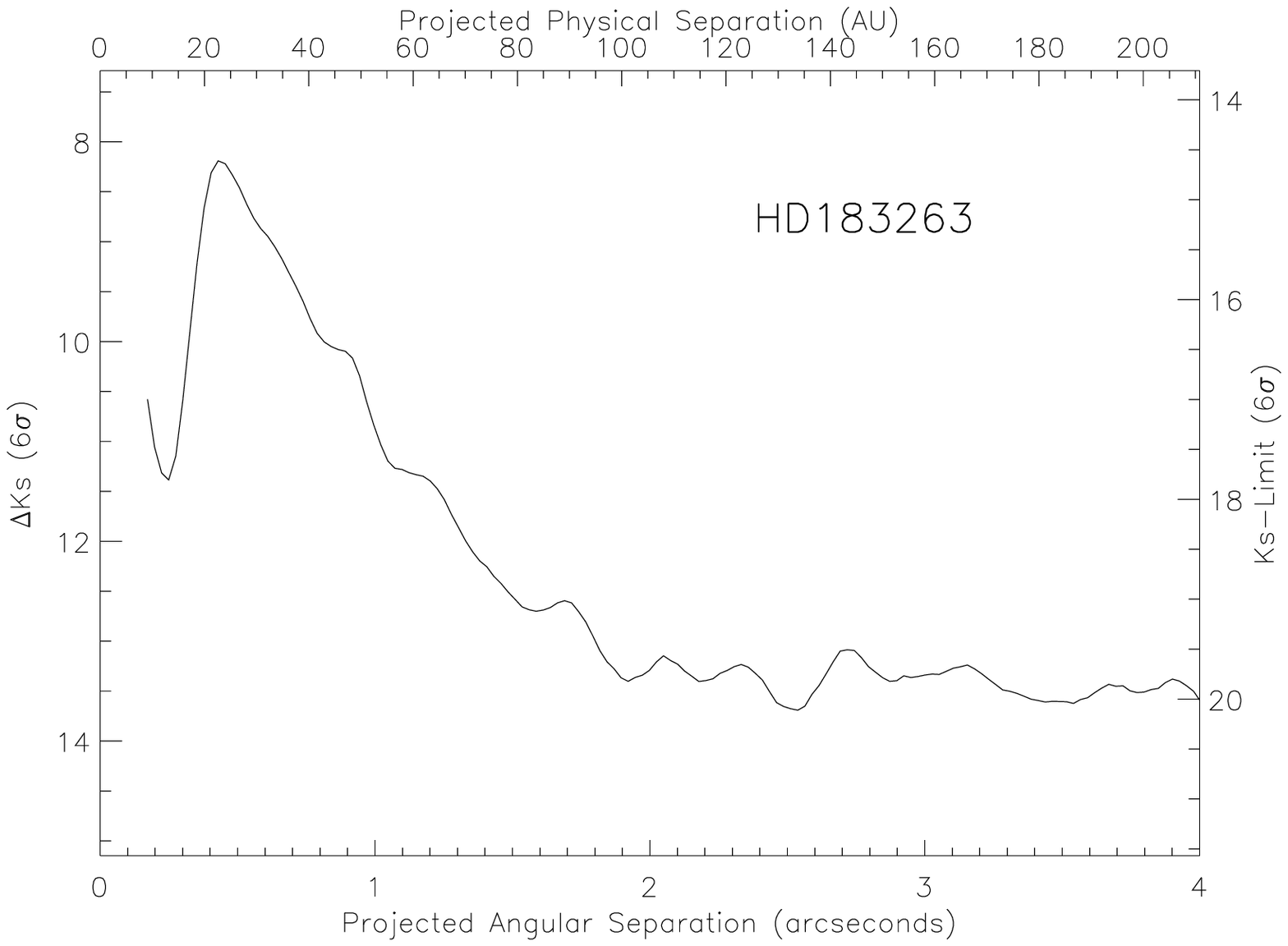}
\hspace{.6cm}
   \includegraphics[width=8cm]{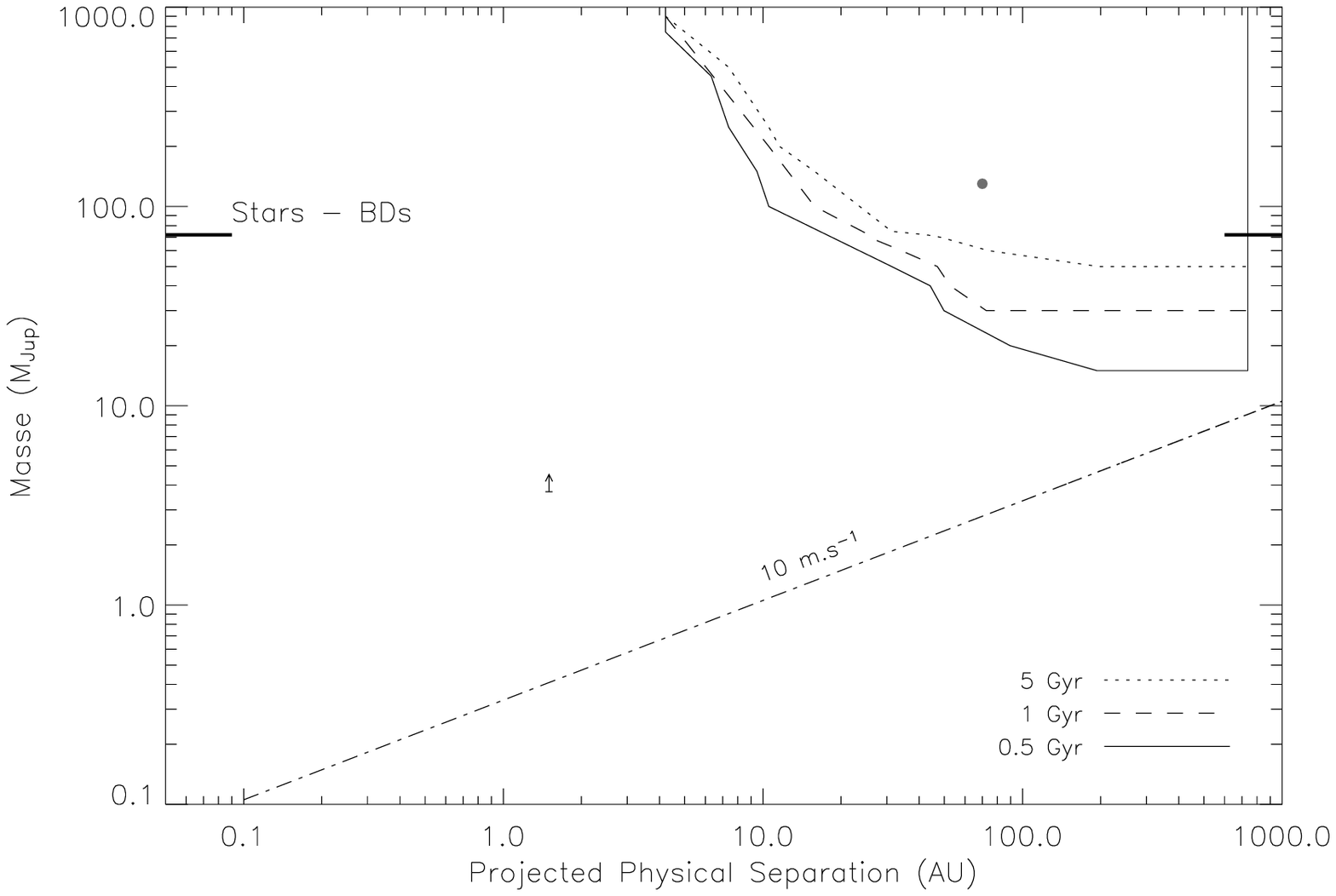}\hspace{.6cm}\\
\caption{Detection Limits of HD\,162020 (July 2005, VLT/NACO, total exposure time of 300s),  HD\,168443 (May 2003, CFHT/PUEO-KIR, total exposure time of 300s), HD\,179949 (November 2003, CFHT/PUEO-KIR, total exposure time of 300s) and HD\,183263 (July 2005, VLT/NACO, total exposure time of 300s). See detail of the detection limit estimation in Section 2.2}
\label{figlim5}
\end{figure*}

\begin{figure*}

 \centering
   \includegraphics[width=8cm]{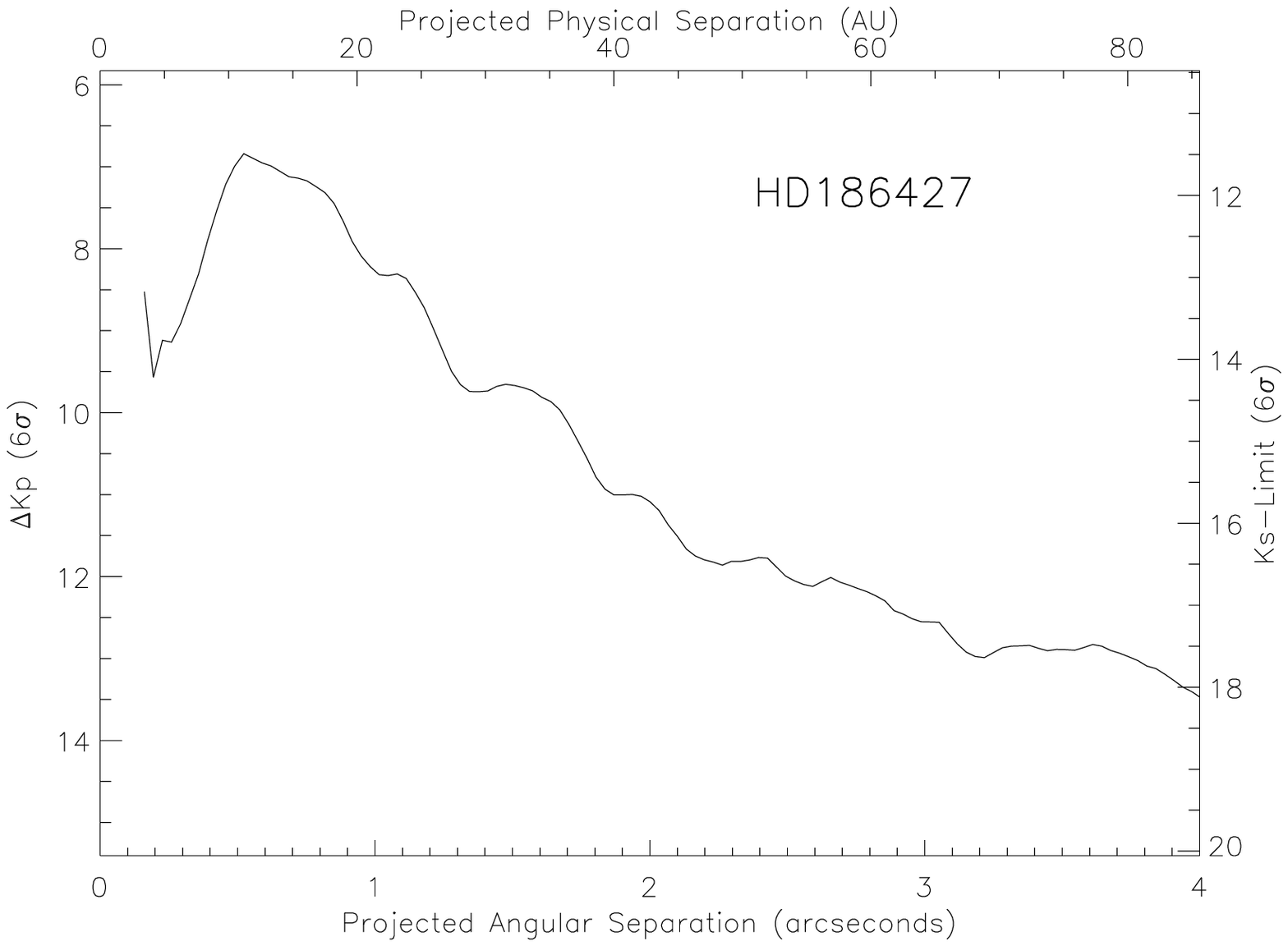}
\hspace{.6cm}
   \includegraphics[width=8cm]{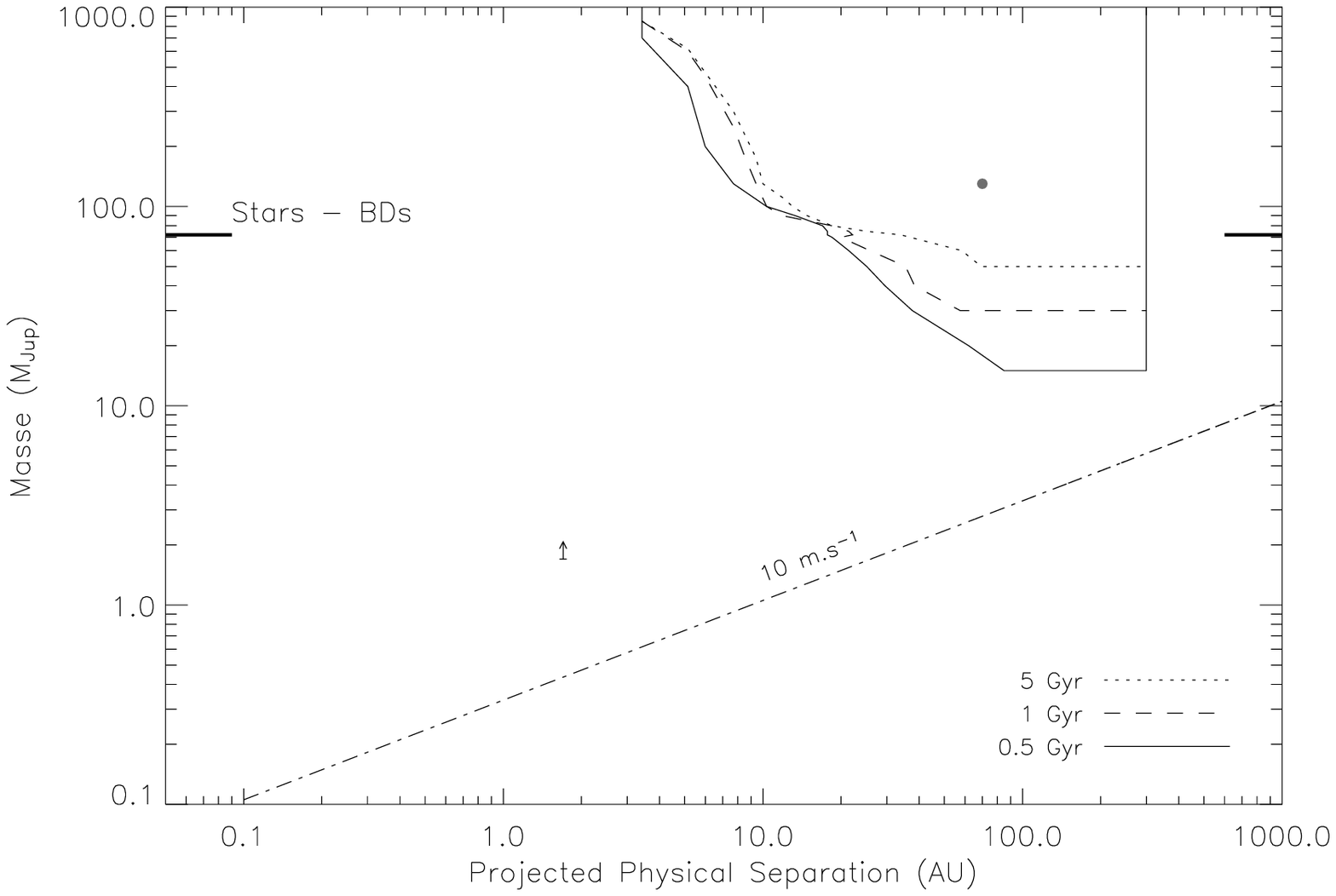}\hspace{.6cm}\\

   \includegraphics[width=8cm]{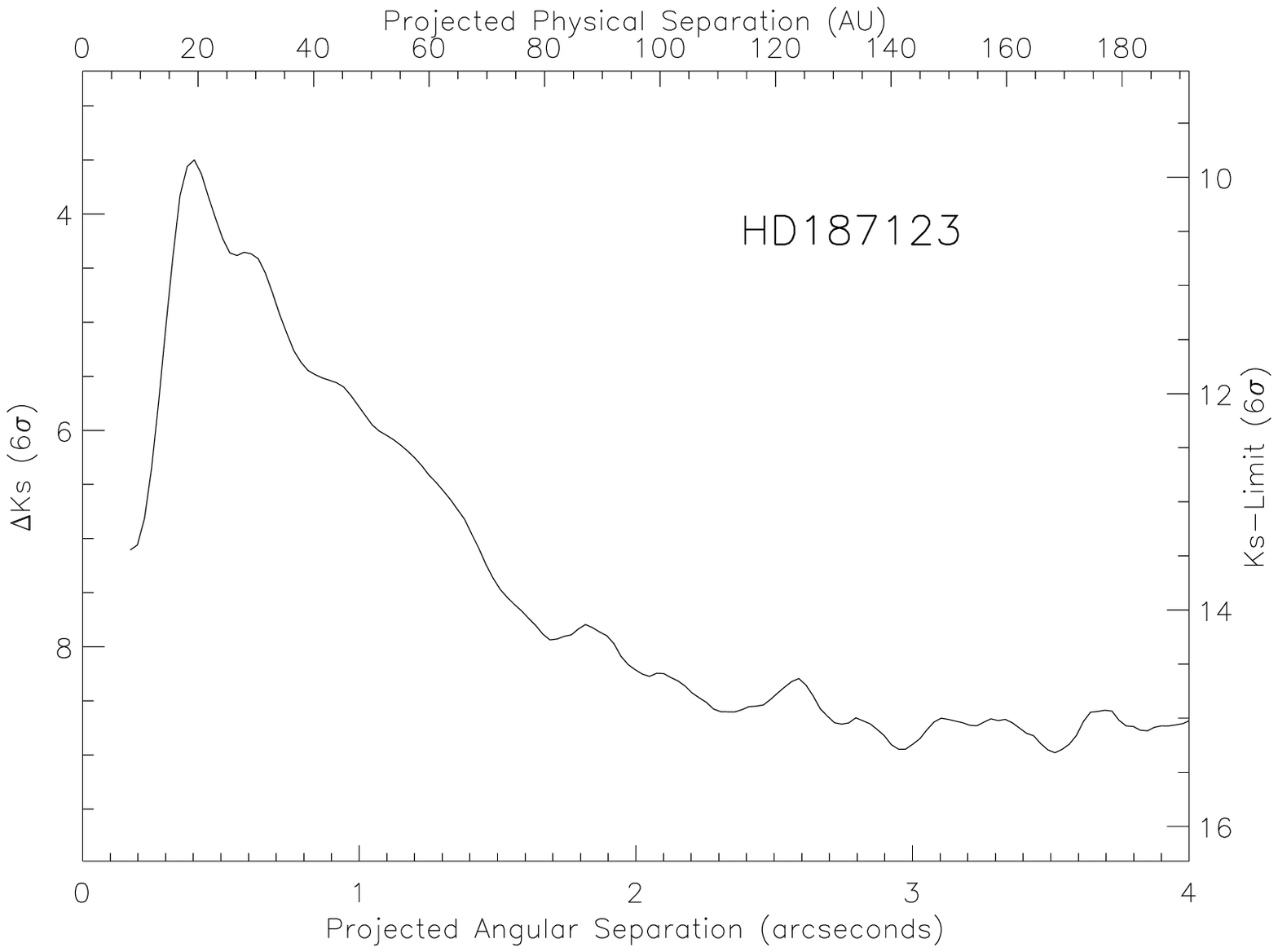}
\hspace{.6cm}
   \includegraphics[width=8cm]{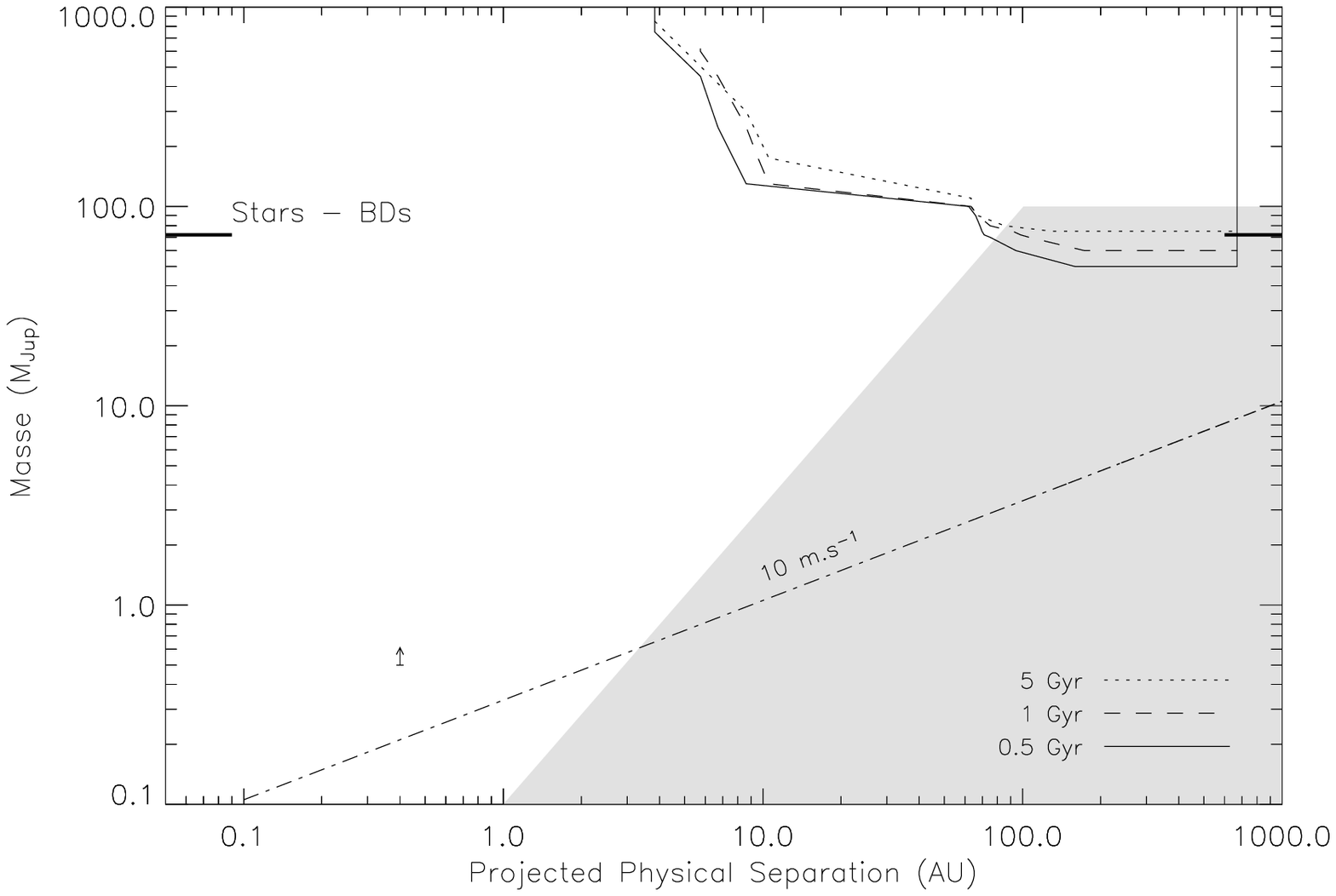}\hspace{.6cm}\\

   \includegraphics[width=8cm]{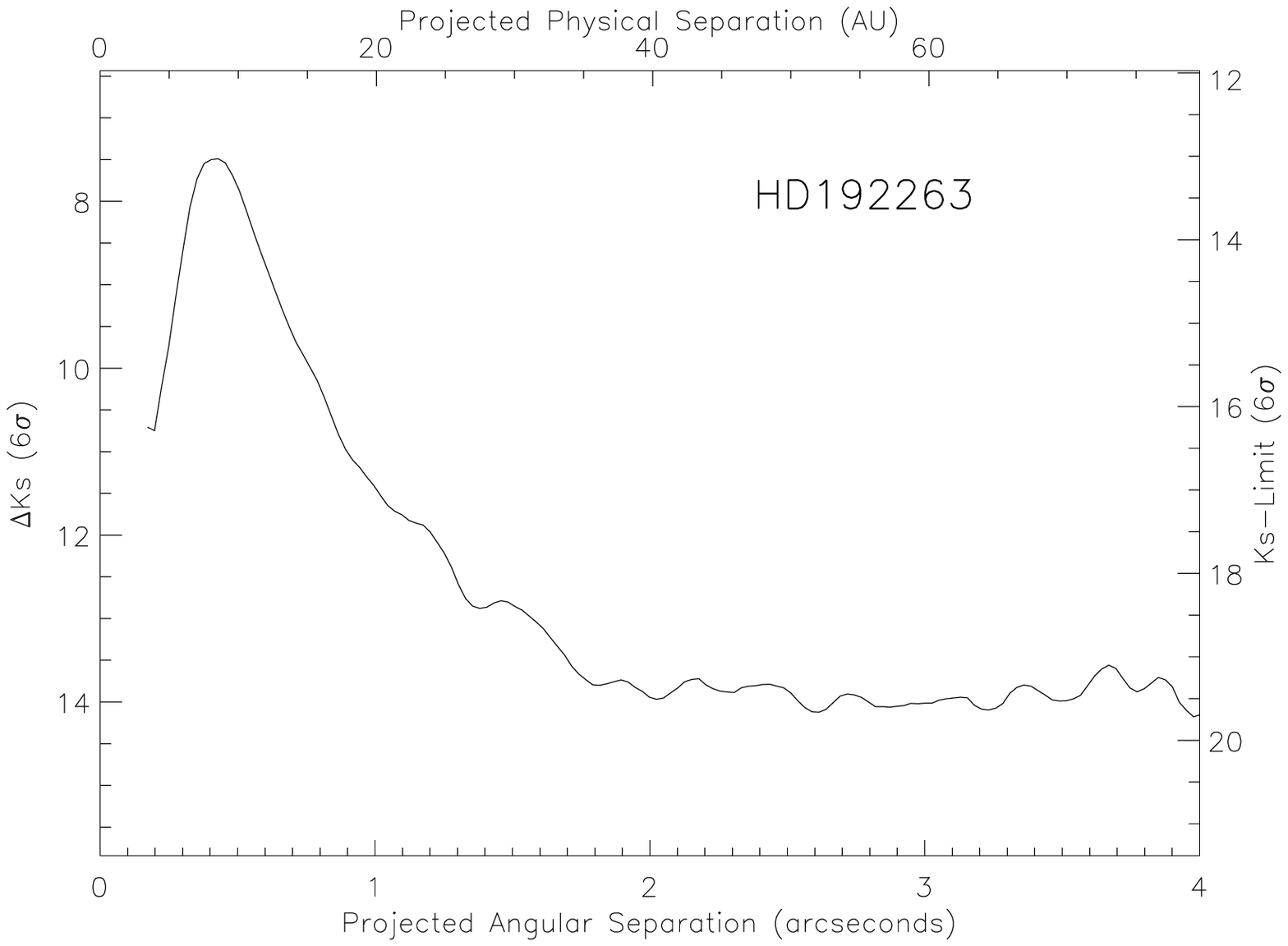}
\hspace{.6cm}
   \includegraphics[width=8cm]{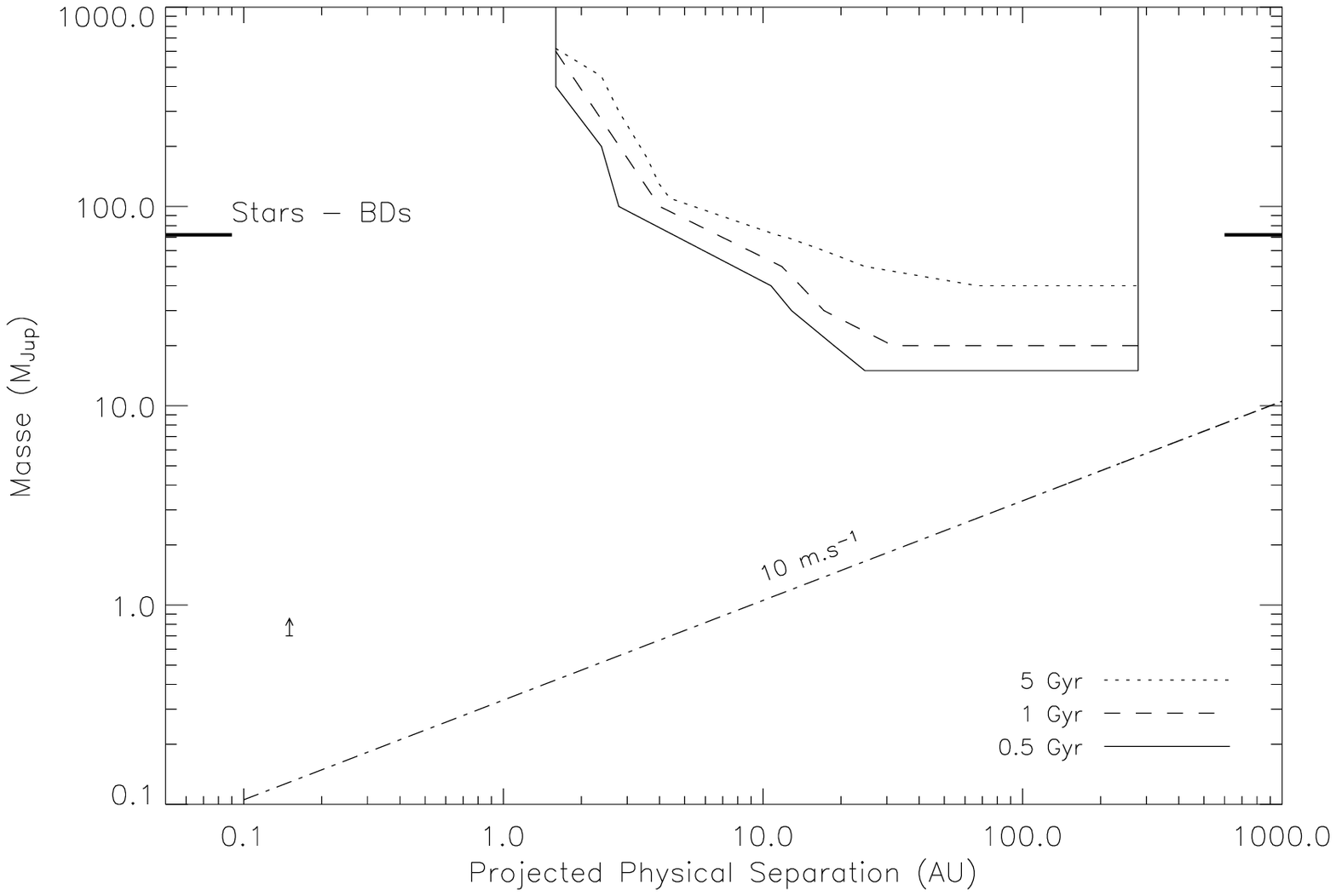}\hspace{.6cm}\\

   \includegraphics[width=8cm]{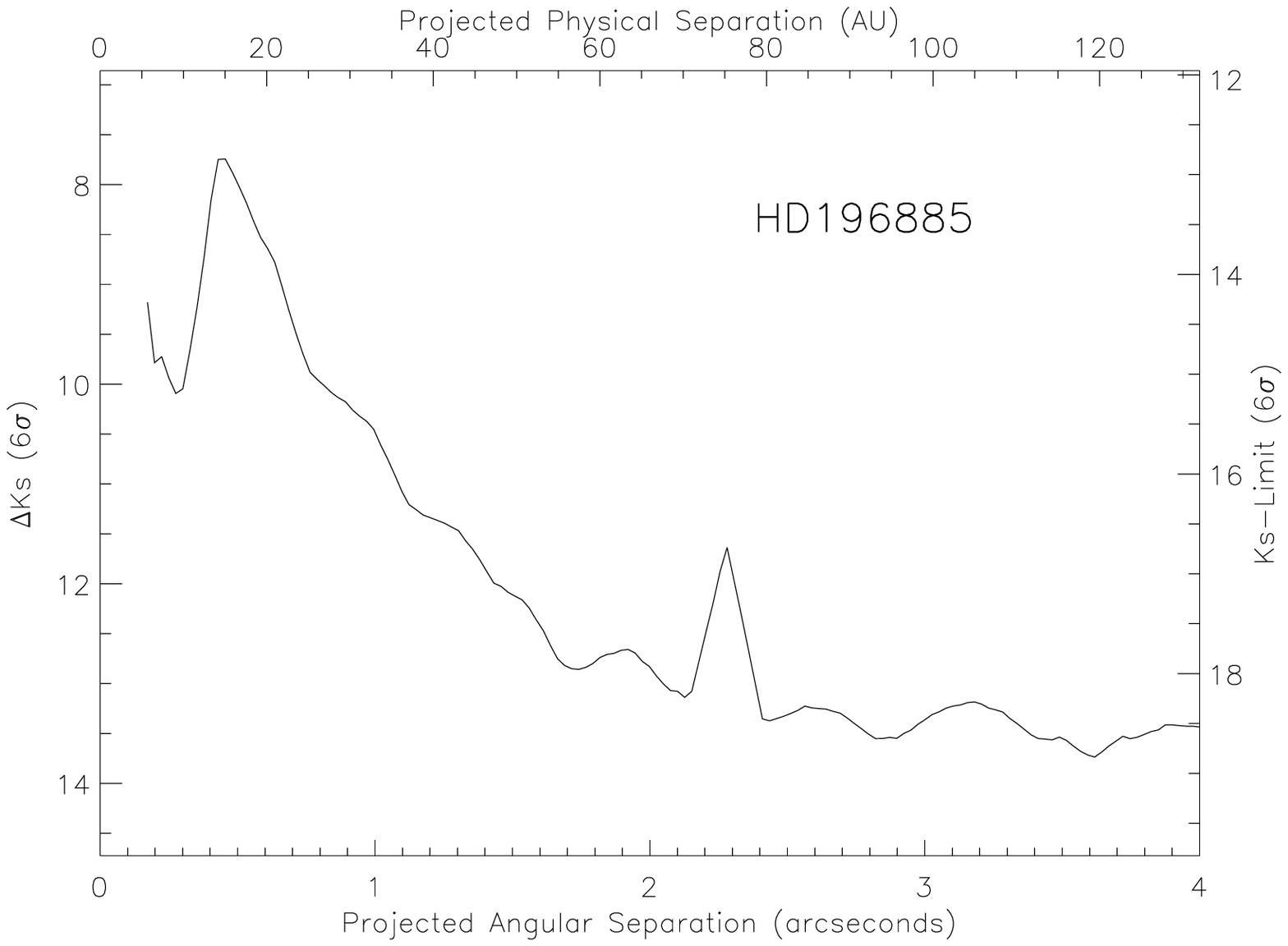}
\hspace{.6cm}
   \includegraphics[width=8cm]{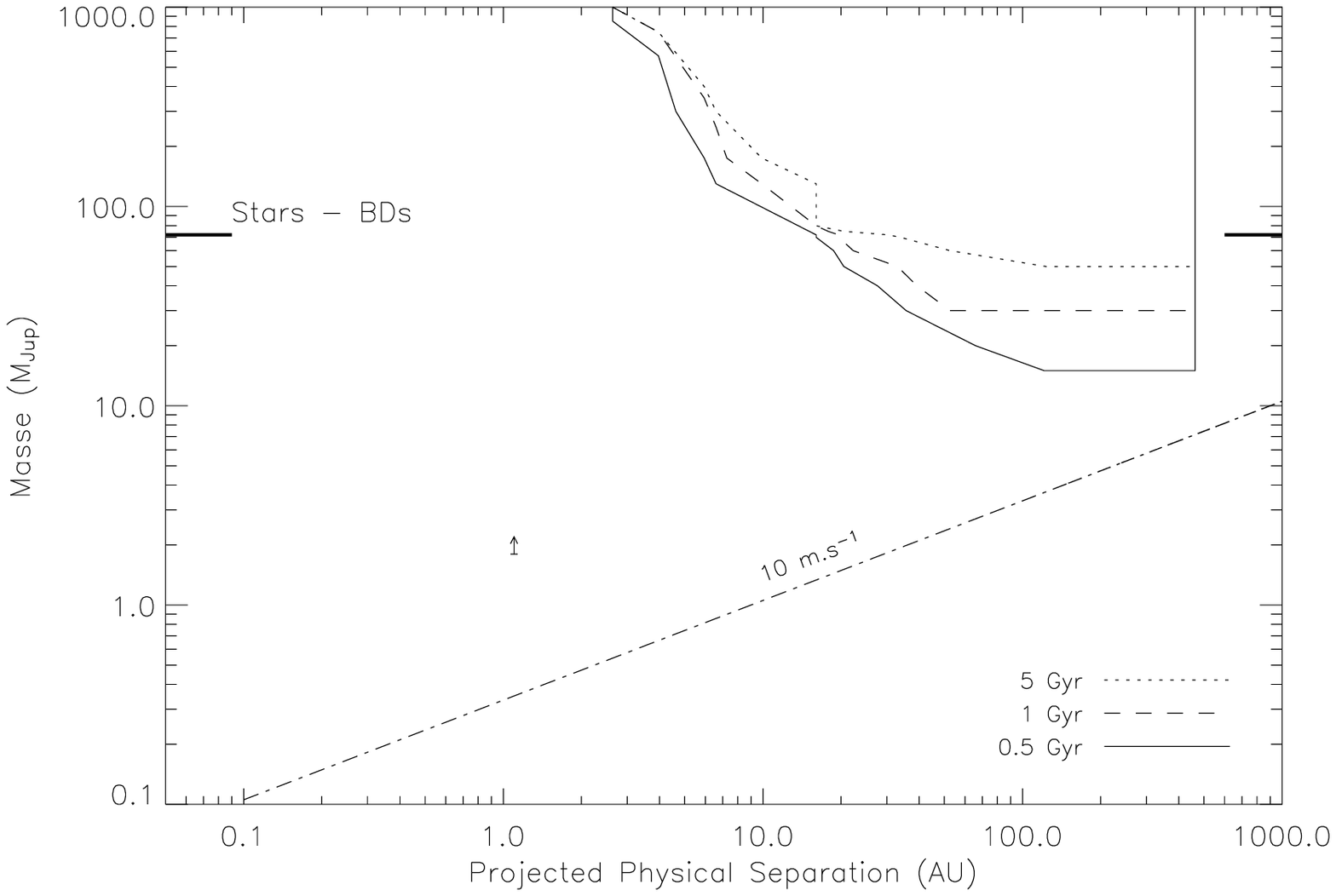}\hspace{.6cm}\\
\caption{Detection Limits of HD\,186123 or 16~Cyg~B (May 2003, CFHT/PUEO-KIR, total exposure time of 204s), HD\,187123 (July 2005, VLT/NACO, total exposure time of 49s), HD\,192263 (November 2003, VLT/NACO, total exposure time of 375s) and HD\,196885 (July 2005, VLT/NACO, total exposure time of 300s). See detail of the detection limit estimation in Section 2.2}
\label{figlim6}
\end{figure*}

\begin{figure*}

\centering

   \includegraphics[width=8.cm]{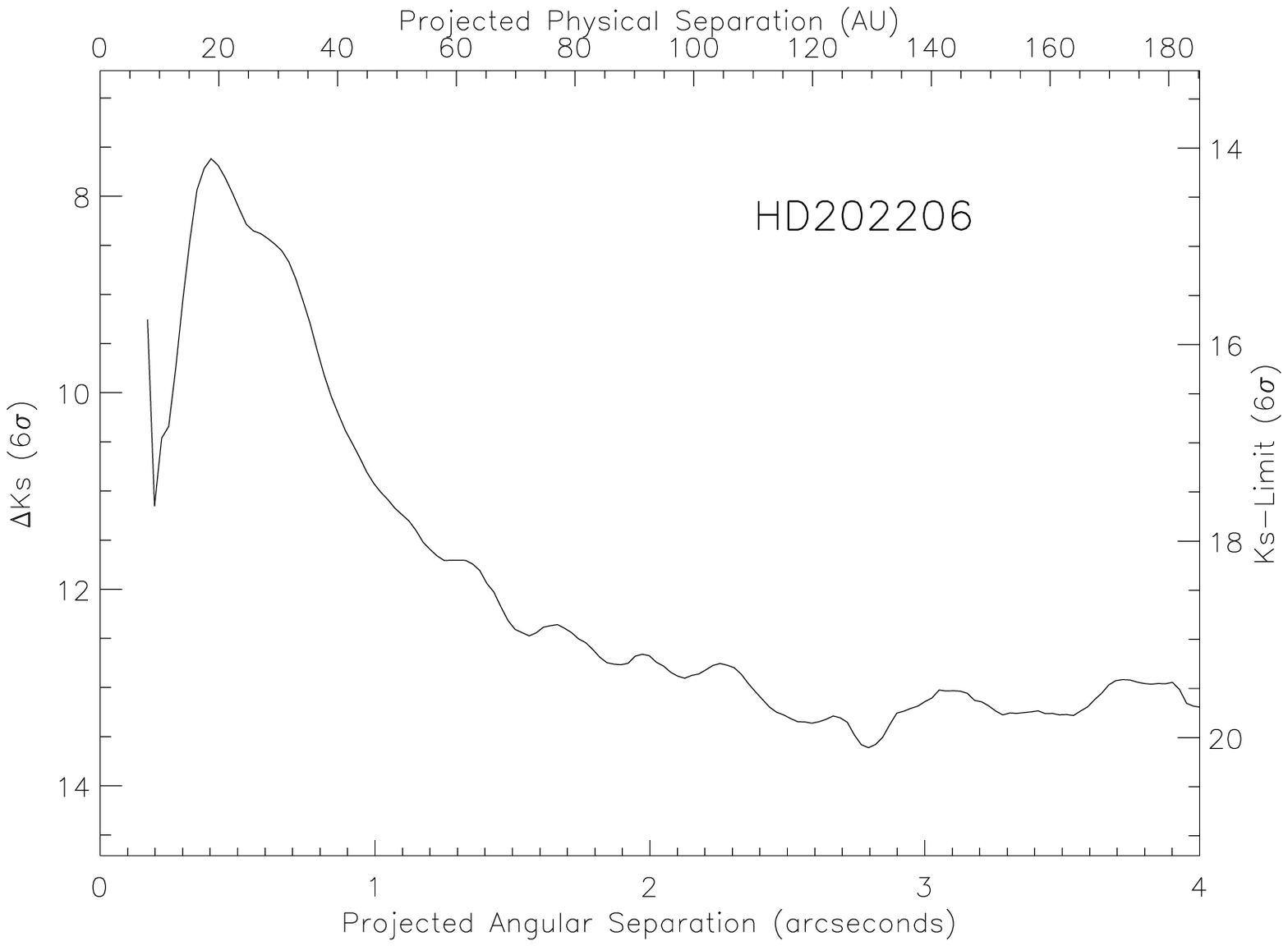}
\hspace{.6cm}
   \includegraphics[width=8.cm]{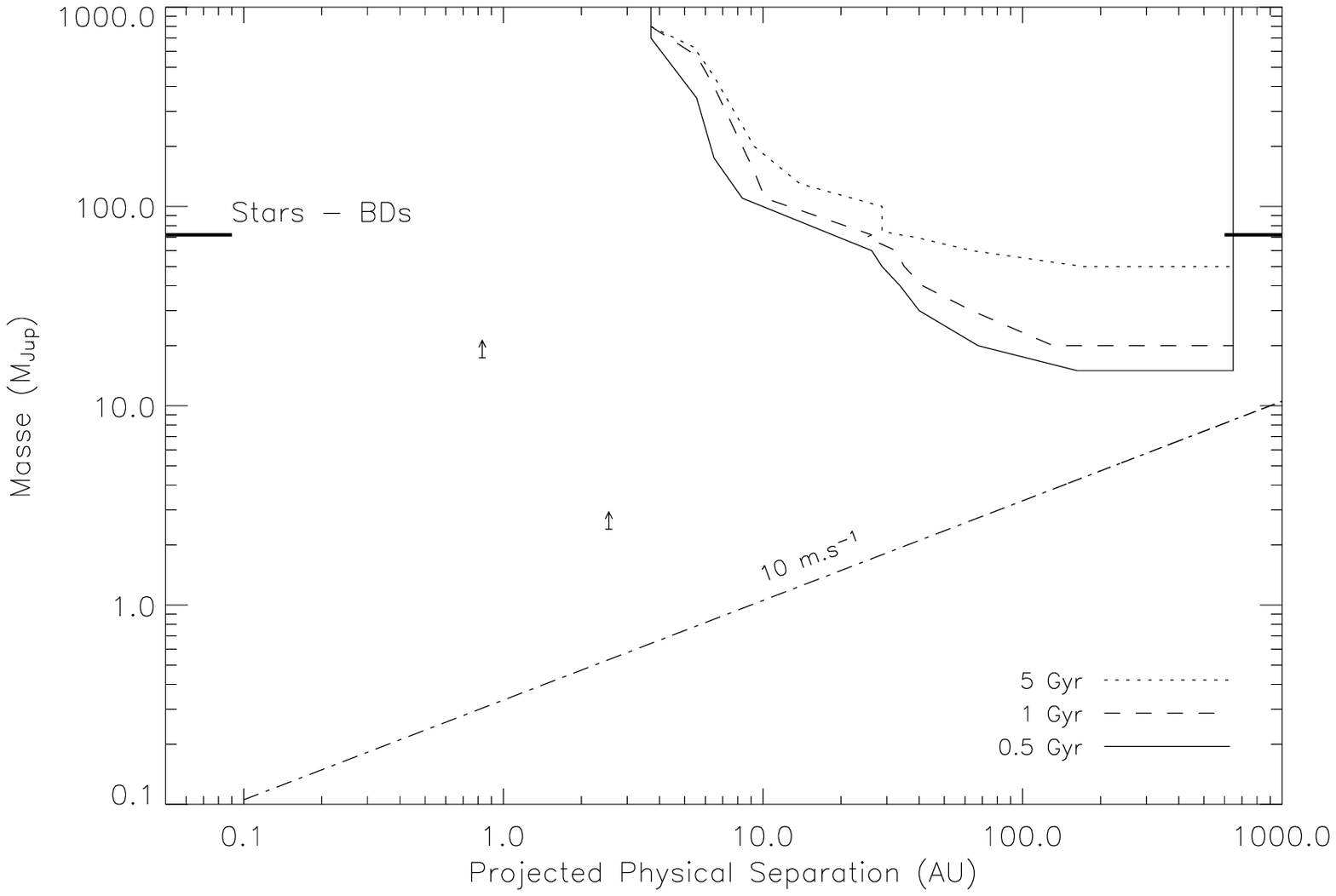}\hspace{.6cm}\\

   \includegraphics[width=8.cm]{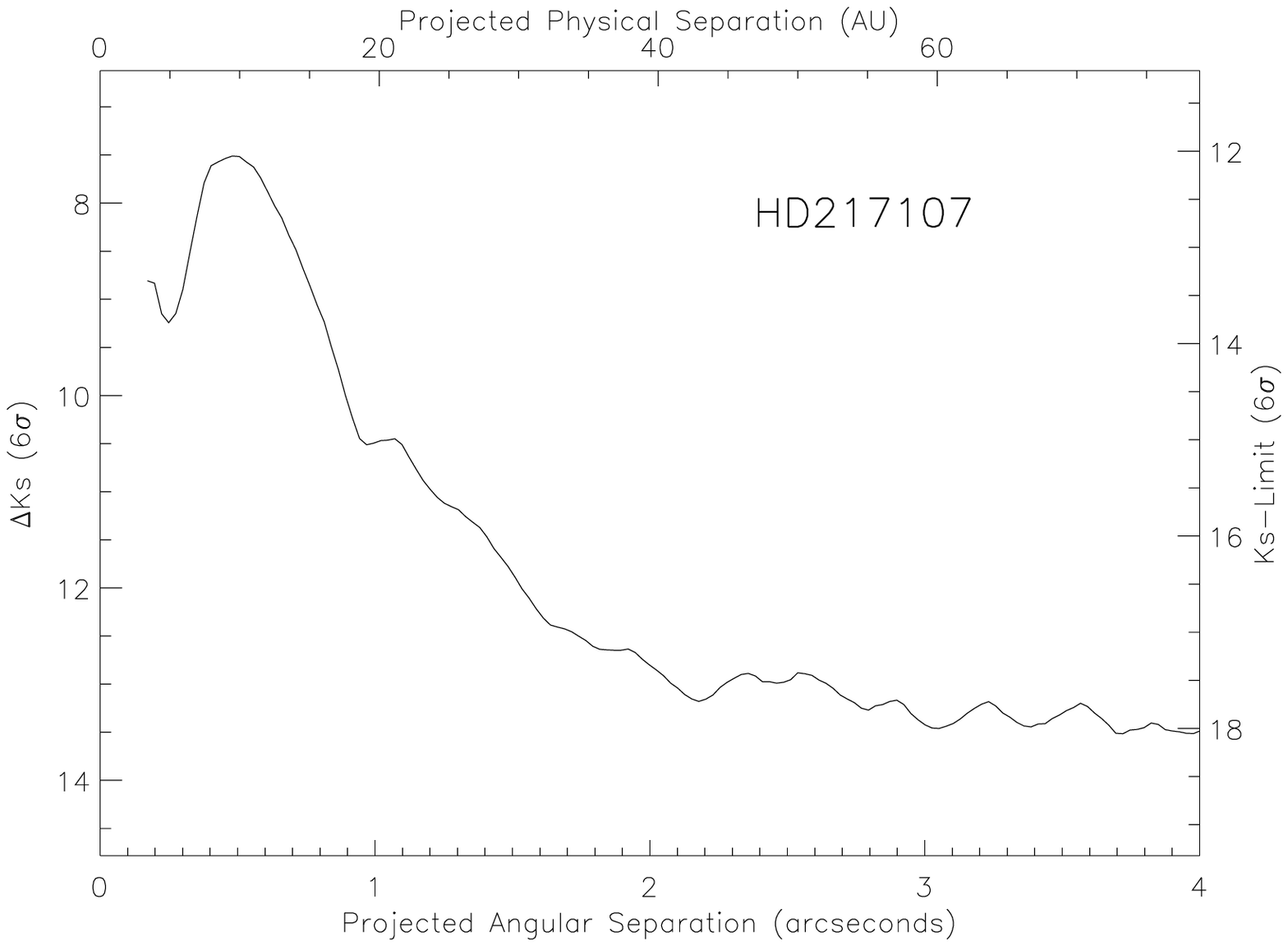}
\hspace{.6cm}
   \includegraphics[width=8.cm]{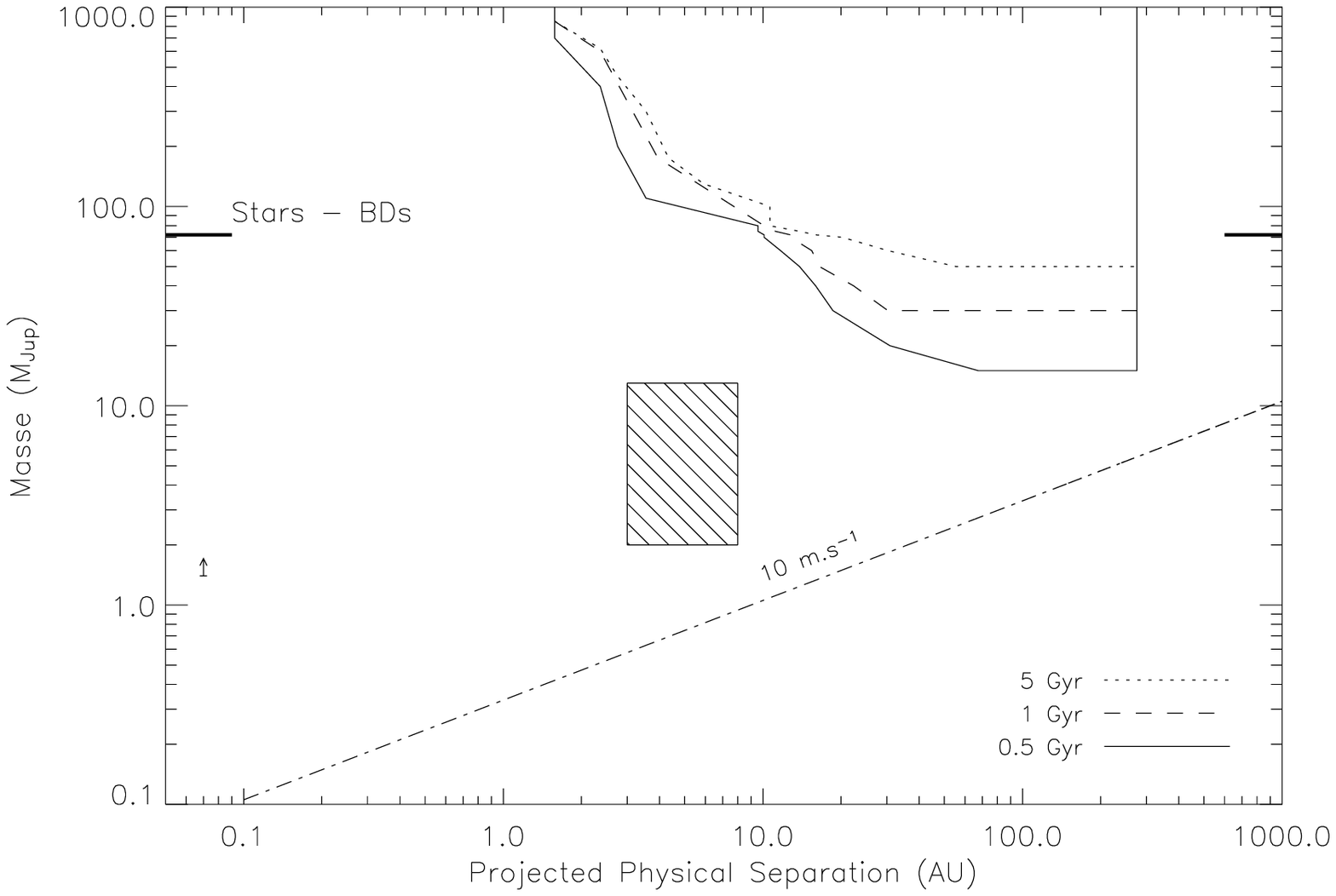}\hspace{.6cm}\\
\caption{Detection Limits of HD\,202206 (July 2005, VLT/NACO, total exposure time of 300s) and  HD\,217107 (July 2005, VLT/NACO, total exposure time of 228s). See detail of the detection limit estimation in Section 2.2. On the detection limit of  HD\,217107, converted in mass, the \textit{dashed box} give the result of the best-fit orbit from (Vogt et al. 2005), which strongly suggests that the third component HD\,217107\,c has an orbit of 3--8 AU and a mass within the planetary regime (less than 13~M$_{\rm{Jup}}$) rather than stellar or brown dwarf. }

\label{figlim7}
   \end{figure*}

The typical detection performances are reported in Table~\ref{tab:perf} for angular separations of 1 and 2~$\!''$. Using the K magnitude and the distance of the stars, we convert the typical contrast into absolute magnitude limit. Finally, based on the predictions of the BA98 (Baraffe et al. 1998) evolutionary model in the stellar regime and the COND03 (Baraffe et al. 2003) model in the substellar regime, the detection performances can be expressed in terms of mass for different ages: 0.5, 1.0, 5.0 and 10.0 Gyr. For stars as young as 0.5~Gyr, PUEO-KIR and NACO were sensitive to companions with masses down to 30 and 15\mjup respectively at 2~$\!''$. The individual detection limits for each planetary host are given in Fig.~\ref{lalim}--\ref{figlim7}.

As expected, the VLT/NACO detection limits compare favourably to those obtained at Keck (Luhman \& Jayawardhana 2002), despite the smaller aperture of the VLT.  They also compare favourably with those obtained by Masciadri et al. (2005), using NACO. In both cases, the surveys were performed using saturated images, whereas our survey uses a coronagraph, which allows deeper observations. The comparison with the simultaneous differential imaging (SDI) technique is not straightforward as the detection limit with SDI does not uniquely translate into a detection limit in terms of contrast and, thus temperature and mass. Close to the star, the detected signal is actually the subtraction of two narrow band images centered inside and outside the CH$_4$ signature. A comparison of the NACO coronagraph detection limit to the SDI one, taking into account the {\it effective} flux that remains after subtraction, shows that, when scaled to similar exposure times, the SDI does not provide a significant gain in the domain accessible to the coronagraph images ($\geq0.35~\!''$). Inside this region, given the NACO PSF, the 5~Gyr-old  objects that we can expect to detect mainly have greater temperatures than 1200~K. Therefore, their spectra present no significant CH$_4$ absorption at SDI bands.

%
\section{Discussion}

\subsection{Stars with drifts and undetected companions}

The 6 stars HD\,13445 (Gl\,86) HD\,28185, HD\,154857, HD\,187123, HD\,202206 and HD\,217107 exhibit a long term RV drift, possibly indicating the presence of long-period companions. The case of  HD\,13445 (Gl\,86) is discussed in another paper (Lagrange et al. 2006). The properties of the 5 other planetary systems are reported in Table~\ref{tab:egp}.

For the 3 systems HD\,28185, HD\,154857 and HD\,187123, we assume, for a rough estimation, that the outer companion, responsible for the RV long term drift, evolves on a circular orbit with a mass negligible compared to the primary ($\rm{M}_{3}<0.1$~M$_{\odot}$). If so, the RV data enable us to exclude a domain of mass as a function of the semi-major axis (see shadowed zones in Fig.~7, 9, 11). Considering now our imaging data, we know that no comoving companion candidates were detected for the 2 stars HD\,28185 and HD\,187123. However, contrary to the RV measurements, the imaging data do not allow us to rule out definitively a domain of mass and semi-major axis. Imaging is indeed only sensitive to projected physical separations. We can only assert that, although we were sensitive down to given minimum semi-major axis (see Table~\ref{tab:egp} for masses of 0.1 and 0.5~M$_{\odot}$), we did not detect any comoving objects that could explain the long term RV drifts of the central stars. In the case of HD\,154857 observed at one epoch, several companion candidates have been detected and future follow-up observations are needed to confirm if they are comoving and if they could be responsible for the drift.

For the two systems HD\,202206 and HD\,217107, recent RV measurements have explained the existence of the long term drift with the detection of a third body in the system. For HD\,202206, Correia et al. (2005) showed that the RV drift is caused by a lower mass planet, clearly beyond the detection performances of our imaging survey. In the case of HD\,217107,  Vogt et al. (2005) identified a significant curvature in the velocities suggesting the existence of a second possible planet, HD\,217107\,c, likely orbiting at 3--8 AU. Our non detection supports the fact that HD\,217107\,c, is unlikely to be a stellar companion.

\begin{table*}[t]
\caption[]{Offset positions of the HD\,1237 B and HD\,27442 B relative to the parent stars.}
\label{tab:astrometry}
\begin{tabular*}{\textwidth}{@{\excs}llllllllllll}
\hline\hline\noalign{\smallskip}
&             &  \multicolumn{4}{c}{\underline{\hspace{2.4cm}OBSERVED\hspace{2.4cm}}}  &   \multicolumn{2}{c}{\underline{\hspace{0.9cm}IF BACKGROUND\hspace{0.9cm}}}    &      &           \\
Name & UT Date      &    $\Delta\alpha$             & $\Delta\delta$              & Separation     &    PA           &    $\Delta\alpha$ (IF BKG)             & $\Delta\delta$ (IF BKG) & $\chi^2$ &   P($\chi^2$)\\ 
             &    &         (mas)                  &    (mas)                 &    (mas)       &  ($^o$)          &    (mas)                 &    (mas)   &   &            \\

\noalign{\smallskip}\hline\noalign{\smallskip}
HD\,1237\,B   & 03/06/2003  & $-319\pm14$    & $3844\pm16$        & $3857\pm15$       & $355.3\pm0.3$ &&&&&\\
              & 10/11/2003  & $-324\pm8$     & $3857\pm13$        & $3871\pm11$       & $355.1\pm0.2$ &   $-41\pm14$      &    $3933\pm17$   & 290  & $<1$e-9 \\
              & 29/07/2005  & $-340\pm8$     & $3870\pm13$        & $3885\pm15$       & $354.9\pm0.3$ &   $621\pm14$      & $4019\pm16$     & 3500 & $<1$e-9           \\
\noalign{\smallskip}\hline\noalign{\smallskip}
HD\,27442\,B  & 17/11/2003  & $7750\pm21$ & $10525\pm29$        &  $13070\pm21$      & $36.3\pm0.2$  &              &               &     &     \\
              & 22/09/2004  & $7796\pm18$ & $10498\pm24$        &  $13076\pm21$      & $36.6\pm0.2$  &  $7751\pm21$ &  $10637\pm29$ & 17  & 3e-4\\
              & 29/07/2005  & $7758\pm18$ & $10501\pm23$        &  $13056\pm21$      & $36.4\pm0.2$  &  $7793\pm21$ &  $10739\pm29$ & 39  & 6e-8\\
\noalign{\smallskip}\hline
\end{tabular*}
\end{table*}

\subsection{A K-dwarf companion to HD\,196885?}

HD\,196885\,A is a F8IV star, part of a wide binary system with the star BD+104351\,B located $\sim190~\!''$ North of HD\,196885\,A. HD\,196885\,A is also known to host a planet (hereafter HD\,196885\,Ab), detected through RV measurements during the Lick RV survey. Although no related publications can be found, the EGP characteristics are reported by the California \& Carnegie Planet Search Team on their webpage\footnote{\emph http://exoplanets.org/esp/hd196885/hd196885.shtml}. HD\,196885\,Ab has a minimum mass of $\rm{M}_{2}\rm{sin}\,\textit{i}=1.84~M_{\rm{Jup}}$, a period $\rm{P}=386.0$ days and an eccentricity $e=0.3$. In addition, at about $0.7~\!''$ from HD\,196885\,A, we have imaged a relatively bright companion candidate (see in Table~3). Based on the 2MASS catalog (Cutri et al. 2003) and the photometry of HD\,196885\,B, we can estimate the probability of finding HD\,196885\,B within 3~$\!''$ of the line of sight of HD\,196885\,A. Within a field of $5\times5$\,deg$^2$, centered around HD\,196885\,A, the density of nIR sources of similar JK magnitudes than HD\,196885\,B is about 19.3 sources/deg$^2$. The probability of detecting HD\,196885\,B within a radius of 3~$\!''$ of A is then $\sim1e-5$, therefore HD\,196885\,B is unlikely to be a background source.

At a distance of 33~pc measured for HD\,196885\,A (Perryman et al. 1997), the JHK absolute photometry of the visual companion HD\,196885\,B would be: M$_{\rm{J}}=9.8\pm0.2$, M$_{\rm{H}}=9.2\pm0.2$ and  M$_{\rm{K}}=8.7\pm0.2$. Using BA98 model predictions, we derive a possible mass for HD\,196885\,B of 0.6~M$_{\odot}$, using an age estimate for the system of 0.5~Gyr (Lambert et al. 2004). Therefore, HD\,196885\,B is likely to be a late-K dwarf orbiting HD\,196885\,A at a projected physical distance of 25~AU. With an estimated period of $\sim90$~yr, we should be able to monitor the orbital motion of B relative to A. Using RV measurements, this will enable us to determine the orbital characteristics of the system and to test how the inner planetary system has been affected by this massive stellar companion. Presently, the planet HD\,196885\,Ab does not show unusual physical properties that could be explained by the presence of HD\,196885\,B. The maximum drift per year, induced by an outer companion of 0.6~M$_{\odot}$ on the 1.3~M$_{\odot}$ star HD\,196885\,A, is about 110~m.s$^{-1}$.yr$^{-1}$ (assuming a circular orbit). Therefore, the influence of HD\,196885\,B should be clearly observed on the RV curve of HD\,196885\,A.
\begin{table}[t]
\caption{Photometry of the 3 systems HD\,1237\,AB, HD\,27442\,AB and HD\,196885\,AB. }
\label{tab:photometry}
\centering
\begin{tabular*}{\columnwidth}{@{\excs}llll}     
\hline\hline\noalign{\smallskip} 
Component         &  J  & H    &  Ks          \\
             &  (mag)         &  (mag)    & (mag)               \\
\noalign{\smallskip} \hline\noalign{\smallskip} 
HD\,1237\,A$^{a}$ &  $5.37\pm0.03$  &  $4.99\pm0.04$ &  $4.86\pm0.03$      \\
HD\,1237\,B$^{b}$ & $11.0\pm0.2$  &  $10.4\pm0.2$ &  $9.9\pm0.2$    \\
\hline
HD\,27442\,A$^{a}$ & $2.4\pm0.2$  & &$1.8\pm0.2$       \\
HD\,27442\,B$^{b}$ & $13.0\pm0.3$ &    &  $12.5\pm0.2$      \\
\noalign{\smallskip} \hline
HD\,196885\,A$^{b}$ & $5.5\pm0.1$  &$5.2\pm0.1$ &$5.1\pm0.1$       \\
HD\,196885\,B$^{b}$ & $9.1\pm0.1$  &$8.5\pm0.1$ &$8.2\pm0.1$       \\
\noalign{\smallskip} \hline
\end{tabular*}
\begin{list}{}{}
\item[$^{\mathrm{a}}$] from the 2MASS All-Sky Catalog of Point Sources (Cutri et al. 2003).
\item[$^{\mathrm{b}}$] from $^{\mathrm{a}}$ and NACO measurements presented in this work.
\end{list}
\end{table}

\subsection{An M-dwarf companion to HD\,1237}  

HD\,1237\,A is a bright G6V dwarf with a space velocity vector, a Lithium abundance and chromospheric activity which indicate that this star is likely a member of the 0.8~Gyr old Super Cluster of the Hyades (see Naef et al. 2001). In the course of the CORALIE RV survey, Naef et al. (2001) discovered a possible substellar companion (hereafter HD\,1237\,Ab), with a minimum mass of $\rm{M}_{2}\rm{sin}\,\textit{i}=1.94~M_{\rm{Jup}}$, a period $\rm{P}=311.29$ days and an eccentricity $e=0.24$. 

The companion HD\,1237\,B has been imaged at a projected physical separation of 68~AU from HD\,1237\,A. Fig.~\ref{figastro} displays, in a ($\Delta\alpha$, $\Delta\delta$) diagram, the offset positions of HD\,1237\,B from A, observed at 3 different epochs. The expected evolution of the relative A-B positions, under the assumption that B is a stationnary background object, is indicated in Fig.~\ref{figastro} and in Table~\ref{tab:astrometry}. With a normalized probability $\rm{P}(\chi ^2)<1e-9$ that HD\,1237\,B is a background stationary object, we find that HD\,1237\,A and B are comoving. At a distance of 17.6~pc (Perryman et al. 1997), the JHK absolute photometry of the comoving companion HD\,1237\,B is: M$_{\rm{J}}=9.8\pm0.2$, M$_{\rm{H}}=9.2\pm0.2$ and  M$_{\rm{K}}=8.7\pm0.2$. By comparison with the predictions of the BA98 model, we derive a possible mass of 0.13~M$_{\odot}$ for an age of 0.8~Gyr. This implies that HD\,1237\,B is likely a mid-M dwarf. Moreover, our astrometric observations marginally reveal an orbital motion of HD\,1237\,B relative to A (see Fig.~\ref{figastro}). Additional observations, coupled to RV measurements, should enable us to derive the orbital properties of HD\,1237\,B and to investigate whether HD\,1237\,B has affected the dynamics of the planet HD\,1237\,Ab. As for the HD\,196885 system, the planet HD\,1237\,Ab does not show any unusual physical properties which could be caused by HD\,1237\,B. The maximum drift per year that should produce HD\,1237\,B on A is 4~m.s$^{-1}$.yr$^{-1}$ (assuming a circular orbit). This presence of this drift will only be detectable on a period of several years since the chromospheric activity of HD\,1237\,A induces a RV residual of 18 m\,s$^{-1}$, which dominates the RV measurements.

\subsection{A possible white dwarf orbiting HD\,27442?}

HD\,27442\,A is a K2IV subgiant star with an age estimate of 10~Gyr (Randich et al. 1999). In the course of the Anglo-Australian RV survey, Butler et al. (2001) have detected a giant planet (hereafter HD\,27442\,Ab) to the star HD\,27442\,A. HD\,27442\,Ab has a minimum mass of $\rm{M}_{2}\rm{sin}i=1.28~M_{\rm{Jup}}$ and orbits around HD\,27442\,A with a period $\rm{P}=423.84$ days and with an eccentricity $e=0.07$. 

At $\sim 13~\!''$ (240~AU in projected physical separation) from HD\,27442\,A, we have detected the faint comoving companion HD\,27442\,B. The normalized probability that HD\,27442\,B is a background star (see Fig.~\ref{figastro} and Table~\ref{tab:astrometry}) is $\rm{P}(\chi ^2)<1e-7$. At 18.2~pc (Perryman et al. 1997), we derive a JK absolute photometry for HD\,27442\,B: M$_{\rm{J}}=11.7\pm0.2$ and M$_{\rm{K}}=11.2\pm0.2$. The position of HD\,27442\,B coincides with that of a $\rm{V} = 12.5$ source, reported in year 1930 in the Washington Visual Double Star Catalog (WDS, Worley \& Douglass 1997) at a relative position of $\Delta=13.7~\!''$ and $\rm{PA}=35^o$ from HD\,27442\,A. We have probably detected the nIR counterpart of this object. To test this hypothesis, we can examine the point where HD\,27442\,B would be expected to appear if it had been a background source. Based on the relative position of B from A in year 1930 and the proper motion of HD\,27442\,A, we find that this background source would be located at $11.4~\!''$ East and $23.5~\!''$ North from HD\,27442\,A. We do not detect any object at such a position in any catalogs or using NTT/Sofi H-band images from the ESO archive. This null detection supports the interpretation that both observations correspond to the same object. 

The visible and nIR photometry of HD\,27442\,B appears incompatible with that expected for any main sequence stars or brown dwarfs at 18.2~pc. This photometry is however consistent with that predicted by the evolutionary model of Bergeron et al. (2001) for white dwarfs with hydrogen- and helium-rich atmospheres. HD\,27442\,B is then most likely to be a white dwarf companion with a mass ranging between 0.3 and 1.2~M$_{\odot}$ and an effective temperature ranging between 9000 and 17000~K. If confirmed, this system would be a similar case to HD\,13445 (Gl\,86)~AB. Gl\,86~A is indeed a K0V star hosting  a 4~$\rm{M}_{\rm{Jup}}$ (minimum mass) planet and a probable 0.5~M$_{\odot}$ white dwarf companion orbiting at about 20~AU. Given a minimum separation of $\sim240$~AU between HD\,27442\,A and B, it is very unlikely that the latter influenced the dynamics of the inner planetary system. The maximum drift induced by a possible 0.3-1.2~M$_{\odot}$ outer companion on the 1.2~M$_{\odot}$ star HD\,27442\,A is less than 2~m.s$^{-1}$.yr$^{-1}$ (assuming a circular orbit). Improved detetection capabilities and RV measurements over several years are required in order to detect the presence of HD\,27442\,B in the RV curve of~A.


\section{Conclusions}

We have conducted a deep coronographic adaptive optics imaging survey of 26 stars with planets detected through radial velocity measurements. The domain investigated typically ranges between $0.1~\!''$ to $15~\!''$ (i.e. about 3 to 500~AU, according to the mean distance of the sample). The survey is sensitive to stellar and substellar companions with masses greater than $30~\rm{M}_{\rm{Jup}}$ (0.5~Gyr) with CFHT and  $15~\rm{M}_{\rm{Jup}}$ (0.5~Gyr) with VLT, at $2~\!''$ ($\sim60$~AU) from the primary star. The main results are that:  
\begin{itemize}
\item[-] 6 stars did no show any companion candidates: HD\,28185, HD\,92788, HD\,95128 (47 UMa), HD\,106252, HD\,186427 (16 Cyg B) and HD\,217107,
\item[-] 8 stars were surrounded by faint sources all identified as background, stationary, objects: HD\,17051, HD\,89744, HD\,130322, HD\,160601, HD\,168443, HD\,187123, HD\,192263  and HD\,202206,
\item[-] 3 stars revealed comoving objects which are likely companions: HD\,1237 , HD\,13445 (Gl\,86)  and HD\,27442. Orbital motion are detected for the companions to HD\,13445 (Gl\,86) and to HD\,1237, confirming the companionship of both objects. A close and bright stellar companion candidate is also detected to the star HD\,196885 with a probability for this object to be a background object which is very low,
\item[-] 8 stars revealed at least one faint companion candidate. These require follow-up observations to identify possible comoving companions. They are: HD\,52265, HD\,114762, HD\,121504, HD\,141937, HD\,154857, HD\,162020, HD\,179949 and HD\,183263.  
\end{itemize}

The five stars HD\,28185, HD\,154857, HD\,187123, HD\,202206 and HD\,217107 exhibit a long trend RV drift. The companion to HD\,13445 (Gl\,86) is a known physical companion, discussed in more detail in a forthcoming paper (Lagrange et al. 2006). The companions HD\,1237\,B and HD\,196885\,B are likely a mid-M dwarf and a late-K dwarf, based on a comparison of their photometry with evolutionary model predictions. Future spectroscopic observations should enable us to confirm the low stellar mass status of both objects. In addition, a dedicated monitoring of their orbital motion, to complement RV measurements, should enable us to derive the orbital parameters of these systems and investigate if these companions have affected the dynamics of the inner planetary systems. Finally, we also found that the comoving companion HD\,27442\,B is likely a white dwarf companion orbiting the 10~Gyr-old subgiant HD\,27442\,A. This would mean that this system would be very similar to the binary system HD\,13445 (Gl\,86)\,AB, with a primary which is a planetary host and the secondary, a $0.5~\rm{M}_{\odot}$ white dwarf located at $\sim20~$AU. Future spectroscopic observations are required to confirm this scenario. 

\begin{acknowledgements}
      We would like to thank the staff of ESO-VLT and CFHT and Gilles Chabrier, Isabelle Baraffe and France Allard for providing the latest update of their evolutionary models. We also acknowledge partial financial support from the {\sl Programmes Nationaux de Plan\'etologie et de Physique Stellaire} (PNP \& PNPS), in France.
\end{acknowledgements}

\end{document}